\documentclass[11pt, letterpaper]{article}

\usepackage{latexsym,multirow}
\usepackage{amssymb,amsmath, bm, nccmath, mathtools}
\usepackage{amsthm}
\usepackage{cases}
\usepackage{bm}

\allowdisplaybreaks

\usepackage[labelfont=bf]{caption}

\usepackage[T1]{fontenc}

\usepackage[backend=bibtex, authordate, giveninits=true,uniquename=mininit,natbib]{biblatex-chicago}
\usepackage{xcolor}
\definecolor{linkcolor}{HTML}{0645AD}
\usepackage[hidelinks]{hyperref}

\usepackage{bookmark}

\usepackage{xr}
\usepackage{dcolumn}
\newcolumntype{.}{D{.}{.}{-1}}
\newcolumntype{d}[1]{D{.}{.}{#1}}
\usepackage{multirow}
\usepackage{booktabs}
\usepackage{makecell}
\usepackage{tabularx}

\theoremstyle{plain}
\newtheorem{result}{Result}

\newtheorem{assumption}{Assumption}
\newtheorem{theorem}{Theorem}

\theoremstyle{plain}

\usepackage{rotating}

\usepackage{arydshln}

\usepackage[compact]{titlesec}

\usepackage{pdfpages}

\newcommand{\blind}{0}

\def\*#1{\mathbf{#1}}

\renewcommand\r{\right}
\renewcommand\l{\left}

\newcommand\E{\mathbb{E}}
\newcommand\Var{{\rm Var}}

\newcommand\cT{\mathcal{T}}

\newcommand\bX{\mathbf{X}}

\newcommand\Cov{\mbox{Cov}}

\newcommand\otau{\overline{\tau}}
\newcommand\obeta{\overline{\beta}}
\newcommand\stau{\tau^{\textsf{SA}}}
\newcommand\ostau{\otau^{\textsf{SA}}}
\newcommand\sbeta{\beta^{\textsf{SA}}}
\newcommand\osbeta{\obeta^{\textsf{SA}}}

\newcommand\wstau{\widehat{\tau}^{\textsf{SA}}}
\newcommand\wostau{\widehat{\otau}^{\textsf{SA}}}
\newcommand\wsbeta{\widehat{\beta}^{\textsf{SA}}}
\newcommand\wosbeta{\widehat{\obeta}^{\textsf{SA}}}

\newcommand\taud{\widehat{\tau}_{\texttt{DID}}}
\newcommand\taus{\widehat{\tau}_{\texttt{s-DID}}}
\newcommand\taue{\widehat{\tau}_{\texttt{e-DID}}}
\newcommand\taudd{\widehat{\tau}_{\texttt{d-DID}}}

\newcommand{\argmin}{\operatornamewithlimits{argmin}}

\newcommand\spacingset[1]{\renewcommand{\baselinestretch}%
{#1}\small\normalsize}

\spacingset{1.35}

\newcommand{\tit}{Using Multiple Pre-treatment Periods to Improve \\
  Difference-in-Differences and Staggered Adoption Designs\thanks{
  The methods proposed in this article can be implemented via the open-source statistical software \textsf{R} package
\texttt{DIDdesign} available at \href{https://github.com/naoki-egami/DIDdesign}{\texttt{https://github.com/naoki-egami/DIDdesign}}.
We are grateful to Edmund Malesky, Cuong Viet Nguyen, and Anh Tran for providing us with
data and answering our questions. We thank Adam Glynn, Chad Hazlett, Shiro Kuriwaki, Ian Lundberg,
John Marshall, Xiang Zhou, and participants of the 2019 Summer Meetings of the Political Methodology
Society and the 2019 American Political Science Association Annual Conference for helpful comments and
discussions.
We also thank the editor and our two anonymous reviewers for
providing us with valuable
comments. }
}
\begin{document}
\if0\blind
\title{\tit}
\author{
Naoki Egami\thanks{
  Assistant Professor, Department of Political Science, Columbia
  University, New York NY 10027.
  Email: \href{mailto:naoki.egami@columbia.edu}{\texttt{naoki.egami@columbia.edu}};
  URL: \href{https://naokiegami.com}{\texttt{https://naokiegami.com}}.
  }
\and
Soichiro Yamauchi\thanks{
  PhD Candidate, Department of Government, Harvard University, Cambridge MA 02138.
  Email: \href{mailto:syamauchi@g.harvard.edu}{\texttt{syamauchi@g.harvard.edu}};
  URL: \href{https://soichiroy.github.io}{\texttt{https://soichiroy.github.io}}
  }
}
\date{
  This version: February 3, 2022\\
  First draft: December 6, 2019
}

\spacingset{1.2}
\maketitle

\begin{abstract}
  While a difference-in-differences (DID) design was originally developed with
  one pre- and one post-treatment period, data from additional
  pre-treatment periods are often available. How can
  researchers improve the DID design with such multiple pre-treatment
  periods under what conditions? We first use potential outcomes to
  clarify three benefits of multiple pre-treatment periods: (1)
  assessing the parallel trends assumption, (2) improving
  estimation accuracy, and (3) allowing
  for a more flexible parallel trends assumption.
  We then propose a new estimator, \textit{double} DID, which combines
  all the benefits through the generalized method of moments and
  contains the two-way fixed effects regression as
  a special case. We show that the double DID requires a weaker
  assumption about outcome trends and is more efficient than existing DID estimators. We also generalize the double DID to the
  staggered adoption design where different units can receive the
  treatment in different time periods. We illustrate the proposed
  method with two empirical applications, covering both the basic DID
  and staggered adoption designs. We offer an open-source \textsf{R} package
  that implements the proposed methodologies.
\end{abstract}

\spacingset{1.5}

\clearpage

\section{Introduction}
\label{sec:intro}
Over the last few decades, social scientists have developed and
applied various approaches to make credible causal
inference from observational data. One of the most popular is a
difference-in-differences (DID) design \citep{bertrand2004much,
  angrist2008mostly}. When the outcome trend of the control group would have been the same as the trend of the outcome in the
treatment group in the absence of the treatment (known as the parallel trends assumption), the DID design
enables scholars to estimate causal effects even in the presence of
time-invariant unmeasured confounding \citep{abadie2005semiparametric}. In its most basic form,  we compare treatment
and control groups over two time periods --- one before and the other
after the treatment assignment.

In practice, it is common to apply the DID method with additional
pre-treatment periods.\footnote{In our literature review of \textit{American
  Political Science Review} and \textit{American Journal of Political Science}
  between 2015 and 2019, we found that about 63\% of the papers that
  use the DID design have more than one pre-treatment period. See
  Appendix~\ref{sec:review} for details about our literature review. \vspace{0.1in}}
However, in contrast to the basic
two-time-period case, there are a number of different ways to analyze
the DID design with multiple pre-treatment periods. One popular approach is
to apply the two-way fixed effects regression to the entire time
periods and supplement it with alternative model specifications by
including time-trends or leads of the treatment variable to
assess possible violations of the parallel trends assumption. Another is to stick with the two-time-period DID and limit the use of additional
pre-treatment periods only to the assessment of pre-treatment
trends.\footnote{For each approach, we provide examples in Appendix~\ref{sec:review}.}
This variation of approaches raises an important practical question: how should
analysts incorporate multiple pre-treatment periods into the DID
design, and under what assumptions? In Section~\ref{sec:benefit}, we begin by examining three benefits of
multiple pre-treatment periods using potential outcomes
\citep{imbens2015causal}: (1) assessing the parallel trends
assumption, (2) improving estimation accuracy, and (3) allowing for a more flexible parallel
trends assumption. While these benefits have been discussed in the
literature, we revisit them to clarify that each benefit requires
different assumptions and estimators.
As a result, in practice, researchers tend to enjoy only a subset of
the three benefits they can exploit from multiple pre-treatment
periods. While our literature review finds that more than 90\% of
papers based on the DID design enjoy at least one of the three
benefits, we also find that only 20\% of the papers enjoy all three benefits.

Our main contribution is to propose a new, simple estimator that
achieves all three benefits together. We use the generalized method of moments (GMM) framework
\citep{hansen1982gmm} to develop the \textit{double
  difference-in-differences} (double DID).
At its core, we combine two popular DID estimators: the standard DID estimator, which relies on the canonical parallel-trends
assumptions, and the sequential DID
estimator \citep[e.g.,][]{lee2016did, mora2019did}, which only requires that the change in the trends
is the same across treatment and control groups (what we call the
\textit{parallel trends-in-trends assumption}).
While each estimator itself is not new, the new combination of the two estimators via the GMM allows us to
optimally exploit the three benefits of multiple pre-treatment periods.

The proposed double DID approach makes several key methodological
contributions. First, we show that the proposed method achieves better theoretical
properties than widely-used DID estimators that constitute
the double DID. Compared to the standard DID estimator and the two-way fixed effects regression, the double DID has smaller standard
errors (i.e., more efficient) and is unbiased under a weaker
assumption. While the former estimators require the parallel trends
assumption, the double DID only requires the parallel trends-in-trends assumption. The double DID also
improves upon the sequential DID estimator, which is inefficient when the
parallel trends assumption holds. By using the GMM theory, we show
that the double DID is more efficient than the sequential DID when the parallel trends assumption holds. Therefore, our proposed
GMM approach enables methodological improvement both in terms of
identification and estimation accuracy.

Second, and most importantly in practice, the double DID blends all the three benefits of
multiple pre-treatment periods within a single framework. Therefore,
instead of using different estimators for enjoying each benefit as
required in existing methods, researchers can use the double DID
approach to exploit all the benefits. Given that only 20\% of papers based on the DID design currently enjoy all the
three benefits, our proposed unified approach to optimally exploit all
the three benefits of multiple pre-treatment periods is essential in practice.

We also propose three extensions of our double DID estimator. First,
we develop the double DID regression, which can incorporate
pre-treatment observed covariates to make the DID design more robust
and efficient (Section~\ref{subsubsec:d-reg}). Second, we allow for
any number of \textit{pre}- and \textit{post}-treatment periods (Section~\ref{subsec:K-did}). While the
parallel trends-in-trends assumption can allow for time-varying unmeasured confounders
that linearly change over time, we show how to further relax the
assumption by accounting for even more flexible forms of time-varying
unmeasured confounding when we have more \textit{pre}-treatment
periods. Because our proposed methods allow for any number of \textit{post}-treatment periods,
researchers can also estimate not only short-term causal effects but
also longer-term causal effects. Finally,
we generalize our double DID estimator to the staggered adoption
design where different units can receive the treatment in different
time periods (Section~\ref{sec:sad}).
This design is increasingly more popular in political science and
social sciences \citep[e.g.,][]{ben2019synthetic, athey2021design, marcus2021role}.

We offer a companion \textsf{R} package \textsf{DIDdesign} that implements the
proposed methods. We illustrate our proposed methods through two empirical
applications. In Section~\ref{subsec:basic-app}, we revisit \citet{malesky2014impact}, which study how
the abolition of elected councils affects local public services. This
serves as an example of the basic DID design where treatment assignment happens
only once. In Appendix~\ref{subsec:sa-app}, we reanalyze \citet{paglayan2019}, which
examines the effect of granting collective bargaining rights to teacher's unions
on educational expenditures and teacher's salaries. This is an example
of the staggered adoption design.

\paragraph{Related Literature.}
This paper builds on the large literature of time-series
cross-sectional data. Generalizing the well-known case of two periods
and two groups \citep[e.g.,][]{abadie2005semiparametric}, recent papers use potential
outcomes to unpack the nonparametric connection between the DID and
two-way fixed effects regression estimators, thereby proposing extensions to relax strong parametric
and causal assumptions \citep[e.g.,][]{strezhnev2018semiparametric, imai2019should, callaway2020difference,
  athey2021design, goodman2021difference,imai2020two}. Our paper also uses potential outcomes to clarify nonparametric foundations on the use
of multiple pre-treatment periods. The key difference is that, while
this recent literature mainly considers identification under the
parallel trends assumption, we study both estimation accuracy and
identification under more flexible assumptions of trends. We do so
both in the basic DID setup and in the staggered adoption design.

Another class of popular methods is the synthetic control
method \citep{abadie2010synthetic} and their recent
extensions \citep[e.g.,][]{xu2017generalized, ben2019synthetic, pang2021bayesian} that estimate a weighted
average of control units to approximate a treated unit.
As carefully noted in those papers, such methodologies require long pre-treatment periods to accurately estimate
a pre-treatment trajectory of the treated unit \citep{abadie2010synthetic}; for example, \citet{xu2017generalized}
recommends collecting more than ten pre-treatment periods. In
contrast, the proposed double DID can be applied as long as there is
more than one pre-treatment period, and is better suited when there
are a small to moderate number of pre-treatment periods.\footnote{In
  our literature review, we found that most DID applications have less than 10
  pre-treatment periods, and the median number of pre-treatment
  periods is $3.5$. See Appendix~\ref{sec:review} for more details.}  However, we
also show in Appendix~\ref{subsec:sa-app} that the double DID can
achieve performance comparable to variants of synthetic control methods even
when there are a large number of pre-treatment periods. We
offer additional discussions about relationships between our proposed
approach and synthetic control methods in Appendix~\ref{sec:compare}.

\section{Three Benefits of Multiple Pre-treatment Periods}
\label{sec:benefit}
The difference-in-differences (DID) design is one of the most widely
used methods to make causal inference from observational studies. The basic DID design consists
of treatment and control groups measured at two time periods, before
and after the treatment assignment. While the basic DID design only requires data from one post- and one pre-treatment period,
additional pre-treatment periods are often available. Unfortunately, however, assumptions behind different uses of
multiple pre-treatment periods have often remained unstated.

In this section, we use potential outcomes to discuss three well-known practical benefits of multiple pre-treatment
periods: (1) assessing the parallel trends assumption, (2) improving
estimation accuracy, and (3) allowing for a more flexible parallel trends assumption.
This section serves as a methodological foundation for developing a
new approach in Sections~\ref{sec:d-did} and~\ref{sec:sad}.

As our running example, we focus on a study of how the abolition of
elected councils affects local public services.
\citet{malesky2014impact} use the DID design to examine the
effect of recentralization efforts in Vietnam. The abolition of
elected councils, the main treatment of interest, was implemented in
2009 in about 12\% of all the communes, which are the smallest administrative
units that the paper considers. For each commune, a variety of outcomes related to public services,
such as the quality of infrastructure, were measured in 2006, 2008, and
2010. With this data, \citet{malesky2014impact} aim to estimate the
causal effect of abolishing elected councils on various measures of
local public services.

\subsection{Setup}
To begin with, let $D_{it}$ denote the binary treatment for unit $i$ in time period $t$ so that
$D_{it} = 1$ if the unit is treated in time period $t$, and $D_{it} = 0$ otherwise.
In this section, we consider two pre-treatment time periods $t \in \{0, 1\}$ and one
post-treatment period $t = 2$. We choose this setup here because it is sufficient for
examining benefits of multiple pre-treatment periods, but we also generalize our methods to an arbitrary number of \textit{pre}- and \textit{post-}
treatment periods (Section~\ref{subsec:K-did}), and to the staggered
adoption design (Section~\ref{sec:sad}). In our example, two pre-treatment
periods are 2006 and 2008, and one post-treatment period is
2010. Thus, the treatment group receives the treatment only at time
$t=2$; $D_{i0} = D_{i1} = 0$ and $D_{i2}  =  1$, whereas units in the
control group never gets treated $D_{i0} = D_{i1} = D_{i2} = 0$.
We refer to the treatment group as $G_i = 1$ and the control group as $G_i = 0.$
Outcome $Y_{it}$ is measured at time $t \in \{0, 1, 2\}.$
In addition to panel data where the same units are measured over time, the DID design accommodates
repeated cross-sectional data, in which different communes are sampled at three time periods.

To define causal effects, we rely on the
potential outcomes framework \citep{imbens2015causal}. For each time period, $Y_{it}(1)$ represents the quality of infrastructure that
commune $i$ would achieve in time period $t$ if commune $i$ had abolished elected
councils. $Y_{it}(0)$ is similarly defined.
For an individual commune, the causal effect of abolishing elected
councils on the quality of infrastructure in time period $t$ is $Y_{it}(1)  - Y_{it}(0)$. As the treatment is assigned in the second time
period, we are interested in estimating a causal effect at time $t=2$,
and a causal effect of interest is formally defined as $Y_{i2}(1) - Y_{i2}(0).$

In the DID design, we are interested in estimating the average treatment effect for
treated units (ATT) \citep{angrist2008mostly}:
\begin{equation}
  \tau = \E[Y_{i2}(1) -  Y_{i2}(0)  \mid G_i = 1], \label{eq:att}
\end{equation}
where the expectation is over units in the treatment group $G_{i} = 1$.

\subsubsection*{DID with One Pre-Treatment Period}
Before we discuss benefits of multiple pre-treatment periods from
Section~\ref{subsec:assess},  we briefly review the DID with one
pre-treatment period to fix ideas for settings with multiple
pre-treatment periods.

In the basic DID design, researchers can identify the ATT based on the widely-used
assumption of \textit{parallel trends} --- if the treatment group had not received the treatment in the second period,
its outcome trend would have been the same as the trend of the outcome in the control group.
\citep{angrist2008mostly}.
\begin{assumption}[Parallel Trends]
  \label{as-parallel}
  \begin{equation}
    \E[Y_{i2}(0) \mid G_i = 1] -  \E[Y_{i1}(0)  \mid G_i = 1]  \  =  \
    \E[Y_{i2}(0) \mid G_i = 0] -  \E[Y_{i1}(0)  \mid G_i = 0]. \label{eq:parallel}
  \end{equation}
\end{assumption}
\noindent The left-hand side of equation~\eqref{eq:parallel} is the trend in outcomes for the treatment group
$G_i = 1$, and the right is the one for
the control group $G_i = 0$. Under the parallel trends assumption, we estimate the ATT via the difference-in-differences estimator.
\begin{equation}
  \widehat{\tau}_{\texttt{DID}} = \l(\frac{\sum_{i\colon G_i = 1} Y_{i2}}{n_{12}} -
  \frac{\sum_{i\colon G_i = 1} Y_{i1}}{n_{11}}\r)  - \l(\frac{\sum_{i\colon G_i = 0} Y_{i2}}{n_{02}} -  \frac{\sum_{i\colon G_i = 0} Y_{i1}}{n_{01}}\r), \label{eq:did-est}
\end{equation}
where $n_{1t}$ and $n_{0t}$ are  the numbers of units in the treatment
and control groups at time $t \in \{1, 2\}$, respectively.

When we analyze panel data, we can compute $\widehat{\tau}_{\texttt{DID}}$
nonparametrically via a linear regression with unit and time fixed
effects. This numerical equivalence in the two-time-period case is
often used to justify the two-way fixed effects regression as the DID design
\citep{angrist2008mostly}. We discuss additional results on
nonparametric equivalence between a regression estimator and the DID
estimator in Appendix~\ref{subsec:did-reg}.

\subsection{Benefit 1: Assessing Parallel Trends Assumption}
\label{subsec:assess}
We now consider how researchers can exploit multiple pre-treatment
periods, while clarifying necessary underlying assumptions.

The first and the most common use of multiple pre-treatment periods is to
assess the identification assumption of parallel trends. As the
validity of the DID design rests on this assumption,
it is critical to evaluate its plausibility in any application. However,
the parallel trends assumption itself involves counterfactual
outcomes, and thus analysts cannot empirically test it
directly.
Instead, we often investigate whether trends for treatment and
control groups are parallel in pre-treatment periods as a placebo test
\citep{angrist2008mostly}.

\begin{figure}[!t]
  \begin{center}
    \includegraphics[width = 0.8\textwidth]{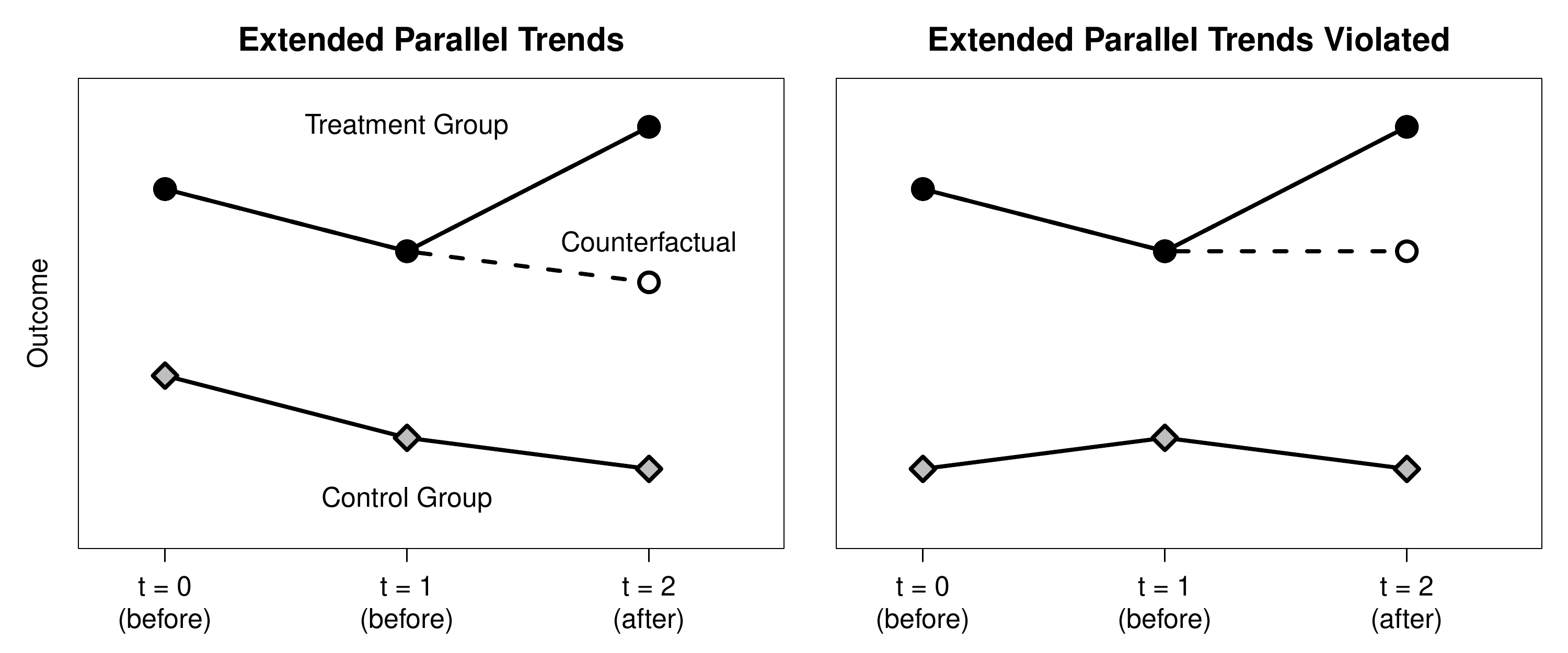}
  \end{center}
  \vspace{-0.25in} \spacingset{1}{\caption{Parallel Pre-treatment
      Trends (left) and Non-Parallel Pre-treatment Trends (right). }\label{fig:e-parallel}}
\end{figure}

Specifically, researchers often estimate the DID for the pre-treatment periods:
\begin{equation}
\l(\frac{\sum_{i\colon G_i = 1} Y_{i1}}{n_{11}} - \frac{\sum_{i\colon G_i = 1}
  Y_{i0}}{n_{10}}\r)  - \l( \frac{\sum_{i\colon G_i = 0} Y_{i1}}{n_{01}} -
\frac{\sum_{i\colon G_i = 0} Y_{i0}}{n_{00}}\r). \label{eq:pre-test}
\end{equation}
We then check whether the DID estimate on
pre-treatment periods is statistically distinguishable from zero. For example, we can
apply the DID estimator to 2006 and 2008 as if 2008 were the
post-treatment period, and assess whether the estimate would be close to
zero. In Figure~\ref{fig:e-parallel}, a DID estimate on the
pre-treatment periods would be close to zero for the left panel,
while it would be negative for the right panel where two groups have
different pre-treatment trends.
In Appendix~\ref{subsec:leads}, we show that a robustness check with leads effects \citep{angrist2008mostly},
which incorporates leads of the treatment variable into the two-way fixed effects
regression and checks whether their coefficients are zero,
is equivalent to this DID on the pre-treatment periods.

The basic idea behind this test is that if trends are parallel from 2006 to
2008, it is more likely that the parallel trends assumption holds for
2008 and 2010. Hence, instead of considering parallel trends only from
2008 to 2010, the test evaluates the two related parallel trends together.
By doing so, this popular test tries to make the DID design falsifiable.

Importantly, this approach does not test the parallel trends assumption itself
(Assumption~\ref{as-parallel}), which is untestable due to
counterfactual outcomes. Instead, it tests the \textit{extended parallel trends} assumption --- the parallel trends hold for
pre-treatment periods, from $t = 0$ to $t = 1$, as well as from a
pre-treatment period $t = 1$ to a post-treatment period $t=2$:
\begin{assumption}[Extended Parallel Trends]
  \label{as-e-parallel}
  \begin{equation}
  \begin{cases}
    \E[Y_{i2}(0) \mid G_i = 1] -  \E[Y_{i1}(0)  \mid G_i = 1]  =
    \E[Y_{i2}(0) \mid G_i = 0] -  \E[Y_{i1}(0)  \mid G_i = 0] \\
    \E[Y_{i1}(0) \mid G_i = 1] -  \E[Y_{i0}(0)  \mid G_i = 1]  =
    \E[Y_{i1}(0) \mid G_i = 0] -  \E[Y_{i0}(0)  \mid G_i = 0]
  \end{cases}
  \label{eq:ex-parallel}
\end{equation}
\vspace{-0.25in}
\end{assumption}
The first line of the extended parallel trends assumption is the same as the standard parallel trends assumption,
and the second line is the parallel trends for pre-treatment periods. Because outcome trends are observable in
pre-treatment periods, the test of pre-treatment trends
(equation~\eqref{eq:pre-test}) directly tests this assumption.

It is important to emphasize that, even if we find the DID estimate on
pre-treatment periods is close to zero, we cannot confirm the extended
parallel trends assumption (Assumption~\ref{as-e-parallel}) or the
parallel trends assumption (Assumption~\ref{as-parallel}). This is because it is still possible that
trends between $t = 1$ (pre-treatment) and $t = 2$ (post-treatment)
are not parallel. Therefore, it is always important to substantively
justify the parallel trends assumption in addition to using this statistical test based on pre-treatment trends.

\subsection{Benefit 2: Improving Estimation Accuracy}
\label{subsec:improve}
As we discussed above, many existing DID studies that utilize the test
of pre-treatment trends can be viewed as the DID design with the
extended parallel trends assumption. However, this extended parallel
trends assumption is often made implicitly, and thus, it is used only
for assessing the parallel trends assumption. Fortunately, if
the extended parallel trends assumption holds, we can also estimate
the ATT with higher accuracy, resulting in smaller standard errors.

This additional benefit becomes clear by simply restating the extended
parallel trends assumption as follows.
\begin{equation}
  \begin{cases}
    \E[Y_{i2}(0) \mid G_i = 1] -  \E[Y_{i1}(0)  \mid G_i = 1]  =
    \E[Y_{i2}(0) \mid G_i = 0] -  \E[Y_{i1}(0)  \mid G_i = 0] \\
    \E[Y_{i2}(0) \mid G_i = 1] -  \E[Y_{i0}(0)  \mid G_i = 1]  =
    \E[Y_{i2}(0) \mid G_i = 0] -  \E[Y_{i0}(0)  \mid G_i = 0].
  \end{cases}
  \label{eq:ex-parallel-2}
\end{equation}

Under the extended parallel trends assumption, there are two natural
DID estimators for the ATT.
\begin{align}
  \widehat{\tau}_{\texttt{DID}} & = \l(\frac{\sum_{i\colon G_i = 1} Y_{i2}}{n_{12}} - \frac{\sum_{i\colon G_i = 1} Y_{i1}}{n_{11}}\r)  - \l(\frac{\sum_{i\colon G_i = 0} Y_{i2}}{n_{02}} -  \frac{\sum_{i\colon G_i = 0} Y_{i1}}{n_{01}}\r), \nonumber\\
  \widehat{\tau}_{\texttt{DID(2,0)}} & = \l( \frac{\sum_{i\colon G_i = 1} Y_{i2}}{n_{12}} -
                         \frac{\sum_{i\colon G_i = 1} Y_{i0}}{n_{10}}\r)
                         - \l( \frac{\sum_{i\colon G_i = 0} Y_{i2}}{n_{02}}
                         -  \frac{\sum_{i\colon G_i = 0}
                         Y_{i0}}{n_{00}}\r). \label{eq:did-02}
\end{align}
Under the extended parallel trends assumption, both estimators are
unbiased and consistent for the ATT. Thus, we can increase estimation accuracy by combining the two estimators,
for example, simply averaging them.
\begin{equation}
  \widehat{\tau}_{\texttt{e-DID}} = \frac{1}{2}
  \widehat{\tau}_{\texttt{DID}}  +  \frac{1}{2}
  \widehat{\tau}_{\texttt{DID(2,0)}}. \label{eq:e-did}
\end{equation}
Intuitively, this extended DID estimator is more efficient because we have more
observations to estimate counterfactual outcomes for
the treatment group $\E[Y_{i2} (0) \mid G_i =  1]$.

In the panel data settings, we show that this extended DID
estimator $\widehat{\tau}_{\texttt{e-DID}}$ is equivalent to the two-way fixed
effects estimator fitted to the three periods $t \in \{0, 1, 2\}$.
\begin{equation}
  Y_{it}  \sim \alpha_i + \delta_t + \beta D_{it}, \label{eq:tfe}
\end{equation}
where $\alpha_i$ is a unit fixed effect, $\delta_t$ is a time fixed
effect, and a coefficient of the treatment variable $\beta$ is
numerically equal to $\widehat{\tau}_{\texttt{e-DID}}$. We also
present more general results about nonparametric relationships between
the extended DID and the two-way fixed effects estimator in
Appendix~\ref{subsec:e-did-reg}.

\subsection{Benefit 3: Allowing For A More Flexible Parallel Trends Assumption}
\label{subsec:allow}

In this section, we consider scenarios in which
the extended parallel trends assumption may not be plausible.
Multiple pre-treatment periods are also useful in accounting for
some deviation from the parallel trends assumption. We discuss a popular generalization of the
difference-in-differences estimator, a \emph{sequential} DID estimator,
which removes bias due to certain violations of the parallel trends
assumption \citep[e.g.,][]{lee2016did, mora2019did}.
We clarify an assumption behind this simple method and relate it to the parallel trends assumption.

To introduce the sequential DID estimator, we begin with the extended
parallel trends assumption. As we described in
Section~\ref{subsec:assess}, when the extended
parallel trends assumption holds, a DID estimator applied to
pre-treatment periods $t = 0$ and $t = 1$ should be zero in expectation.
In contrast, when trends of treatment and control groups are not
parallel, a DID estimate on pre-treatment periods would be
non-zero.
The sequential DID estimator uses this DID estimate from
pre-treatment periods to adjust for bias in the standard DID estimator. In
particular, it subtracts the DID estimator on pre-treatment periods
from the standard DID estimator that uses pre- and
post-treatment periods $t = 1$ and $t = 2.$
\begin{align}
  \widehat{\tau}_{\texttt{s-DID}} & =  \l\{ \l(\frac{\sum_{i\colon G_i = 1} Y_{i2}}{n_{12}} -
     \frac{\sum_{i\colon G_i = 1} Y_{i1}}{n_{11}}\r)  - \l(\frac{\sum_{i\colon
     G_i = 0} Y_{i2}}{n_{02}} -  \frac{\sum_{i\colon G_i = 0}
     Y_{i1}}{n_{01}}\r) \r\} \nonumber\\
  & \qquad  -  \l\{ \l(\frac{\sum_{i\colon G_i = 1} Y_{i1}}{n_{11}} -
     \frac{\sum_{i\colon G_i = 1} Y_{i0}}{n_{10}}\r)  - \l(\frac{\sum_{i\colon
     G_i = 0} Y_{i1}}{n_{01}} -  \frac{\sum_{i\colon G_i = 0}
     Y_{i0}}{n_{00}}\r) \r\}, \label{eq:s-did}
\end{align}
where the first four terms are equal to the standard DID estimator
(equation~\eqref{eq:did-est}), and the last four terms are the DID
estimator applied to pre-treatment periods $t = 0$ and $t = 1$
(equation~\eqref{eq:pre-test}).

This sequential DID estimator requires the \textit{parallel trends-in-trends}
assumption --- in the absence of the treatment, the change in the outcome trends
of the treatment group is equal to the change in the outcome trends of
the control group \citep[e.g.,][]{mora2019did}. While the parallel
trends assumption requires that the outcome trends themselves are the
same across the treatment and control groups, the \textit{parallel trends-in-trends}
assumption only requires the change in trends over time to be the
same. Formally, the parallel trends-in-trends assumption can be
written as follows.
\begin{assumption}[Parallel Trends-in-Trends]
  \label{as-parallel-tit}
  \begin{eqnarray}
    && \underbrace{
       \big\{ \E[Y_{i2}(0) \mid G_{i} = 1]  - \E[Y_{i1}(0) \mid G_i = 1] \big\}
       }_{\text{\normalfont Trend of the treatment group from $t=1$ to $t=2$}}
       - \underbrace{
       \big\{ \E[Y_{i1}(0) \mid G_{i} = 1] - \E[Y_{i0}(0) \mid G_{i} = 1] \big\}
       }_{\text{\normalfont Trend of the treatment group from $t=0$ to $t=1$}}
       \nonumber \\[10pt]
    & = &
          \underbrace{
          \big\{ \E[Y_{i2}(0) \mid G_{i} = 0]  - \E[Y_{i1}(0) \mid G_{i} = 0] \big\}
          }_{\text{\normalfont Trend of the control group from $t=1$ to $t=2$}}
          - \underbrace{
          \big\{ \E[Y_{i1}(0) \mid G_{i} = 0] - \E[Y_{i0}(0) \mid G_{i} = 0] \big\}
          }_{\text{\normalfont Trend of the control group from $t=0$ to $t=1$}}. \hspace{0.2in}
          \label{eq:parallel-tit}
  \end{eqnarray}
\end{assumption}
\noindent Here, the left-hand side represents how the outcome trends
of the treatment group change between (from $t=0$ to $t=1$) and (from
$t=1$ to $t=2$). The right-hand side quantifies the same change in
the outcome trends for the control group.

\begin{figure}[!t]
  \begin{center}
    \includegraphics[width = 0.9\textwidth]{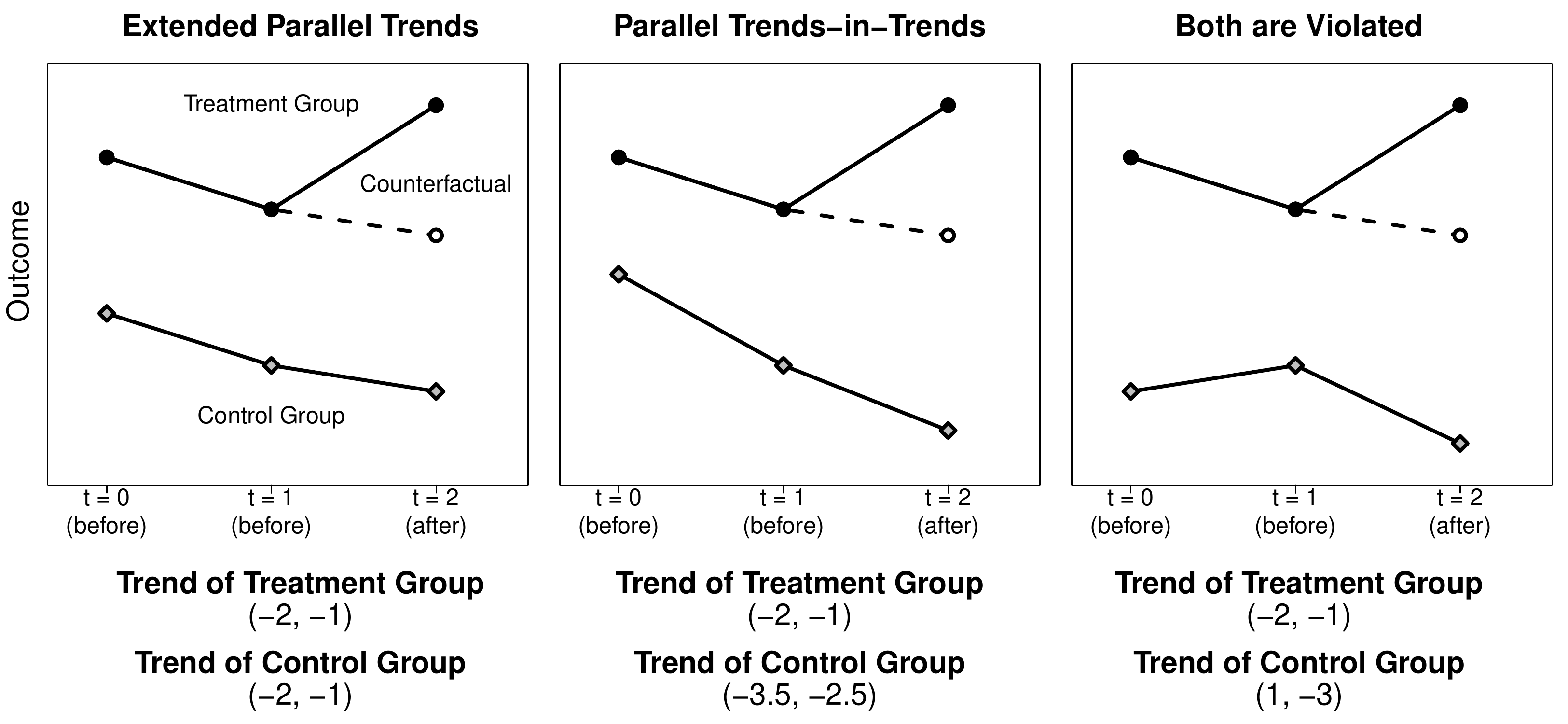}
  \end{center}
  \vspace{-0.25in} \spacingset{1}{\caption{Comparing Extended
      Parallel Trends and Parallel Trends-in-Trends Assumptions.
      \textit{Note:} Below each panel, we report the trends of the
      control potential outcomes for the treatment and
      control groups. The first
      and second elements show the outcome trends (from $t = 0$ to
      $t=1$) and (from $t=1$ to $t=2$), respectively.
      The extended parallel trends assumption (left
      panel) means that
      the outcome trends are the same across the treatment and control
      groups for both (from $t = 0$ to $t=1$) and (from $t=1$ to
      $t=2$). The parallel trends-in-trends assumption
      (middle panel) only requires its change over time is the same
      across the treatment and control groups; $(- 1) - (- 2) = (-2.5)
      - (-3.5) = 1$. Both assumptions are violated in the right panel.}\label{fig:s-parallel}}
\end{figure}

We also emphasize an alternative way to interpret the parallel
trends-in-trends assumption. Unlike the parallel trends assumption
that assumes the time-invariant unmeasured confounding, the parallel trends-in-trends
assumption can account for \textit{linear time-varying} unmeasured
confounding --- unobserved confounding increases or decreases over
time but with some constant rate. We provide examples and formal
justification of this interpretation in Appendix~\ref{subsubsec:alt-int-ptt}.

Figure~\ref{fig:s-parallel} visually illustrates that the parallel trends-in-trends assumption holds even when
the trends of the treatment and control groups are not parallel, as
long as its change over time is the same. Under the parallel trends-in-trends assumption, the sequential DID estimator is unbiased
and consistent for the ATT. Importantly, the extended parallel trends
assumption is stronger than the parallel trends-in-trends assumption, and
thus, the sequential DID estimator is unbiased and consistent for
the ATT under the extended parallel trends assumption as well.

We demonstrate that a common robustness
check of including group- or unit-specific time trends
\citep{angrist2008mostly} is nonparametrically equivalent to
the sequential DID estimator (see Appendix~\ref{subsec:s-did-reg}).
Within the potential outcomes framework, we clarified that
these common techniques are justified under the parallel
trends-in-trends assumption.

\section{Double Difference-in-Differences}
\label{sec:d-did}
We saw in the previous section that multiple pre-treatment periods provide the three related benefits.
We have clarified that each benefit requires
different assumptions and estimators, and as a result,
in practice, researchers tend to enjoy only a subset of the three
benefits.
In this section, we propose a new, simple estimator, which we call the \textit{double difference-in-differences} (double DID), that
blends all the three benefits of multiple pre-treatment periods in a
single framework. Here, we introduce the double DID with settings with
two pre-treatment periods.

We also provide three extensions. First, we propose the double DID
regression to include observed pre-treatment covariates
(Section~\ref{subsubsec:d-reg}). Second, we generalize the proposed method to any
number of \textit{pre}- and \textit{post}-treatment periods in the DID
design (Section~\ref{subsec:K-did}). Finally, we extend it to the
staggered adoption design, where the timing of the treatment assignment can
vary across units (Section~\ref{sec:sad}).

\subsection{Double DID via Generalized Method of Moments}
\label{subsec:b-d-did}
We propose the double DID estimator within a framework of the generalized method of moments (GMM)
\citep{hansen1982gmm}. In particular, we combine the standard DID
estimator and the sequential DID estimator via the GMM:
\begin{equation}
  \widehat{\tau}_{\texttt{d-DID}} = \argmin_{\tau}
  \begin{pmatrix}
    \tau - \widehat{\tau}_{\texttt{DID}}\\[-5pt]
    \tau - \widehat{\tau}_{\texttt{s-DID}}
  \end{pmatrix}^\top
  \mathbf{W}
  \begin{pmatrix}
    \tau - \widehat{\tau}_{\texttt{DID}}\\[-5pt]
    \tau - \widehat{\tau}_{\texttt{s-DID}}
  \end{pmatrix} \label{eq:d-did}
  \vspace{0.05in}
\end{equation}
where $\mathbf{W}$ is a weight matrix of dimension $2 \times 2$.

The important property of the proposed double DID estimator is
that it contains all of the popular estimators that we considered in
the previous sections as special cases. Table~\ref{tab:d-did}
illustrates that a particular choice of the weight matrix
$\mathbf{W}$ recovers the standard DID, the extended DID, and the
sequential DID estimators, respectively.

Using the GMM theory, we can estimate the optimal weight matrix
$\widehat{\*W}$ such that asymptotic standard errors of the double DID
estimator are minimized, which we describe in detail in Section~\ref{subsubsec:step2}. Therefore, users
do not need to manually pick the weight matrix $\mathbf{W}$.

\begin{table}[!t]
  \centering
  \begin{tabular}{cccc}
    \toprule
    & Standard DID & Extended DID & Sequential DID \\
    \midrule
    Weight Matrix &
                    \multirow{2}{*}{$ \begin{pmatrix}
                        1 & 0 \\
                        0 & 0
                      \end{pmatrix}
                            $}
                   & \multirow{2}{*}{$ \begin{pmatrix}
                       3 & 0 \\
                       0 & -1
                     \end{pmatrix}
                           $}
                   & \multirow{2}{*}{$ \begin{pmatrix}
                       0 & 0 \\
                       0 & 1
                     \end{pmatrix}
                           $}\\[-3pt]
    $\mathbf{W}$ & & &\\[3pt]
    \bottomrule
  \end{tabular}
  \caption{Double DID as Generalization of Popular DID Estimators.}\label{tab:d-did}
\end{table}

We emphasize that the double DID estimator provides a unifying
framework to consider identification assumptions and to estimate
treatment effects within the framework of the GMM. The double DID estimator proceeds with the following two steps.

\subsubsection{Step 1:  Assessing Underlying Assumptions}
The first step is to assess the underlying assumptions. We use this
first step to adaptively choose the weight matrix $\mathbf{W}$ in the
second step. In this first step, we check the extended parallel trends assumption by applying the DID
estimator on pre-treatment periods (equation~\eqref{eq:pre-test}) and testing whether the estimate is
statistically distinguishable from zero at a conventional
level. To take into account correlated errors, we
cluster standard errors at the level of treatment assignment.

Importantly, this step of the double DID can be viewed as the
over-identification test in the GMM framework
\citep{hansen1982gmm}, which tests whether all the moment
conditions are valid. In the context of the double DID estimator, we assume that the sequential DID estimator is correctly specified and
test the null hypothesis that the standard DID estimator is correctly
specified. Then, the null hypothesis of the over-identification test
becomes exactly the same as testing whether an estimate of the DID estimator applied to pre-treatment
periods is equal to zero.

\textbf{Equivalence Approach.}
We note that the standard hypothesis testing approach has a risk of conflating evidence for
parallel trends and statistical inefficiency. For example, when sample
size is small, even if pre-treatment trends of the treatment and
control groups differ, a test of the difference might not be
statistically significant due to large standard error, and analysts
might ``pass'' the pre-treatment-trends test. To mitigate such
concerns, we also incorporate an equivalence approach
\citep[e.g.,][]{hartman2018equi} in which we evaluate the null
hypothesis that trends of two groups are \textit{not} parallel in
pre-treatment periods.\footnote{\citet{xu2020practical} propose a
  similar test for a different class of estimators, what they refer to as ``counterfactual
  estimators.''} By using this approach, researchers can
``pass''  the pre-treatment-trends test only when estimated pre-treatment
trends of the two groups are similar with high accuracy, thereby
avoiding the aforementioned common mistake.
To facilitate the interpretation of the equivalence
confidence interval, we report the standardized interval, which can be interpreted as the standard deviation from the baseline
control mean. We provide technical details in Appendix~\ref{sec:equiv} and provide an empirical example in
Section~\ref{subsec:basic-app}.

\subsubsection{Step 2:  Estimation of the ATT}
\label{subsubsec:step2}
The second step is estimation of the ATT.
When the extended parallel trends assumption is plausible, we
estimate the optimal weight matrix $\widehat{\mathbf{W}}$ building on the theory
of the efficient GMM \citep{hansen1982gmm}.
Specifically, the optimal weight matrix that
minimizes the variance of the estimator is given by the inverse of the variance-covariance
matrix of the two DID estimators:
\begin{equation}
  \widehat{\mathbf{W}}  =
  \begin{pmatrix}
    \widehat{\Var}(\widehat{\tau}_{\texttt{DID}}) &
    \widehat{\Cov}(\widehat{\tau}_{\texttt{DID}}, \widehat{\tau}_{\texttt{s-DID}})\\
    \widehat{\Cov}(\widehat{\tau}_{\texttt{DID}},
    \widehat{\tau}_{\texttt{s-DID}}) & \widehat{\Var}(\widehat{\tau}_{\texttt{s-DID}})
  \end{pmatrix}^{-1}
  \label{eq:optimalW}
\end{equation}
While the double DID approach can
take any weight matrix, this optimal weight matrix allows us to compute the weighted average
of the standard DID and the sequential DID estimator such that the
resulting variance is the smallest. In particular, when this optimal
weight matrix is used, the double DID estimator can be explicitly written as
\begin{equation}
  \widehat{\tau}_{\texttt{d-DID}} = w_1 \widehat{\tau}_{\texttt{DID}}
  + w_2 \widehat{\tau}_{\texttt{s-DID}} \label{eq:d-did-w}
\end{equation}
where $w_1 + w_2 = 1$, and
\begin{align*}
w_1  & = \cfrac{\widehat{\Var}(\widehat{\tau}_{\texttt{s-DID}}) -
             \widehat{\Cov}(\widehat{\tau}_{\texttt{DID}}, \widehat{\tau}_{\texttt{s-DID}})}{\widehat{\Var}(\widehat{\tau}_{\texttt{DID}})
             + \widehat{\Var}(\widehat{\tau}_{\texttt{s-DID}}) - 2\widehat{\Cov}(\widehat{\tau}_{\texttt{DID}},
             \widehat{\tau}_{\texttt{s-DID}})}, \\
w_2  &=  \cfrac{\widehat{\Var}(\widehat{\tau}_{\texttt{DID}}) -
             \widehat{\Cov}(\widehat{\tau}_{\texttt{DID}}, \widehat{\tau}_{\texttt{s-DID}})}{\widehat{\Var}(\widehat{\tau}_{\texttt{DID}})
             + \widehat{\Var}(\widehat{\tau}_{\texttt{s-DID}}) - 2\widehat{\Cov}(\widehat{\tau}_{\texttt{DID}},
             \widehat{\tau}_{\texttt{s-DID}})}.
\end{align*}
By pooling information from
both the standard DID and sequential DID, the asymptotic variance of
the double DID is smaller than or equal to variance of either the
standard and sequential DIDs. This is analogous to Bayesian
hierarchical models where pooling information from multiple groups
makes estimation more accurate than separate estimation based on each group.

In addition, because the extended DID is a
special case of the double DID (as described in
Table~\ref{tab:d-did}), the asymptotic variance of the double DID is
also smaller than or equal to variance of the extended DID. Therefore,
$\Var(\taudd) \leq \mbox{min}(\Var(\taud), \Var(\taus), \Var(\taue)).$
We provide the proof
in Appendix~\ref{sec:d-d-did}.

Following \citet{bertrand2004much}, we estimate the variance-covariance matrix of $\widehat{\tau}_{\texttt{DID}}$ and $\widehat{\tau}_{\texttt{s-DID}}$ via block-bootstrap where the block is taken at the level of treatment assignment.
Specifically, we obtain a pair of two estimators $\{\widehat{\tau}^{(b)}_{\texttt{DID}}, \widehat{\tau}^{(b)}_{\texttt{s-DID}}\}$ for $b = 1, \ldots, B$ with $B$ number of bootstrap iterations, and compute the empirical variance-covariance matrix.
Given an estimate of the weight matrix (equation~\eqref{eq:optimalW}), we obtain the double DID estimate as a
weighted average (equation~\eqref{eq:d-did-w}). We can obtain the variance estimate of $\taudd$ by following the standard efficient GMM variance formula:
\begin{equation*}
  \widehat{\text{Var}}(\taudd) = (\bm{1}^{\top}\widehat{\*W} \bm{1})^{-1},
\end{equation*}
where $\bm{1}$ is a two-dimensional vector of ones.

\paragraph{Remark.}
Under the extended parallel trends assumption, both the standard
DID and the sequential DID estimator are consistent for the ATT, and thus, any weighted
average is a consistent estimator. But the optimal weight matrix
(equation~\eqref{eq:optimalW}) chooses the most efficient
estimator among all consistent estimators. As we clarify more below, we do not use the
weighted average of the standard DID and the sequential DID when the extended parallel
trends assumption is violated. \qed

When only the parallel trends-in-trends assumption
is plausible, the double DID contains one moment condition $\tau
- \widehat{\tau}_{\texttt{s-DID}} = 0$, and thus, it reduces to the
sequential DID estimator. This is equivalent to choosing the weight
matrix $\mathbf{W}$ with $W_{11} = W_{12} = W_{21} = 0$ and $W_{22} =
1$ (the third column in Table~\ref{tab:d-did}).

When both assumptions are implausible, there is no credible estimator
for the ATT
without making further stringent assumptions. However, when there
are more than two pre-treatment periods, researchers can also use the
proposed generalized $K$-DID (discussed in
Section~\ref{subsec:K-did}) to further relax the parallel trends-in-trends assumption.

\subsection{Double DID Enjoys Three Benefits}
The proposed double DID estimator naturally enjoys the three benefits of multiple pre-treatment periods within a unified framework.
\paragraph{1. Assessing Underlying Assumptions}
The double DID incorporates the assessment of
underlying assumptions in its first step as the over-identification
test. When the trends in pre-treatment periods are not parallel,
researchers have to pay the most careful attention to research
design and use domain knowledge to assess the parallel trends-in-trends assumption.

\paragraph{2. Improving Estimation Accuracy}
When the extended parallel trends assumption holds, researchers can
combine two DIDs with equal weights (i.e., the extended DID estimator, which
is numerically equivalent to the two-way fixed effects regression)
to increase estimation accuracy (Section~\ref{subsec:improve}). In this
setting, the double DID further improves estimation accuracy
because it selects the optimal weights as the GMM estimator. In Section~\ref{sec:sim}, we use simulations to show that the double DID achieves smaller standard errors
than the extended DID estimator.

\paragraph{3. Allowing For A More Flexible Parallel Trends Assumption}
Under the parallel trends-in-trends assumption, the double DID
estimator converges to the sequential DID estimator. However, when the extended
parallel trends assumption holds, the double DID uses optimal weights
and is not equal to the sequential DID. Thus, the double DID estimator
avoids a dilemma of the sequential DID --- it is consistent under a
weaker assumption of the parallel trends-in-trends but is less
efficient when the extended parallel trends assumption holds. By naturally
changing the weight matrix in the GMM framework, the double DID achieves high estimation accuracy
under the extended parallel trends assumption and, at the same time,
allows for more flexible time-varying unmeasured confounding under the
parallel trends-in-trends assumption.

\subsection{Extensions}
\subsubsection{Double DID Regression}
\label{subsubsec:d-reg}
Like other DID estimators, the double DID estimator has a nice connection
to a regression approach. We propose the double DID regression with
which researchers can include other pre-treatment covariates $\bX_{it}$ to make the DID design more robust and
efficient. We provide technical details in Appendix~\ref{sec:d-reg}.

\subsubsection{Generalized $K$-Difference-in-Differences}
\label{subsec:K-did}
We generalize the proposed method to \emph{any} number of pre-
and post-treatment periods in Appendix~\ref{sec:general}, which we call $K$-difference-in-differences ($K$-DID).
This generalization has two central benefits. First, it enables researchers to use longer \textit{pre}-treatment
periods to allow for even more flexible forms of unmeasured
time-varying confounding beyond the linear time-varying unmeasured
confounding under  the parallel trends-in-trends assumption
(Assumption~\ref{as-parallel-tit}).
$K$-DID allows for time-varying unmeasured confounding that follows a $(K-1)$th order polynomial function when researchers have $K$ pre-treatment periods.
We can view the double DID as a special case of $K$-DID because in the
double DID we have $K = 2$ pre-treatment periods, and it can allow for
unmeasured confounding that follows ($2 - 1 = 1$)st order polynomial
function (i.e., a linear function).

Second, we also allow for any number of
\textit{post}-treatment periods so that researchers can
estimate not only short-term causal effects but also longer-term
causal effects. This generalization can be crucial in many
applications because treatment effects might not have an immediate
impact on the outcome.

\subsection{Empirical Application}
\label{subsec:basic-app}
\citet{malesky2014impact} utilize the basic DID design to study how the
abolition of elected councils affects local public services in
Vietnam. To estimate the causal effects of the institutional change,
the original authors rely on data from 2008 and 2010,
which are before and after the abolition of elected councils in
2009. Then, they supplement the main analysis by assessing trends in
pre-treatment periods from 2006 to 2008. In this section, we apply the
proposed method and illustrate how to improve this basic DID design.

Although \citet{malesky2014impact} employ the exact same DID design to all of
the thirty outcomes they consider, each outcome might require
different assumptions, as noted in the original paper. Here, we focus on reanalyzing three outcomes that have different
patterns of pre-treatment periods. By doing so, we clarify how researchers can use the
double DID method to transparently assess underlying assumptions and
employ appropriate DID estimators under different settings. We provide
an analysis of all thirty outcomes in Appendix~\ref{subsec:ap_app}.

\subsubsection{Visualizing and Assessing  Underlying Assumptions}\label{sec:app-equivalence}
\begin{figure}[!t]
  \begin{center}
    \includegraphics[width=\textwidth]{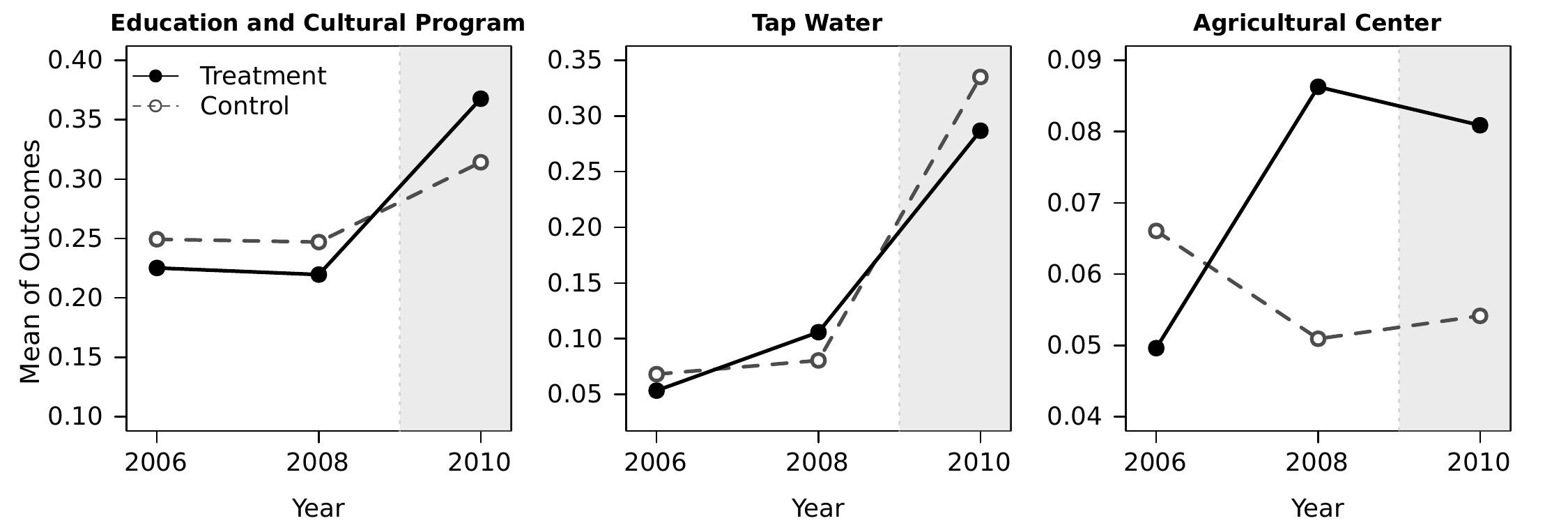}
  \end{center}
  \vspace{-0.25in} \spacingset{1}{\caption{Visualizing Trends of
      Treatment and Control Groups. \textit{Note:} We report trends
      for the treatment group (black solid line with solid circles) and the
      control group  (gray dashed line with hollow circles). Two
      pre-treatment periods are 2006 and 2008. One post-treatment
      period, 2010, is indicated by the gray shaded area.}\label{fig:assess}}
\end{figure}

The first step of the DID design is to visualize trends of treatment and
control groups. Figure~\ref{fig:assess} shows trends of three
different outcomes: ``Education and Cultural Program,''  ``Tap Water,'' and ``Agricultural
Center.''\footnote{\spacingset{1}{\footnotesize
    See Appendix~\ref{subsec:ap_app} for definitions.}} Although the original analysis uses the same DID
design for all of them, they have distinct trends in the pre-treatment periods.
The first outcome of ``Education and Cultural Program'' has parallel trends in pre-treatment periods. For the other two outcomes, trends do not look parallel in either of the cases.
While the trends for the second outcome (``Tap Water'') have similar
directions, trends for the third outcome (``Agricultural Center'') have opposite signs.
This visualization of trends serves as a transparent first
step to assess the underlying assumptions necessary for the DID estimation.

The next step is to formally assess underlying assumptions.
As in the original study, it is common to incorporate additional covariates to make the parallel trends assumption more plausible. Based on detailed domain knowledge,
\citet{malesky2014impact} include four control variables: area size of each commune, population size,
whether national-level city or not, and regional fixed effects.
Thus, we assess the conditional extended parallel trends assumption by
fitting the DID regression to
pre-treatment periods from 2006 to 2008, where $\bX_{it}$ includes the
four control variables.
If the conditional extended parallel trends assumption holds, estimates of the DID regression on pre-treatment trends should be close to zero.

While a traditional approach is to assess whether estimates are
statistically distinguishable from zero with the conventional 5\% or
10\% level, we also report results based on an equivalence approach that we recommend in Section~\ref{sec:d-did}.
Specifically, we compute the 95\% standardized equivalence confidence interval,
which quantifies the smallest equivalence range supported by the
observed data \citep{hartman2018equi}. In the context of this
application, the equivalence confidence interval is standardized based
on the mean and standard deviation of the control group in 2006. For example,
if the 95\% standardized equivalence confidence interval is $[-\nu, \nu],$ this
means that the equivalence test rejects the hypothesis that the DID
estimate (standardized with respect to the baseline control outcome) on pre-treatment periods is larger than $\nu$ or smaller than
$-\nu$ at the 5\% level. Thus, the conditional extended parallel trends
assumption is more plausible when the equivalence confidence interval is shorter.

The results are summarized in Table~\ref{tab:assess}.
Standard errors are computed via block-bootstrap at the district level, where we take 2000 bootstrap iterations.
For the first outcome, as the graphical
presentation in Figure~\ref{fig:assess} suggests, a statistical test
suggests that the extended parallel trends assumption is plausible.

For the second outcome, the test of the parallel trends reveals that the parallel trends
assumption is less plausible for this outcome than for the first
outcome. Finally, for the third outcome, both traditional and equivalence approaches provide little
evidence for parallel trends, as graphically clear in Figure~\ref{fig:assess}. Although we only have two pre-treatment periods as in the original analysis, if more than two
pre-treatment periods are available, researchers can assess the
extended parallel trends-in-trends assumption in a similar way by
applying the sequential DID estimator to pre-treatment periods. Upon
assessing the underlying parallel trends assumptions, we now proceed
to estimation of the ATT via the double DID.

\begin{table}[!t]
  \centering
  \begin{tabular}{lccccc}
  \toprule
  & Estimate & Std. Error & p-value & 95\% Std. Equivalence CI\\
  \midrule
    \makecell[cl]{Education and \\[-5pt] \quad Cultural Program}
    & $-0.007$ & 0.096 & 0.940 & $[-0.166, 0.166]$\\
    Tap Water & 0.166 & 0.083 & 0.045 & $[-0.302, 0.302]$\\
    Agricultural Center & 0.198 & 0.082 & 0.015 & $[-0.332, 0.332]$\\
  \bottomrule
  \end{tabular}
  \spacingset{1}{\caption{Assessing Underlying Assumptions Using the
      Pre-treatment Outcomes. \textit{Note:}
      We evaluate the conditional extended parallel trends
      assumption for three different outcomes. The table reports DID
      estimates on pre-treatment trends, standard errors,
      p-values, and the 95\% standardized equivalence confidence intervals.}\label{tab:assess}}
\end{table}

\subsubsection{Estimating Causal Effects}
\begin{figure}[!t]
  \begin{center}
    \includegraphics[width=\textwidth]{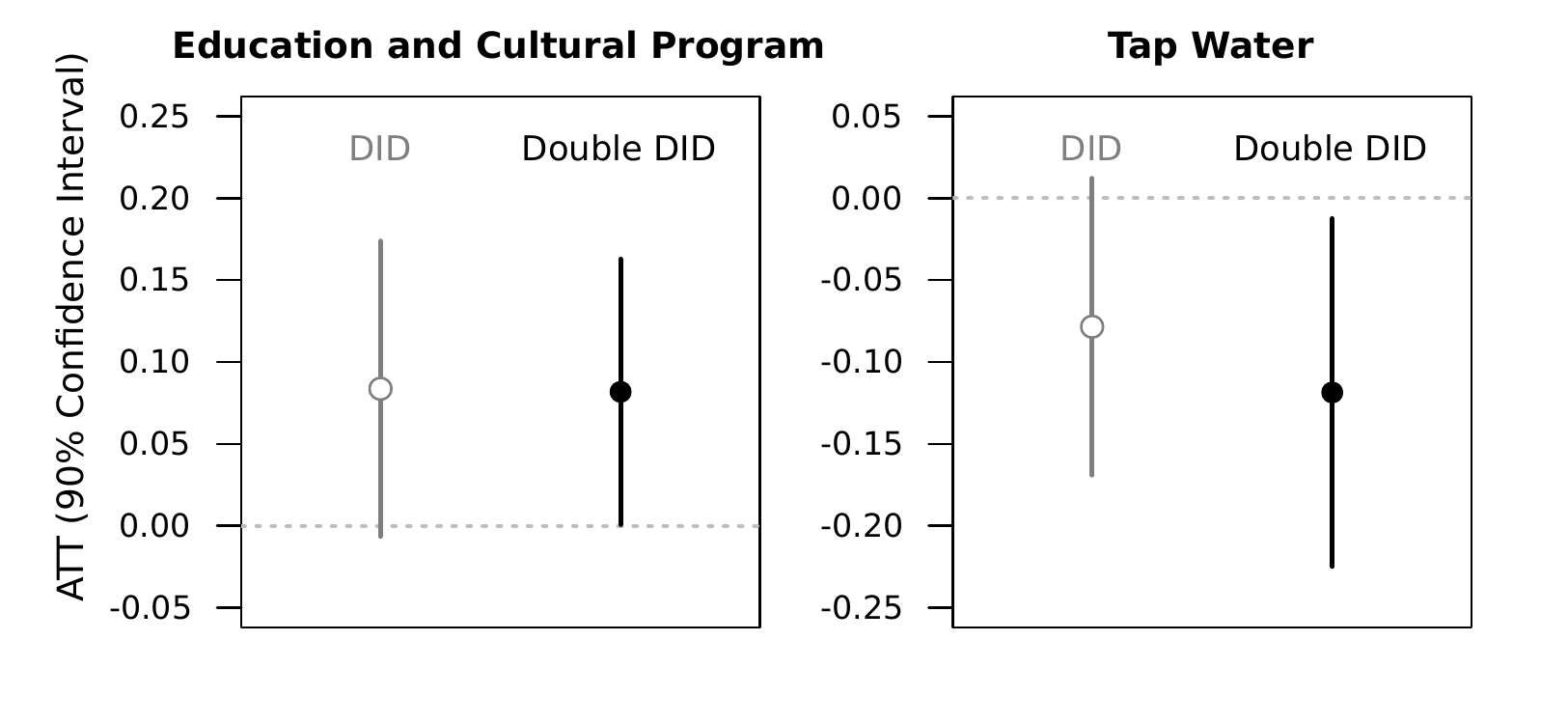}
  \end{center}
  \vspace{-0.25in} \spacingset{1}{\caption{Estimating Causal Effects
      of Abolishing Elected Councils.
      \textit{Note:} We compare estimates from the standard DID and the proposed double DID.
    }\label{fig:estimate}}
\end{figure}

Within the double DID framework, we select appropriate DID estimators
after the empirical assessment of underlying assumptions.
For the first outcome, diagnostics in the previous section suggest that the extended parallel trends assumption is plausible.
In such settings, the double DID is expected to produce similar point estimates with smaller standard errors compared to the conventional DID estimator.
The first plot of Figure~\ref{fig:estimate} clearly shows this pattern.
In the figure, we report point estimates as well as 90\% confidence intervals following the original paper (see Figure 3 in \cite{malesky2014impact}).
Using the standard DID estimator, the original estimate of
the ATT on ``Education and Cultural Program'' was $0.084$
(90\% CI = [$-0.006, 0.174$]).
Using the double DID estimator, an estimate is instead
$0.082$ (90\% CI = [$0.001, 0.163$]). By using the double DID
estimator, we shrink standard errors by about 10\%.
Although we only have two pre-treatment periods here, when there are more pre-treatment periods, efficiency gain of the double DID can be even larger.

For the second outcome, we did not have enough evidence to support the extended parallel trends assumption.
Thus, instead of using the standard DID as in the original analysis,
we rely on the parallel trends-in-trends assumption. In this case, the double DID estimates the ATT by allowing for linear time-varying unmeasured confounding in contrast to the standard DID that assumes constant unmeasured confounders.
The second plot of Figure~\ref{fig:estimate} shows the important difference between the two methods. While the standard DID estimates is $-0.078$ (90\% CI = [$-0.169, 0.012$]), the
double DID estimate is $-0.119$ (90\% CI = [$-0.225, -0.012$]).
Given that the extended parallel trends assumption is not plausible, this result
suggests that the standard DID suffers from substantial bias (the
bias of $0.04$ corresponds to more than 50\% of the original point estimate).
By incorporating non-parallel pre-treatment trends, the double DID shows that
the original DID estimate was underestimated by a large amount.

Finally, for the third outcome, the previous
diagnostics suggest that the extended parallel trends assumption is
implausible. It is possible to use the double DID under the parallel
trends-in-trends assumption. However, trends of treatment and control
groups have opposite signs, implying the double DID
estimates are highly sensitive to the parallel trends-in-trends
assumption. Given that the parallel trends-in-trends assumption is
also difficult to justify here, there is no credible estimator of the ATT
without making additional stringent assumptions. While we
focused on the three outcomes here, the double DID improves upon the
standard DID in a similar way for the other outcomes as well (see Appendix~\ref{subsec:ap_app}).

\section{Staggered Adoption Design}\label{sec:SAD}
\label{sec:sad}
In this section, we extend the proposed double DID estimator to
the staggered adoption design where the timing of the treatment assignment can
vary across units \citep{strezhnev2018semiparametric, ben2019synthetic, athey2021design}.

\subsection{The Setup and Causal Quantities of Interest}\label{subsec:setup-sad}
\begin{figure}[!t]
  \begin{center}
    \includegraphics[width = 0.7\textwidth]{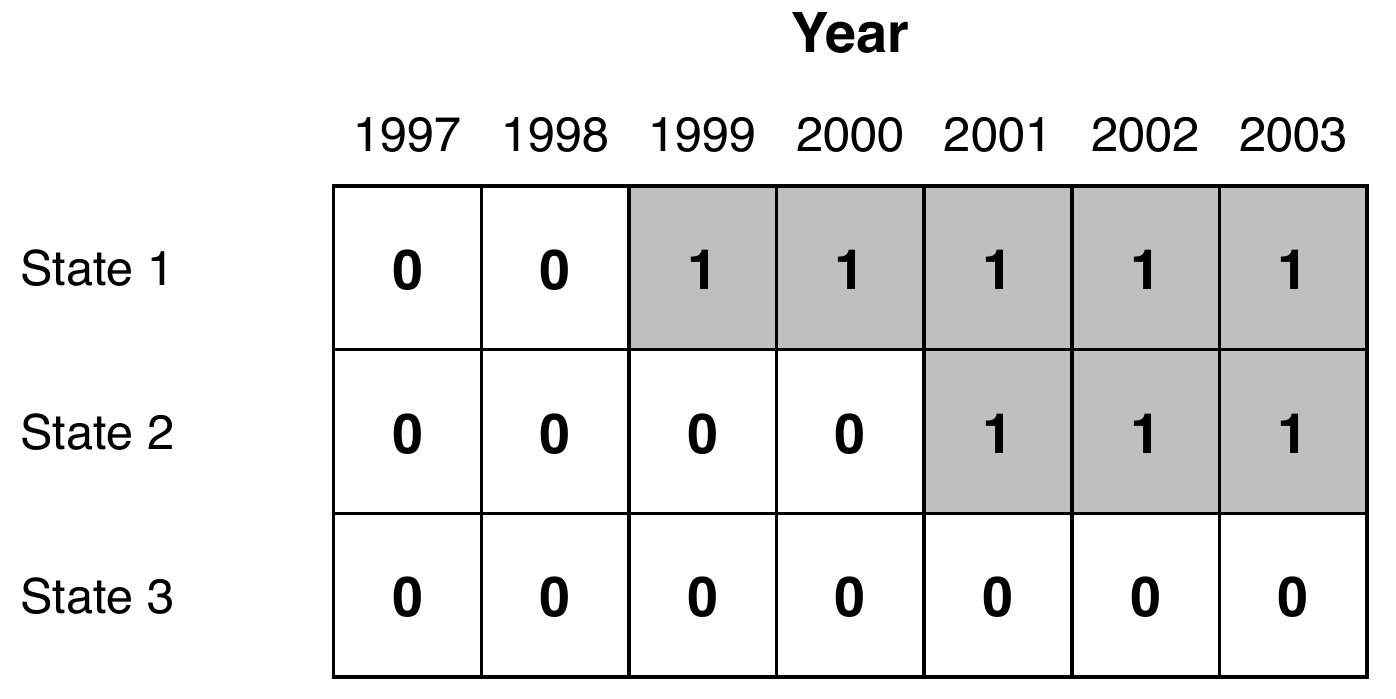}
  \end{center}
  \vspace{-0.25in} \spacingset{1}{\caption{Example of the Staggered Adoption
      Design. \textit{Note:} We use gray cells of ``1'' to denote the
      treated observation and use white cells of ``0'' to denote the
      control observation. }\label{fig:SA}}
\end{figure}

In the staggered adoption (SA) design, different units can receive the
treatment in different time periods. Once they receive the treatment, they remain exposed to the treatment
afterward. Therefore, $D_{it} = 1$ if $D_{im} = 1$ where $m < t.$
We can thus summarize information about the treatment assignment by the timing of the
treatment $A_i$ where $A_i \equiv \mbox{min} \ \{t: D_{it} =
1\}.$ When unit $i$ never receives the treatment until the end of time
$T$, we let $A_i = \infty.$ For example, in many applications where researchers are
interested in the causal effect of state- or local-level policies,
units adopt policies in different time points and remain
exposed to such policies once they introduce the policies.
In Appendix~\ref{subsec:sa-app}, we provide its example based on \citet{paglayan2019}. See Figure~\ref{fig:SA} for
visualization of the SA design.

Following the recent literature on the SA design,
we make two standard assumptions in the SA design: no anticipation assumption and
invariance to history assumption \citep{imai2019should, athey2021design}. This implies that, for unit $i$ in period $t$, the
potential outcome $Y_{it}(1)$ represents the outcome of
unit $i$ that would realize in period $t$ if unit $i$ receives the
treatment at or before period $t$. Similarly, $Y_{it}(0)$ represents the outcome of
unit $i$ that would realize in period $t$ if unit $i$ does not receive
the treatment by period $t$. Finally, we generalize group indicator $G$ as follows.
\begin{equation}
  G_{it} \ = \
  \begin{cases}
    \ \ 1 & \mbox{if  } A_i = t \\
    \ \ 0 & \mbox{if  } A_i > t \\
    \ \ -1 & \mbox{if  } A_i < t
  \end{cases}\label{eq:group-sad}
\end{equation}
where $G_{it} = 1$ represents units who receive the treatment at time
$t$, and $G_{it} = 0$ ($G_{it} = -1$) indicates units who receive
the treatment after (before) time $t$.

Under the SA design, the \textit{staggered adoption ATT} (SA-ATT) at time
$t$ is defined as follows.
\begin{equation*}
  \stau(t) = \E[Y_{it}(1) -  Y_{it}(0)  \mid G_{it} = 1],
\end{equation*}
which represents the causal effect of the treatment in period $t$ on units with
$G_{it} = 1$, who receive the treatment at time $t.$ This is a
straightforward extension of the standard ATT (equation~\eqref{eq:att})
in the basic DID setting. Researchers might also be interested in
the \textit{time-average staggered adoption ATT} (time-average SA-ATT).
\begin{equation*}
  \ostau = \sum_{t \in \cT} \pi_t \stau(t),
\end{equation*}
where $\cT$ represents a set of the time periods for which
researchers want to estimate the ATT.
For example, if a researcher is interested in estimating the ATT for the entire sample periods,
one can take  $\cT = \{1, \ldots, T\}$.
The SA-ATT in period $t$, $\stau(t),$ is weighted by the proportion of units who receive the treatment at time
$t$: $\pi_t = \sum_{i=1}^n \mathbf{1}\{A_i = t\}/\sum_{i=1}^n
\mathbf{1}\{A_i \in \cT\}$.

\subsection{Double DID for Staggered Adoption Design}
\label{subsec:sa-d-did}
Under what assumptions can we identify the SA-ATT and the time-average
SA-ATT? Here, we first extend the standard DID estimator under the
parallel trends assumption and the sequential DID estimator under the parallel trends-in-trends
assumption to the SA design. Formally, we define the standard DID
estimator for the SA-ATT at time $t$ as \\[-3pt]
\begin{equation*}
\wstau_{\texttt{DID}}(t) = \l(\frac{\sum_{i\colon G_{it} = 1} Y_{it}}{n_{1t}} -
  \frac{\sum_{i\colon G_{it} = 1} Y_{i,t-1}}{n_{1,t-1}}\r)  -
  \l(\frac{\sum_{i\colon G_{it} = 0} Y_{it}}{n_{0t}} -
  \frac{\sum_{i\colon G_{it} = 0} Y_{i,t-1}}{n_{0,t-1}}\r), \vspace{10pt}
\end{equation*}
which is consistent for the SA-ATT under the following parallel trends assumption in period $t$ under the SA design:
\begin{equation*}
  \E[Y_{it}(0) \mid G_{it} = 1] -  \E[Y_{i,t-1}(0)  \mid G_{it} = 1]  \  =  \
  \E[Y_{it}(0) \mid G_{it} = 0] -  \E[Y_{i,t-1}(0)  \mid G_{it} = 0]. \label{eq:parallel-sad}
\end{equation*}

Similarly, we can define the sequential DID estimator for the SA-ATT
at time $t$ as
\begin{align*}
\wstau_{\texttt{s-DID}}(t) & =  \l\{ \l(\frac{\sum_{i\colon G_{it} = 1} Y_{it}}{n_{1t}} -
     \frac{\sum_{i\colon G_{it} = 1} Y_{i,t-1}}{n_{1,t-1}}\r)  - \l(\frac{\sum_{i\colon
     G_{it} = 0} Y_{it}}{n_{0t}} -  \frac{\sum_{i\colon G_{it} = 0}
     Y_{i,t-1}}{n_{0,t-1}}\r) \r\} \nonumber\\
  &\qquad  -  \l\{ \l(\frac{\sum_{i\colon G_{it} = 1} Y_{i,t-1}}{n_{1,t-1}} -
     \frac{\sum_{i\colon G_{it} = 1} Y_{i,t-2}}{n_{1,t-2}}\r)  - \l(\frac{\sum_{i\colon
     G_{it} = 0} Y_{i,t-1}}{n_{0,t-1}} -  \frac{\sum_{i\colon G_{it} = 0}
     Y_{i,t-2}}{n_{0,t-2}}\r) \r\},
\end{align*}
which is consistent for the SA-ATT under the following parallel
trends-in-trends assumption in period $t$ under the SA design:
\begin{align*}
  & \{\E[Y_{it}(0) \mid G_{it} = 1] -  \E[Y_{it}(0)  \mid G_{it} = 0]\}
                     - \{\E[Y_{i,t-1}(0) \mid G_{it} = 1] -  \E[Y_{i,t-1}(0)  \mid G_{it} = 0]\}
                     \nonumber \\
  &= \{\E[Y_{i,t-1}(0) \mid G_{it} = 1] -  \E[Y_{i,t-1}(0)  \mid G_{it} = 0]\}  -  \{\E[Y_{i,t-2}(0) \mid G_{it} = 1] -  \E[Y_{i,t-2}(0)  \mid G_{it} = 0]\}. \label{eq:parallel-tit-sad}
\end{align*}

Finally, combining the standard and sequential DID estimators, we can extend the double DID to the SA design as follows.
\begin{equation*}
  \wstau_{\texttt{d-DID}}(t) = \argmin_{\stau(t)}
  \begin{pmatrix}
    \stau(t) - \wstau_{\texttt{DID}}(t)\\[-5pt]
    \stau(t) - \wstau_{\texttt{s-DID}}(t)
  \end{pmatrix}^\top
  \mathbf{W}(t)
  \begin{pmatrix}
    \stau(t) - \wstau_{\texttt{DID}}(t)\\[-5pt]
    \stau(t) - \wstau_{\texttt{s-DID}}(t)
  \end{pmatrix} \label{eq:d-did-t-sad}
  \vspace{0.05in}
\end{equation*}
where $\mathbf{W}(t)$ is a weight matrix. Under the SA design,
similar to the basic design, the standard DID and sequential DID
estimators are special cases of our proposed double DID estimator with
specific choices of the  weight matrix. As in
Section~\ref{subsec:b-d-did}, we can estimate the optimal weight
matrix $\widehat{\*W}(t)$ (details below), and thus, users do not need
to choose it manually.

Like the basic double DID estimator in
Section~\ref{subsec:b-d-did}, the double DID for the SA design also consists of
two steps. The first step is to assess the underlying
assumptions using the standard DID for the SA design with two points
$\{t-1, t-2\}$ for units that are not yet treated at time
$t-1$, that is, $\{i: G_{it} \geq 0\}$. This is a generalization of
the pre-treatment-trends test in the basic DID
setup (Section~\ref{subsec:assess}). The second step is to estimate
the SA-ATT at time $t$. When only the parallel trends-in-trends
assumption is plausible, we choose weight matrix $\mathbf{W}(t)$ where
$\mathbf{W}(t)_{11} = \mathbf{W}(t)_{12} = \mathbf{W}(t)_{21} = 0$ and
$\mathbf{W}(t)_{22} = 1$, which converges to the sequential DID under
the SA design. When the extended parallel trends assumption is plausible, we use the optimal
weight matrix defined as $\widehat{\mathbf{W}}(t) = \widehat{\Var} (\wstau_{(1:2)}(t))^{-1}$ where $\Var(\cdot)$ is the
variance-covariance matrix and $\wstau_{(1:2)}(t)  =  (\wstau_{\texttt{DID}}(t),
\wstau_{\texttt{s-DID}}(t))^\top.$ This optimal weight matrix provides
us with the most efficient estimator (i.e., the smallest standard
error). We provide further details on the implementation in Appendix~\ref{sec:K-sad}.

To estimate the time-average SA-DID, we extend the double DID as
follows.
\begin{equation*}
  \wostau_{\texttt{d-DID}} = \argmin_{\ostau}
  \begin{pmatrix}
    \ostau - \wostau_{\texttt{DID}}\\[-5pt]
    \ostau - \wostau_{\texttt{s-DID}}
  \end{pmatrix}^\top
  \overline{\*W}
  \begin{pmatrix}
    \ostau - \wostau_{\texttt{DID}}\\[-5pt]
    \ostau - \wostau_{\texttt{s-DID}}
  \end{pmatrix} \label{eq:d-did-o-sad}
  \vspace{0.05in}
\end{equation*}
where $\wostau_{\texttt{DID}}$ and $\wostau_{\texttt{s-DID}}$ are time-averages of the DID and sequential DID estimators,
\begin{equation*}
  \wostau_{\texttt{DID}} =  \sum_{t \in \cT} \pi_t
  \wstau_{\texttt{DID}}(t), \hspace{0.2in} \mbox{and} \hspace{0.2in}
  \wostau_{\texttt{s-DID}} =  \sum_{t \in \cT} \pi_t \wstau_{\texttt{s-DID}}(t).
\end{equation*}
The optimal weight matrix $\widehat{\overline{\*W}}$ is equal to $\widehat{\Var}
(\wostau_{(1:2)})^{-1}$ where
$\wostau_{(1:2)}  =  (\wostau_{\texttt{DID}},
\wostau_{\texttt{s-DID}})^\top.$

\section{Concluding Remarks}
\label{sec:con}
While the most basic form of the DID only requires two time periods
--- one before and the other after treatment assignment, researchers can often collect data from several additional
pre-treatment periods in a wide range of applications. In this article, we show that such multiple
pre-treatment periods can help improve the basic DID design and the
staggered adoption design in three ways: (1) assessing
underlying assumptions about parallel trends, (2) improving estimation
accuracy, and (3) enabling more flexible DID estimators. We use the
potential outcomes framework to clarify assumptions required to enjoy each benefit.

We then propose a simple method, the double DID, to
combine all three benefits within the GMM framework.  Importantly,
the double DID contains the popular two-way fixed effects regression
and nonparametric DID estimators as special cases, and it uses the GMM
to further improve with respect to identification and estimation
accuracy. Finally, we generalize the double
DID estimator to the staggered adoption design where the timing of the
treatment assignment can vary across units.

\vspace{0.3in}
\spacingset{1.55}
\pdfbookmark[1]{References}{References}

\printbibliography

\clearpage
\appendix

\spacingset{1}

\setcounter{table}{0}
\setcounter{equation}{1}
\setcounter{figure}{0}
\setcounter{assumption}{0}
\renewcommand {\thetable} {A\arabic{table}}
\renewcommand {\thefigure} {A\arabic{figure}}
\renewcommand {\theassumption} {A\arabic{assumption}}
\renewcommand {\theequation} {A.\arabic{equation}}

\section{Literature Review}
\label{sec:review}

\subsection{Papers in \textit{APSR} and \textit{AJPS}}
We conduct a review of the literature to assess current practices of the difference-in-differences (DID) design.
Specifically, we search articles published in \textit{American
  Political Science Review} and \textit{American Journal of Political
  Science} from 2015 to 2019. Some of the papers we reviewed were
accepted in 2019 and were officially published in 2020.
Using Google Scholar, we find articles that contains any of the following keywords: ``two-way fixed effect'', ``two-way fixed effects'', ``difference in difference'' or ``difference in differences.''
We then manually select articles from the list that uses the basic DID
design and the staggered adoption design (see the main text for details about the first two design). 
This procedure left us with a total of 25 articles, 11 from APSR and 14 from AJPS.
Table~\ref{appx_tab:APSR} and \ref{appx_tab:AJPS} show the articles in the list published in APSR and AJPS, respectively.

To determine the number of pre-treatment periods, we manually assess the listed articles.
Among the 25 articles, 20 articles use the basic DID design, and 5 articles use the staggered adoption design.
When a paper uses the basic DID design, we can determine the length of
the pre-treatment periods from the data description and the time of
the treatment assignment. On the other hand, the pre-treatment periods for the staggered adoption and the general design
are set to the total number of time-periods available in the data, as
the length of pre-treatment periods varies across units.

We found that most DID applications have less than 10
pre-treatment periods. The median number of pre-treatment periods is $3.5$ and, the mean number
of pre-treatment periods is about $6$ after removing one unique study
that has more than $100$ pre-treatment periods.

\subsection{Examples of Two Common Approaches}
As we wrote in Section~\ref{sec:intro}, there are several different
popular ways to analyze the DID design with multiple pre-treatment
periods. One common approach is
to apply the two-way fixed effects regression to the entire time
periods, and supplement it with alternative model specifications by
including time-trends or leads of the treatment variable to
assess possible violations of the parallel trends assumption. Examples
include \citet{dube2013cross, truex2014returns,
  earle2015productivity, hall2016systemic, larreguy2017effect}.
Another is to stick with the two-time-period DID and limit the use of additional
pre-treatment periods only to the assessment of pre-treatment
trends. Examples include \citet{ladd2009exploiting, bechtel2011lasting,
  bullock2011more, keele2013much, garfias2018elite}. Note that we list
exemplary papers here and thus, we also include papers from journals
other than APSR and AJPS.

{\small
\begin{table}
\centering
\caption{DID papers on APSR.}\label{appx_tab:APSR}
\begin{tabularx}{\textwidth}{lrX}
\toprule
\textbf{Authors} & \textbf{Year} & \textbf{Title}\\
\midrule
O'brien, D. Z., \& Rickne J. & 2016 & Gender Quotas And Women's Political Leadership\\
Garfias, F. & 2018 & Elite Competition and State Capacity Development: Theory and Evidence From Post-Revolutionary Mexico.\\
Martin, G. J., \& Mccrain, J. & 2019 & Local News And National Politics\\
\makecell[tl]{Blom-Hansen, J., Houlberg, K., \\ \quad Serritzlew, S., \& Treisman, D.} & 2016 &
  Jurisdiction Size and Local Government Policy Expenditure: Assessing The Effect of Municipal Amalgamation\\
Clinton, J. D., \& Sances, M. W. & 2018 &
  The Politics of Policy: The Initial Mass Political Effects of Medicaid Expansion in The States\\
\makecell[tl]{Malesky, E. J. , Nguyen, C. V.,\\ \quad \& Tran, A.}
  & 2014 &
  The Impact of Recentralization on Public Services: A Difference-in-Differences Analysis of the Abolition  of Elected Councils in Vietnam.\\
\makecell[tl]{Larsen, M. V., Hjorth, F.,\\ \quad Dinesen, P. T.,\\ \quad \& Sønderskov, K. M.}  & 2019 &
  When Do Citizens Respond Politically to  The Local Economy? Evidence From Registry Data on Local Housing Markets\\
Becher, M., \& González, I. M. & 2019 & Electoral Reform and Trade-Offs in Representation\\
Selb, P., \& Munzert, S. & 2018 & Examining A Most Likely Case for Strong Campaign Effects\\
\makecell[tl]{Enos, R. D., Kaufman, A. R.,\\ \quad \& Sands, M. L.} & 2019 & Can Violent Protest Change Local Policy Support?\\
Vasiliki Fouka & 2019 & How Do Immigrants Respond to Discrimination?\\
\bottomrule
\end{tabularx}
\end{table}
}

{\small
\begin{table}
\caption{DID papers on AJPS.}\label{appx_tab:AJPS}
\centering
\begin{tabularx}{\textwidth}{lrX}
\toprule
  \textbf{Authors} & \textbf{Year} & \textbf{Title} \\
  \midrule
  \makecell[tl]{Bechtel, M. M., Hangartner, D., \\ \quad \& Schmid, L.} & 2016 & Does compulsory voting increase support for leftist policy? \\
  Bisgaard, M., \& Slothuus, R. & 2018 &
    Partisan elites as culprits? How party cues shape partisan perceptual gaps.\\
  Bischof, D., \& Wagner, M. & 2019 &  Do voters polarize when radical parties enter parliament?\\
  \makecell[tl]{Dewan, T., Meriläinen, J., \\ \quad \& Tukiainen, J.} & 2020 &
    Victorian voting: The origins of party orientation and class alignment.\\
  Earle, J. S., \& Gehlbach, S. & 2015 &
    The Productivity Consequences of Political Turnover: Firm-Level Evidence from Ukraine's Orange Revolution. \\
  Enos, R. D. & 2016 &
    What the demolition of public housing teaches us about the impact of racial threat on political behavior. \\
  Gingerich, D. W. & 2019 &
    Ballot Reform as Suffrage Restriction: Evidence from Brazil's Second Republic. \\
  Hainmueller, J, \& Hangartner, D. & 2019 &
    Does direct democracy hurt immigrant minorities?
    Evidence from naturalization decisions in Switzerland.\\
  Holbein, J. B., \& Hillygus, D. S. & 2016 &
    Making young voters: the impact of preregistration on youth turnout.\\
  Jäger, K. & 2020 &
    When Do Campaign Effects Persist for Years? Evidence from a Natural Experiment. \\
  \makecell[tl]{Lindgren, K. O., Oskarsson, S.,\\ \quad \& Dawes, C. T.} & 2017 &
    Can Political Inequalities Be Educated Away? Evidence from a Large‐Scale Reform. \\
  Lopes da Fonseca, M. & 2017 &
    Identifying the source of incumbency advantage through a constitutional reform. \\
  Paglayan, AS. & 2019 & Public-Sector Unions and the Size of Government\\
  Pardos‐Prado, S., \& Xena, C. & 2019 & Skill specificity and attitudes toward immigration. \\
  \bottomrule
\end{tabularx}
\end{table}
}

\clearpage

\section{Comparison with Three Existing Methods}
\label{sec:compare}
This section clarifies relationships between our proposed double DID and
three existing methods: the two-way fixed effects estimator, the
sequential DID estimator, and synthetic control methods.

\subsection{Relationship with Two-Way Fixed Effects Estimator}
While we contrast the double DID with the two-way fixed effects
estimator throughout the paper, we summarize our discussion here. First, in the
basic DID design, the two-way fixed effects estimator is a special
case of the double DID with a specific choice of the weight matrix
$\mathbf{W}$ (see Table~\ref{tab:d-did}). Therefore, whenever the two-way fixed effects estimator is
consistent for the ATT, the double DID is a more efficient, consistent
estimator of the ATT. This is because the double DID can choose the
optimal weight matrix via the GMM, while the two-way fixed effects
uses the pre-determined equal weights over time. Second, in the
SA design, a large number of recent papers show that
the widely-used two-way fixed effects estimator are in general
inconsistent for the ATT due to treatment effect heterogeneity and
implicit parametric assumptions
\citep{strezhnev2018semiparametric, athey2021design, imai2020two, sun2020estimating}. In contrast, the proposed double DID
in the SA design generalizes nonparametric DID estimators to allow for
treatment effect heterogeneity, and thus, it does not suffer from the
same problem.

\subsection{Relationship with Sequential DID Estimator}
Our double DID estimator contains the sequential DID estimator
\citep[e.g.,][]{lee2016did, mora2019did} as a special case. Our
proposed double DID improves over the sequential DID estimator in two ways. First, when the parallel trends
assumption holds, the double DID optimally combine the standard DID
and the sequential DID to improve efficiency, and it is not equal to the sequential
DID. Therefore, it avoids a dilemma of the sequential DID --- it is
consistent under the parallel trends-in-trends assumption (weaker than
the parallel trends assumption), but is less efficient when the
parallel trends assumption holds. Second, while the sequential DID estimator has only been available for the basic DID
design where treatment assignment happens only once, we generalize it
to the staggered adoption design and further incorporate it into our
staggered-adoption double DID estimator (Section~\ref{sec:sad}).

\subsection{Relationship with Synthetic Control Methods}
\label{subsec:scm}
Another relevant popular class of methods is the synthetic control
methods. While the method was originally designed to estimate the
causal effect on a \textit{single} treated unit, recent extensions allow for multiple treated
units and the staggered adoption design \citep[e.g.,][]{xu2017generalized, avi2018augmented, hazlett2018trajectory, athey2021matrix}.
Despite a wide variety of
innovative extensions, they all share the same core feature: they
require long pre-treatment periods to accurately estimate a
pre-treatment trajectory of the treated units. For example, \citet{xu2017generalized}
recommends collecting more than ten pre-treatment periods. In
contrast, the proposed double DID can be applied as long as there are
more than one pre-treatment periods, and is better suited when there
are a small to moderate number of pre-treatment periods.

When there are a large number of pre-treatment periods
(i.e., long enough to apply the synthetic control methods), we
recommend to apply both the synthetic control methods and proposed
double DID, and evaluate robustness across those approaches. This is
important because they rely on different identification assumptions.
In fact, we show in Section~\ref{subsec:sa-app}, the
double DID can recover credible estimates similar to more flexible
variants of synthetic control methods even when there are a large
number of pre-treatment periods. This robustness provides researchers
with additional credibility for their causal estimates and underlying assumptions.

\clearpage
\section{Nonparametric Equivalence to Regression Estimators}
\label{sec:reg}
In this section, we provide results on the nonparametric connection between
regression estimators and the three DID estimators we discussed in the
paper. This section provides methodological foundations for our main
methodological contributions, which we prove in Sections~\ref{sec:general} and~\ref{sec:K-sad}.

\subsection{Standard DID}
\label{subsec:did-reg}

In practice, we can compute the DID estimator via a linear regression.
We regress the outcome $Y_{it}$ on an intercept, treatment group indicator $G_{i}$, time
indicator $I_t$ (equal to $1$ if post-treatment and $0$
otherwise) and the interaction between the treatment group indicator and the time indicator $G_i \times I_t$.
\begin{equation}
  Y_{it}  \sim \alpha + \theta G_i +  \gamma I_t + \beta (G_i \times I_t), \label{eq:linear-did}
\end{equation}
where $(\alpha, \theta, \gamma, \beta)$ are corresponding
coefficients. In this case, a coefficient of the
interaction term $\beta$ is numerically equal to the DID estimator, $\widehat{\tau}_{\texttt{DID}}$.
Importantly, the linear regression is used
here only to compute the nonparametric DID estimator
(equation~\eqref{eq:did-est}), and thus it does not require any
parametric modeling assumption such as constant treatment effects.
Furthermore, when we analyze panel data in
which the same units are observed repeatedly over time, we obtain exactly the same
estimate via a linear regression with unit and time fixed
effects. This numerical equivalence in the two-time-period case is
often the justification of the two-way fixed effects regression as the DID design
\citep{angrist2008mostly}.
The above equivalence is formally shown below for completeness.

\subsubsection{Repeated Cross-Sectional Data}
For the later use in this Appendix,
we report the well-known result that the standard DID estimator $\widehat{\tau}_{\texttt{DID}}$
(equation~\eqref{eq:did-est}) is equivalent to coefficient
$\widehat{\beta}$ in the regression estimator
(equation~\eqref{eq:linear-did}) \citep{abadie2005semiparametric}.

We define $O_{it}$ to be an indicator variable taking the value $1$
when individual $i$ is observed in time period $t$. Using this
notation, we prove the following result.

\begin{result}[Nonparametric Equivalence of the Standard DID and Regression Estimator]
\label{rc-para}
We write the linear regression estimator (equation~\eqref{eq:linear-did}) as a solution to the following least squares problem.
\begin{equation*}
  (\widehat{\alpha}, \widehat{\theta}, \widehat{\gamma}, \widehat{\beta}) = \argmin \sum^{n}_{i=1}\sum_{t=1}^2
  O_{it}\Big\{Y_{it} - \alpha - \theta G_{i} - \gamma I_{t} - \beta (G_{i}\times I_{t})\Big\}^{2}.
\end{equation*}

Then, $\widehat{\tau}_{\texttt{DID}} = \widehat{\beta}.$
\end{result}

\paragraph{Proof.}
By solving the least squares problem, we obtain the following solutions:
\begin{align*}
\widehat{\alpha} &= \frac{\sum_{i\colon G_i=0}Y_{i1}}{n_{01}} \\
\widehat{\theta} &= \frac{\sum_{i\colon G_i=1}Y_{i1}}{n_{11}} - \frac{\sum_{i\colon G_i=0}Y_{i1}}{n_{01}}\\
\widehat{\gamma} &= \frac{\sum_{i\colon G_i=0}Y_{i2}}{n_{02}} - \frac{\sum_{i\colon G_i=0}Y_{i1}}{n_{01}}\\
  \widehat{\beta} &  = \l(\frac{\sum_{i\colon G_i = 1} Y_{i2}}{n_{12}} -
  \frac{\sum_{i\colon G_i = 1} Y_{i1}}{n_{11}}\r)  - \l(\frac{\sum_{i\colon G_i = 0} Y_{i2}}{n_{02}} -  \frac{\sum_{i\colon G_i = 0} Y_{i1}}{n_{01}}\r),
\end{align*}
which completes the proof.
\hfill\qed\bigskip

\subsubsection{Panel Data}
Again, for the later use in the Appendix,
we report the well-known result that the standard DID estimator $\widehat{\tau}_{\texttt{DID}}$
(equation~\eqref{eq:did-est}) is equivalent to coefficient
$\widehat{\beta}$ in the two-way fixed effects regression estimator in
the panel data setting \citep{abadie2005semiparametric}.

\begin{result}[Nonparametric Equivalence of the Standard DID and
  Two-way Fixed Effects Regression Estimator]\label{tfe-para}

We can write the two-way fixed effects regression estimator as a solution to the following least squares problem.
\begin{equation*}
  (\widehat{\alpha}, \widehat{\delta}, \widehat{\beta})
  = \argmin \sum^{n}_{i=1}\sum^{2}_{t=1}(Y_{it} - \alpha_{i} - \delta_{t} - \beta D_{it})^{2}.
\end{equation*}
  Then, $\widehat{\tau}_{\texttt{DID}} = \widehat{\beta}.$
\end{result}
\paragraph{Proof.}
First we define the demeaned treatment and outcome variables,
$\overline{Y}_i = \sum_{t=1}^2Y_{it}/2$,
$\overline{Y}_t = \sum_{i=1}^nY_{it}/n$,
$\overline{Y} = \sum_{i=1}^n\sum_{t=1}^2Y_{it}/2n$,
$\overline{D}_i = \sum_{t=1}^2D_{it}/2$,
$\overline{D}_t = \sum_{i=1}^nD_{it}/n$, and
$\overline{D} = \sum_{i=1}^n\sum_{t=1}^2D_{it}/2n$.

Given these transformed variables, we can transform the least squares problem into a well-known demeaned form.
\begin{equation*}
  \widehat{\beta}  = \argmin_{\beta} \sum^{n}_{i=1}\sum^{2}_{t=1} (\widetilde{Y}_{it} - \beta \widetilde{D}_{it})^{2}
\end{equation*}
where $\widetilde{Y}_{it} = Y_{it} - \overline{Y}_{i} - \overline{Y}_{t} + \overline{Y}$
and $\widetilde{D}_{it} = D_{it} - \overline{D}_{i} - \overline{D}_{t}
+ \overline{D}$.
Using this notation, we can express $\widehat{\beta}$ as
\begin{equation*}\label{eq:fe2-vcovform}
  \widehat{\beta} =
  \frac{\sum^{n}_{i=1}\sum^{2}_{t=1} \widetilde{D}_{it}\widetilde{Y}_{it}}
  {\sum^{n}_{i=1}\sum^{2}_{t=1}\widetilde{D}^{2}_{it}}
\end{equation*}
where $\widetilde{D}_{it}$ takes the following form,
\begin{equation*}
\widetilde{D}_{it} =
\begin{cases}
1/2 \cdot n_{0} / n     & \text{ if } G_{i} = 1, t = 2\\
-(1/2) \cdot n_{0} / n  & \text{ if } G_{i} = 1, t = 1\\
-(1/2) \cdot n_{1} / n  & \text{ if } G_{i} = 0, t = 2\\
1/2 \cdot n_{1} / n  & \text{ if } G_{i} = 0, t = 1,
\end{cases}
\end{equation*}
where $n_1 =  \sum_{i=1}^n G_{i}$  and $n_0 =  \sum_{i=1}^n(1-G_i)$. Then, the numerator can be written as
\begin{align*}
\sum^{n}_{i=1}\sum^{2}_{t=1} \widetilde{D}_{it}\widetilde{Y}_{it}
  &=   \frac{n_{0}}{2n} \bigg\{\sum^{n}_{i=1}G_{i}\widetilde{Y}_{i2} - \sum^{n}_{i=1}G_{i}\widetilde{Y}_{i1}\bigg\}
    - \frac{n_{1}}{2n}
    \bigg\{\sum^{n}_{i=1}(1 - G_{i})\widetilde{Y}_{i2} - \sum^{n}_{i=1}(1 - G_{i})\widetilde{Y}_{i1}\bigg\}
\end{align*}
and the denominator is given as
\begin{equation*}
\sum^{n}_{i=1}\sum^{2}_{t=1} \widetilde{D}^{2}_{it}  = 2n_{1}\bigg(\frac{n_{0}}{2n}\bigg)^{2} + 2n_{0}\bigg(\frac{n_{1}}{2n}\bigg)^{2}= \frac{n_{1}n_{0}}{2n}.
\end{equation*}
Combining both terms, we get
\begin{align*}
\widehat{\beta} &  = \frac{\sum^{n}_{i=1}\sum^{2}_{t=1} \widetilde{D}_{it}\widetilde{Y}_{it}}
{\sum^{n}_{i=1}\sum^{2}_{t=1}\widetilde{D}^{2}_{it}}\\
&=
\frac{1}{n_{1}}
\bigg\{
\sum^{n}_{i=1}G_{i}\widetilde{Y}_{i2} - \sum^{n}_{i=1}G_{i}\widetilde{Y}_{i1}
\bigg\}
- \frac{1}{n_{0}}
\bigg\{
\sum^{n}_{i=1}(1 - G_{i})\widetilde{Y}_{i2} - \sum^{n}_{i=1}(1 - G_{i})\widetilde{Y}_{i1}
\bigg\}\\
&=
\frac{1}{n_{1}}
\sum^{n}_{i=1}G_{i}(Y_{i2} - Y_{i1})
- \frac{1}{n_{0}}
\sum^{n}_{i=1}(1 - G_{i})(Y_{i2} - Y_{i1})\\
&  =  \widehat{\tau}_{\texttt{DID}},
\end{align*}
which concludes the proof.
\hfill\qed\bigskip

\subsection{Extended DID}
\label{subsec:e-did-reg}
\subsubsection{Repeated Cross-Sectional Data}
We consider a case in which there are two pre-treatment periods $t =
\{0, 1\}$ and one post-treatment period $t = 2$. 
Using this notation, we report the following result.
\begin{result}[Nonparametric Equivalence of the Extended DID and Regression Estimator]
  We focus on a linear regression estimator that is a solution to the following least
  squares problem.
  \begin{equation*}
    (\widehat{\theta}, \widehat{\gamma}, \widehat{\beta}) = \argmin \sum^{n}_{i=1}\sum_{t=0}^2
    O_{it}\l(Y_{it} - \theta G_{i} - \gamma_t  - \beta D_{it} \r)^{2}.
  \end{equation*}
  Then, $\widehat{\beta} = \lambda \widehat{\tau}_{\texttt{DID}} +
  (1-\lambda) \widehat{\tau}_{\texttt{DID(2,0)}}$ where
  \begin{align*}
    \lambda & = \cfrac{n_{11}n_{01} (n_{10} +  n_{00})}{n_{11}n_{01}
               (n_{10} +  n_{00}) + n_{10}n_{00} (n_{11} +  n_{01})},\\
   1 - \lambda & = \cfrac{n_{10}n_{00} (n_{11} +  n_{01})}{n_{11}n_{01}
               (n_{10} +  n_{00}) + n_{10}n_{00} (n_{11} +  n_{01})}.
  \end{align*}
  When the sample size of each group is fixed over time, i.e., $n_{11}
  = n_{10}$ and  $n_{01} =  n_{00}$, $\lambda = 1/2$ and therefore,
  $\widehat{\beta}$ is equivalent to the extended DID estimator of
  equal weights in equation~\eqref{eq:e-did}.
\end{result}

\paragraph{Proof.}
By solving the least squares problem, we obtain
\begin{align*}
\widehat{\theta} &= \lambda \l(\frac{\sum_{i\colon
                   G_i=1}Y_{i1}}{n_{11}} - \frac{\sum_{i\colon
                   G_i=0}Y_{i1}}{n_{01}}\r)  +
                   ( 1- \lambda) \l(\frac{\sum_{i\colon
                   G_i=1}Y_{i0}}{n_{10}} - \frac{\sum_{i\colon G_i=0}Y_{i0}}{n_{00}}\r)\\
\widehat{\gamma}_2 &= \frac{\sum_{i\colon G_i=0}Y_{i2}}{n_{02}} \\
\widehat{\gamma}_1 &= \frac{\sum_{i\colon G_i=1}Y_{i1} +
                     \sum_{i\colon G_i=0}Y_{i1}}{n_{11} + n_{01}} -
                     \frac{n_{11}}{n_{11} + n_{01}}
                     \widehat{\theta}\\
\widehat{\gamma}_0 &= \frac{\sum_{i\colon G_i=1}Y_{i0} +
                     \sum_{i\colon G_i=0}Y_{i0}}{n_{10} + n_{00}} -
                     \frac{n_{10}}{n_{10} + n_{00}}  \widehat{\theta}\\
  \widehat{\beta} &  = \lambda \l\{\l(\frac{\sum_{i\colon G_i = 1} Y_{i2}}{n_{12}} -
  \frac{\sum_{i\colon G_i = 1} Y_{i1}}{n_{11}}\r)  -
                    \l(\frac{\sum_{i\colon G_i = 0} Y_{i2}}{n_{02}} -
                    \frac{\sum_{i\colon G_i = 0} Y_{i1}}{n_{01}}\r) \r\}\\
  & \qquad + (1- \lambda) \l\{\l(\frac{\sum_{i\colon G_i = 1} Y_{i2}}{n_{12}} -
  \frac{\sum_{i\colon G_i = 1} Y_{i0}}{n_{10}}\r)  -
    \l(\frac{\sum_{i\colon G_i = 0} Y_{i2}}{n_{02}} -
    \frac{\sum_{i\colon G_i = 0} Y_{i0}}{n_{00}}\r) \r\},
\end{align*}
which completes the proof.
\hfill\qed\bigskip

\subsubsection{Panel Data}
We report that the extended DID estimator $\widehat{\tau}_{\texttt{e-DID}}$
(equation~\eqref{eq:e-did}) (equal weights: $\lambda = 1/2$) is equivalent to the estimated coefficient
$\widehat{\beta}$ in the two-way fixed effects regression estimator in
the panel data setting with $t = \{0, 1, 2\}$. 

\begin{result}[Nonparametric Equivalence of the Extended DID and Two-way Fixed Effects Regression Estimator]
  We can write the two-way fixed effects regression estimator as a solution to
  the following least squares problem.
  \begin{equation*}
    (\widehat{\alpha}, \widehat{\delta}, \widehat{\beta})
    = \argmin \sum^{n}_{i=1}\sum^{2}_{t=0}(Y_{it} - \alpha_{i} - \delta_{t} - \beta D_{it})^{2}.
  \end{equation*}
  Then, $\widehat{\tau}_{\texttt{e-DID}} = \widehat{\beta}.$
\end{result}
\paragraph{Proof.}
First we define $\overline{Y}_i = \sum_{t=0}^2Y_{it}/3$,
$\overline{Y}_t = \sum_{i=1}^nY_{it}/n$, $\overline{Y} =
\sum_{i=1}^n\sum_{t=0}^2Y_{it}/3n$, $\overline{D}_i = \sum_{t=0}^2D_{it}/3$,
$\overline{D}_t = \sum_{i=1}^nD_{it}/n$, and $\overline{D} =
\sum_{i=1}^n\sum_{t=0}^2D_{it}/3n.$ Then, we can write the two-way fixed effects estimator as a two-way demeaned estimator,
\begin{align*}
\widehat{\beta} & =  \argmin_{\beta}
                  \sum^{n}_{i=1}\sum^{2}_{t=0}(\widetilde{Y}_{it} -
                  \beta \widetilde{D}_{it})^{2}  =  \frac{\sum^{n}_{i=1}\sum^{2}_{t=0}
    \widetilde{D}_{it}\widetilde{Y}_{it}} {\sum^{n}_{i=1}\sum^{2}_{t=0}\widetilde{D}^{2}_{it}},
\end{align*}
as in Result~\ref{tfe-para}, where $\widetilde{Y}_{it} = Y_{it} - \overline{Y}_{i} - \overline{Y}_{t} + \overline{Y}$ and
$\widetilde{D}_{it} = D_{it} - \overline{D}_{i} - \overline{D}_{t} +
\overline{D}$. Importantly, $\widetilde{D}_{it}$ takes the following form:
\begin{equation*}
  \widetilde{D}_{it} =
  \begin{cases}
    2/3 \cdot n_{0} / n  & \text{ if } G_{i} = 1, t =2 \\
    - 1/3 \cdot n_{0} / n  & \text{ if } G_{i} = 1, t =0,1\\
    - 2/3 \cdot n_{1} / n   & \text{ if } G_{i} = 0, t =2\\
    1/3 \cdot n_{1} / n & \text{ if } G_{i} = 0, t =0,1,
  \end{cases}
\end{equation*}
where $n_1 =  \sum_{i=1}^n G_{i}$  and $n_0 =  \sum_{i=1}^n(1-G_i)$. Then, the numerator can be written as
\begin{align*}
  & \sum^{n}_{i=1}\sum^{2}_{t=0} \widetilde{D}_{it}\widetilde{Y}_{it}\\
  &=
    \sum^{n}_{i=1}G_{i}
    \bigg(\frac{2n_{0}}{3n}\bigg)\widetilde{Y}_{i2}
    - \sum^{n}_{i=1}\sum_{t=0}^1
    G_{i}\bigg(\frac{n_{0}}{3n}\bigg)\widetilde{Y}_{it} + \sum^{n}_{i=1}(1 - G_{i})
    \bigg(\frac{-2n_{1}}{3n}\bigg)\widetilde{Y}_{i2}
    + \sum^{n}_{i=1}\sum_{t=0}^1 (1 - G_{i})
    \bigg(\frac{n_{1}}{3n}\bigg)\widetilde{Y}_{it}\\
  &=
    \sum^{n}_{i=1}G_{i}\bigg(\frac{n_{0}}{3n}\bigg)\{
    \widetilde{Y}_{i2} - \widetilde{Y}_{i1} \} +
    \sum^{n}_{i=1}G_{i}\bigg(\frac{n_{0}}{3n}\bigg)
    \{ \widetilde{Y}_{i2} - \widetilde{Y}_{i0}\} \\
  &\qquad - \l\{
    \sum^{n}_{i=1}(1 - G_{i})\bigg(\frac{n_{1}}{3n}\bigg)\{
    \widetilde{Y}_{i2} - \widetilde{Y}_{i1} \}
    +
    \sum^{n}_{i=1}(1 - G_{i})\bigg(\frac{n_{1}}{3n}\bigg)\{
    \widetilde{Y}_{i2} - \widetilde{Y}_{i0} \}\r\}\\
& = \frac{n_0}{3n} \l\{\sum^{n}_{i=1}G_{i} \{Y_{i2} - Y_{i1} \} +
    \sum^{n}_{i=1}G_{i}\{ Y_{i2} - Y_{i0}\}
  \r\} - \frac{n_1}{3n} \l\{
    \sum^{n}_{i=1}(1 - G_{i})  \{Y_{i2} - Y_{i1} \}
    +  \sum^{n}_{i=1}(1 - G_{i}) \{Y_{i2} - Y_{i0} \}\r\}.
\end{align*}
The denominator can be written as
\begin{align*}
\sum^{n}_{i=1}\sum^{2}_{t=0}\widetilde{D}^{2}_{it} &= \frac{n_{0}n_{1}}{n}\cdot\frac{2}{3}.
\end{align*}
Combining the two terms, we have
\begin{align*}
\widehat{\beta}
  &=
    \frac{1}{2n_1} \l\{\sum^{n}_{i=1}G_{i} \{Y_{i2} - Y_{i1} \} +
    \sum^{n}_{i=1}G_{i}\{ Y_{i2} - Y_{i0}\}
  \r\} \\
  &\qquad - \frac{1}{2n_0} \l\{
    \sum^{n}_{i=1}(1 - G_{i})  \{Y_{i2} - Y_{i1} \}
    +  \sum^{n}_{i=1}(1 - G_{i}) \{Y_{i2} - Y_{i0} \}\r\}\\
&=
    \frac{1}{2} \l\{\frac{1}{n_1} \sum^{n}_{i=1}G_{i} \{Y_{i2} -
  Y_{i1} \} - \frac{1}{n_0} \sum^{n}_{i=1}(1 - G_{i})  \{Y_{i2} -
  Y_{i1} \}\r\}  \\
& \qquad  +  \frac{1}{2} \l\{\frac{1}{n_1} \sum^{n}_{i=1}G_{i} \{Y_{i2} -
  Y_{i0} \} - \frac{1}{n_0} \sum^{n}_{i=1}(1 - G_{i})  \{Y_{i2} -
  Y_{i0} \}\r\}\\
&= \frac{1}{2}\widehat{\tau}_{\texttt{DID}} + \frac{1}{2}\widehat{\tau}_{\texttt{DID(2,0)}}.
\end{align*}
By solving the least squares problem, we also obtain
\begin{eqnarray*}
  \widehat{\alpha}_i \ & = & \ \overline{Y}_i - \overline{Y} -
                           \overline{Y}_{t=0} + \widehat{\beta}
                           (\overline{D} - \overline{D}_{t=0})
                           \nonumber \\
  \widehat{\delta}_t \ & = & \ \overline{Y}_t - \overline{Y}_{t=0} +
                           \widehat{\beta} (\overline{D}_{t=0} -
                           \overline{D}_{t}) \nonumber
\end{eqnarray*}
\hfill\qed\bigskip

\subsection{Sequential DID}
\label{subsec:s-did-reg}
The sequential DID estimator is connected to a widely used
regression estimator. In particular, the sequential DID
estimator (equation~\eqref{eq:s-did}) can be computed as a linear regression in which we replace the outcome $Y_{it}$
with a transformed outcome. In panel data, we replace the original
outcome with its first difference $Y_{it} - Y_{i,t-1}$ so that we use
changes instead of levels. In repeated cross-sectional data, we use
the following linear regression.
\begin{equation}
  \Delta Y_{it}  \sim \alpha_s + \theta_s G_i +  \gamma_s I_t + \beta_s (G_i \times
  I_t), \label{eq:linear-sdid}
\end{equation}
where $\Delta Y_{it}  =  Y_{it} - (\sum_{i\colon G_i=1}
Y_{i,t-1})/n_{1, t-1}$ if $G_i = 1$ and $\Delta Y_{it}  = Y_{it} -
(\sum_{i\colon G_i=0} Y_{i,t-1})/n_{0, t-1} $  if $G_i = 0$.
Coefficients are denoted by $(\alpha_s, \theta_s, \gamma_s, \beta_s)$. In this case, a coefficient in front of the
interaction term $\beta_s$ is numerically identical to the sequential DID
estimator. We provide the proof of this equivalence for both panel and repeated
cross-sectional data settings below.

\subsubsection{Repeated Cross-Sectional Data}
We clarify that the sequential DID estimator $\widehat{\tau}_{\texttt{s-DID}}$
(equation~\eqref{eq:s-did}) is equivalent to a coefficient in a
regression estimator with transformed outcomes.

\begin{result}[Nonparametric Equivalence of the Sequential DID and Regression Estimator]
  We focus on a linear regression estimator with a transformed outcome.
  \begin{equation*}
    (\widehat{\alpha}, \widehat{\theta}, \widehat{\gamma}, \widehat{\beta}) = \argmin \sum^{n}_{i=1}\sum_{t=1}^2
    O_{it}\Big\{\Delta Y_{it} - \alpha - \theta G_{i} - \gamma I_{t} - \beta (G_{i}\times I_{t})\Big\}^{2},
  \end{equation*}
  where
\begin{equation*}
  \Delta Y_{it} =
  \begin{cases}
    Y_{i2} -  \frac{\sum_{i\colon G_i=1}Y_{i1}}{n_{11}}  & \text{ if } G_{i} = 1, t =2 \\
    Y_{i1} -  \frac{\sum_{i\colon G_i=1}Y_{i0}}{n_{10}}  & \text{ if } G_{i} = 1, t =1\\
    Y_{i2} -  \frac{\sum_{i\colon G_i=0}Y_{i1}}{n_{01}}   & \text{ if } G_{i} = 0, t =2\\
    Y_{i1} -  \frac{\sum_{i\colon G_i=0}Y_{i0}}{n_{00}} & \text{ if } G_{i} = 0, t =1.
  \end{cases}
\end{equation*}
Then, $\widehat{\tau}_{\texttt{s-DID}} = \widehat{\beta}.$
\end{result}
\paragraph{Proof.}
Using Result~\ref{rc-para}, we obtain
\begin{align*}
  \widehat{\beta} &  = \l(\frac{\sum_{i\colon G_i = 1} \Delta Y_{i2}}{n_{12}} -
                    \frac{\sum_{i\colon G_i = 1} \Delta Y_{i1}}{n_{11}}\r)  - \l(\frac{\sum_{i\colon G_i = 0} \Delta Y_{i2}}{n_{02}} -  \frac{\sum_{i\colon G_i = 0} \Delta Y_{i1}}{n_{01}}\r)\\
& = \l\{ \l(\frac{\sum_{i\colon G_i = 1} Y_{i2}}{n_{12}} -
     \frac{\sum_{i\colon G_i = 1} Y_{i1}}{n_{11}}\r)  - \l(\frac{\sum_{i\colon
     G_i = 0} Y_{i2}}{n_{02}} -  \frac{\sum_{i\colon G_i = 0}
     Y_{i1}}{n_{01}}\r) \r\} \\
  & \qquad \qquad -  \l\{ \l(\frac{\sum_{i\colon G_i = 1} Y_{i1}}{n_{11}} -
     \frac{\sum_{i\colon G_i = 1} Y_{i0}}{n_{10}}\r)  - \l(\frac{\sum_{i\colon
     G_i = 0} Y_{i1}}{n_{01}} -  \frac{\sum_{i\colon G_i = 0}
     Y_{i0}}{n_{00}}\r) \r\},
\end{align*}
which completes the proof.
\hfill\qed\bigskip

Next, we clarify that the sequential DID estimator $\widehat{\tau}_{\texttt{s-DID}}$
(equation~\eqref{eq:s-did}) is also equivalent to a coefficient in a
regression estimator with group-specific time trends. \cite{mora2019did} derive similar results by making the
parametric assumption of the conditional expectations. We prove
nonparametric equivalence without making any assumptions about conditional expectations.

\begin{result}[Nonparametric Equivalence of the Sequential DID and
  Regression Estimator with Group-Specific Time Trends]
  We focus on a linear regression estimator with group-specific time trends.
  \begin{equation*}
    (\widehat{\theta}, \widehat{\gamma}, \widehat{\beta}) = \argmin \sum^{n}_{i=1}\sum_{t=0}^2
    O_{it}\Big\{Y_{it} - \theta_0 G_{i} - \theta_1
    (G_{i} \times t) - \gamma_t - \beta D_{it} \Big\}^{2}.
  \end{equation*}
Then, $\widehat{\tau}_{\texttt{s-DID}} = \widehat{\beta}.$
\end{result}

\paragraph{Proof.}
By solving the least squares problem, we obtain
\begin{align*}
\widehat{\theta}_0 &= \frac{\sum_{i\colon G_i=1}Y_{i0}}{n_{10}} - \frac{\sum_{i\colon G_i=0}Y_{i0}}{n_{00}}\\
\widehat{\theta}_1 &= \l(\frac{\sum_{i\colon
                   G_i=1}Y_{i1}}{n_{11}} - \frac{\sum_{i\colon
                   G_i=0}Y_{i1}}{n_{01}}\r)  -
                     \l(\frac{\sum_{i\colon
                   G_i=1}Y_{i0}}{n_{10}} - \frac{\sum_{i\colon G_i=0}Y_{i0}}{n_{00}}\r)\\
\widehat{\gamma}_2 &= \frac{\sum_{i\colon G_i=0}Y_{i2}}{n_{02}}, \ \
                     \widehat{\gamma}_1 = \frac{\sum_{i\colon
                                          G_i=0}Y_{i1}}{n_{01}}, \ \ \widehat{\gamma}_0 = \frac{\sum_{i\colon G_i=0}Y_{i0}}{n_{00}} \\
  \widehat{\beta} &  = \l\{ \l(\frac{\sum_{i\colon G_i = 1} Y_{i2}}{n_{12}} -
                     \frac{\sum_{i\colon G_i = 1} Y_{i1}}{n_{11}}\r)  - \l(\frac{\sum_{i\colon
                     G_i = 0} Y_{i2}}{n_{02}} -  \frac{\sum_{i\colon G_i = 0}
                     Y_{i1}}{n_{01}}\r) \r\} \\
                   & \qquad \qquad -  \l\{ \l(\frac{\sum_{i\colon G_i = 1} Y_{i1}}{n_{11}} -
                     \frac{\sum_{i\colon G_i = 1} Y_{i0}}{n_{10}}\r)  - \l(\frac{\sum_{i\colon
                     G_i = 0} Y_{i1}}{n_{01}} -  \frac{\sum_{i\colon G_i = 0}
                     Y_{i0}}{n_{00}}\r) \r\},
\end{align*}
which completes the proof.
\hfill\qed\bigskip

\subsubsection{Panel Data}
We clarify that the sequential DID estimator $\widehat{\tau}_{\texttt{s-DID}}$
(equation~\eqref{eq:s-did}) is equivalent to a coefficient in the
two-way fixed effects regression estimator with transformed outcomes.

\begin{result}[Nonparametric Equivalence of the Sequential DID and Two-way Fixed Effects Regression Estimator]
  We focus on the two-way fixed effects regression estimator with transformed outcomes.
  \begin{equation*}
    (\widehat{\alpha}, \widehat{\delta}, \widehat{\beta})
    = \argmin \sum^{n}_{i=1}\sum^{2}_{t=1}(\Delta Y_{it} - \alpha_{i} - \delta_{t} - \beta D_{it})^{2},
  \end{equation*}
  where $\Delta Y_{it} = Y_{it} - Y_{i,t-1}$. Then, $\widehat{\tau}_{\texttt{s-DID}} = \widehat{\beta}.$
\end{result}
\paragraph{Proof.}
As in Result~\ref{tfe-para}, we can focus on the demeaned form.
\begin{equation*}
\widehat{\beta} = \argmin \sum^{n}_{i=1}\sum^{2}_{t=1}  (\widetilde{\Delta Y}_{it} - \beta \widetilde{D}_{it})^{2},
\end{equation*}
where $\widetilde{\Delta Y}_{it} = \Delta Y_{it} - \overline{\Delta
  Y}_{i} - \overline{\Delta Y}_{t} + \overline{\Delta Y}$,
$\overline{\Delta Y}_i = \sum_{t=1}^2 \Delta Y_{it}/2$,
$\overline{\Delta Y}_t = \sum_{i=1}^n \Delta Y_{it}/n$, and
$\overline{\Delta Y} =
\sum_{i=1}^n\sum_{t=1}^2 \Delta Y_{it}/2n$. Similarly, $\widetilde{D}_{it} = D_{it} - \overline{D}_{i} - \overline{D}_{t} +
\overline{D}$, $\overline{D}_i = \sum_{t=1}^2D_{it}/2$,
$\overline{D}_t = \sum_{i=1}^nD_{it}/n$, and $\overline{D} =
\sum_{i=1}^n\sum_{t=1}^2D_{it}/2n.$ Using Result~\ref{tfe-para},
\begin{align*}
\widehat{\beta}
&= \frac{1}{n_{1}}\sum^{n}_{i=1}G_{i}(\Delta Y_{i2} - \Delta Y_{i1})
- \frac{1}{n_{0}}\sum^{n}_{i=1}(1 - G_{i})(\Delta Y_{i2} - \Delta Y_{i1})  \\
&=
\bigg\{
\frac{1}{n_{1}}\sum^{n}_{i=1}G_{i}(Y_{i2} - Y_{i1})
-\frac{1}{n_{0}}\sum^{n}_{i=1}(1 - G_{i})(Y_{i2} - Y_{i1})
\bigg\}\\
&\qquad\quad-
\bigg\{
\frac{1}{n_{1}}\sum^{n}_{i=1}G_{i}(Y_{i1} - Y_{i0})
- \frac{1}{n_{0}}\sum^{n}_{i=1}(1 - G_{i})(Y_{i1} - Y_{i0})
\bigg\}\\
& \equiv \widehat{\tau}_{\texttt{s-DID}},
\end{align*}
which concludes the proof.
\hfill\qed\bigskip

Next, we clarify that the sequential DID estimator $\widehat{\tau}_{\texttt{s-DID}}$
(equation~\eqref{eq:s-did}) is also equivalent to a coefficient in the
two-way fixed effects regression estimator with individual-specific
time trends. 

\begin{result}[Nonparametric Equivalence of the Sequential DID and
  Two-way Fixed Effects Regression Estimator with Individual-Specific
  Time Trends]
  We focus on the two-way fixed effects regression estimator with
  individual-specific time trends
  \begin{equation*}
    (\widehat{\alpha}, \widehat{\xi}, \widehat{\delta}, \widehat{\beta})
    = \argmin \sum^{n}_{i=1}\sum^{2}_{t=0}(Y_{it} - \alpha_{i} -
    (\xi_{i} \times t)- \delta_{t} - \beta D_{it})^{2}.
  \end{equation*}
  Then, $\widehat{\tau}_{\texttt{s-DID}} = \widehat{\beta}.$
\end{result}
\paragraph{Proof.}
By solving the least squares problem, we obtain that
\begin{align*}
 \sum_{\mathclap{i\colon G_i=1}} Y_{i2}
  &= (\widehat{\beta} + \widehat{\gamma}_2) n_1
    + \sum_{\mathclap{i\colon G_i=1}}\widehat{\alpha}_{i}
    + 2\sum_{\mathclap{i\colon G_i=1}} \widehat{\xi}_i,
 \quad
 \sum_{\mathclap{i\colon G_i=0}} Y_{i2}
  = \widehat{\gamma}_2 n_0
    + \sum_{\mathclap{i\colon G_i=0}} \widehat{\alpha}_i
    + 2 \sum_{\mathclap{i\colon G_i=0}} \widehat{\xi}_i\\
 \sum_{\mathclap{i\colon G_i=1}} Y_{i1}
  &= \widehat{\gamma}_1 n_1
    + \sum_{\mathclap{i\colon G_i=1}} \widehat{\alpha}_i
    + \sum_{\mathclap{i\colon G_i=1}} \widehat{\xi}_i,
  \quad\qquad\quad\
 \sum_{\mathclap{i\colon G_i=0}} Y_{i1}
  = \widehat{\gamma}_1 n_0
    + \sum_{\mathclap{i\colon G_i=0}} \widehat{\alpha}_i
    + \sum_{\mathclap{i\colon G_i=0}} \widehat{\xi}_i\\
 \sum_{\mathclap{i\colon G_i=1}} Y_{i0}
  &= \widehat{\gamma}_0 n_1
    + \sum_{\mathclap{i\colon G_i=1}} \widehat{\alpha}_i,
   \quad\qquad\qquad\qquad\quad
 \sum_{\mathclap{i\colon G_i=0}} Y_{i0}
  = \widehat{\gamma}_0 n_0
    + \sum_{\mathclap{i\colon G_i=0}} \widehat{\alpha}_i.
\end{align*}
Therefore, we get
\begin{align*}
  \widehat{\beta} &  = \l\{ \l(\frac{\sum_{i\colon G_i = 1} Y_{i2}}{n_{1}} -
                     \frac{\sum_{i\colon G_i = 1} Y_{i1}}{n_{1}}\r)  - \l(\frac{\sum_{i\colon
                     G_i = 0} Y_{i2}}{n_{0}} -  \frac{\sum_{i\colon G_i = 0}
                     Y_{i1}}{n_{0}}\r) \r\} \\
                   & \qquad \qquad -  \l\{ \l(\frac{\sum_{i\colon G_i = 1} Y_{i1}}{n_{1}} -
                     \frac{\sum_{i\colon G_i = 1} Y_{i0}}{n_{1}}\r)  - \l(\frac{\sum_{i\colon
                     G_i = 0} Y_{i1}}{n_{0}} -  \frac{\sum_{i\colon G_i = 0}
                     Y_{i0}}{n_{0}}\r) \r\},
\end{align*}
which completes the proof.
\hfill\qed\bigskip

\subsubsection{Alternative Interpretation of Parallel Trends-in-Trends
  Assumption}
\label{subsubsec:alt-int-ptt}
We emphasize an alternative way to interpret the parallel
trends-in-trends assumption. Unlike the parallel trends assumption
that assumes the time-invariant unmeasured confounding, the parallel trends-in-trends
assumption can account for \textit{linear time-varying} unmeasured
confounding --- unobserved confounding increases or decreases over
time but with some constant rate. For example, researchers might be worried that some treated communes have higher motivation for
reforms, which is not measured, and the infrastructure qualities differ between
treated and control communes due to this unobserved motivation. The parallel trends
assumption means that the difference in the infrastructure qualities
due to this unobserved confounder does not grow or decline over time. In contrast, the parallel trends-in-trends assumption accommodates a simple yet important case in which
the unobserved difference in the infrastructure qualities does grow or
decline with some fixed rate, which analysts do not need to
specify.
This interpretation comes from the following equivalent representation
of the parallel trends-in-trends assumption.
\begin{align}
  & \underbrace{\{\E[Y_{i2}(0) \mid G_i = 1] -  \E[Y_{i2}(0)  \mid
    G_i = 0]\}}_{\text{\normalfont Bias at $t=2$}}
    - \underbrace{\{\E[Y_{i1}(0) \mid G_i = 1] -  \E[Y_{i1}(0)  \mid
    G_i = 0]\}}_{\text{ Bias at $t=1$}}
    \nonumber \\
  & =  \underbrace{\{\E[Y_{i1}(0) \mid G_i = 1] -  \E[Y_{i1}(0)  \mid G_i = 0]\}}_{\text{ Bias at $t=1$}}
    -  \underbrace{\{\E[Y_{i0}(0) \mid G_i = 1] -  \E[Y_{i0}(0)  \mid G_i = 0]\}}_{\text{ Bias at $t=0$}}. \label{eq:parallel-tit2}
\end{align}
The difference between the mean potential outcome $Y_{it}(0)$ for the
treated and control group at time $t$, $\E[Y_{it}(0) \mid
G_{i} = 1] - \E[Y_{it}(0) \mid G_{i} = 0]$, is often called \emph{bias} (or selection bias) in the literature
\citep[e.g.,][]{heckman1998characterizing, cunningham2021causal}. Equation~\eqref{eq:parallel-tit2}
shows that the parallel trends-in-trends assumption allows for a linear change in bias over time, whereas the bias is assumed
to be constant over time in the extended parallel trends
assumption. This representation is useful when we generalize our
results to $K$ pre-treatment periods where $K > 2$. Importantly, equation~\eqref{eq:parallel-tit} and equation~\eqref{eq:parallel-tit2}
are equivalent, and therefore, researchers can choose whichever
interpretation easy for them to evaluate in each application.

\subsection{Connection to the Leads Test}
\label{subsec:leads}
Here we formally prove the connection between the test of
pre-treatment periods discussed in Section~\ref{subsec:assess} and the
well known leads test \citep{angrist2008mostly}. The leads test
includes $D_{i, t+1}$ into a linear regression and  check whether a
coefficient of $D_{i, t+1}$ is zero.

\subsubsection{Repeated Cross-Sectional Data}
In the repeated cross-sectional data setting, the leads test considers
the following linear regression.
  \begin{equation*}
    (\widehat{\theta}, \widehat{\gamma}, \widehat{\beta}, \widehat{\zeta}) = \argmin \sum^{n}_{i=1}\sum_{t=0}^1
    O_{it}\l( Y_{it} - \theta G_{i} - \gamma_t  -
    \beta D_{it} - \zeta D_{i, t+1} \r)^{2}.
  \end{equation*}
  Then, because $D_{it} = 0$ for all units in $t  = \{0, 1\}$, this
  least squares problem is the same as
  \begin{equation*}
    (\widehat{\theta}, \widehat{\gamma}, \widehat{\zeta}) = \argmin \sum^{n}_{i=1}\sum_{t=0}^1
    O_{it}\l( Y_{it} - \theta G_{i} - \gamma_t - \zeta D_{i, t+1} \r)^{2}.
  \end{equation*}
  Finally, using Result~\ref{rc-para}, we have
\begin{align*}
  \widehat{\zeta} &  = \l(\frac{\sum_{i\colon G_i = 1} Y_{i1}}{n_{11}} -
                    \frac{\sum_{i\colon G_i = 1} Y_{i0}}{n_{10}}\r)  - \l(\frac{\sum_{i\colon G_i = 0} Y_{i1}}{n_{01}} -  \frac{\sum_{i\colon G_i = 0} Y_{i0}}{n_{00}}\r),
\end{align*}
which is the standard DID estimator to the pre-treatment periods $t =
0, 1.$  \qed

\subsubsection{Panel Data}
In the panel data setting, the leads test considers the following two-way fixed effects regression.
\begin{equation*}
  (\widehat{\alpha}, \widehat{\delta}, \widehat{\beta}, \widehat{\zeta})
  = \argmin \sum^{n}_{i=1}\sum_{t=0}^{1}(Y_{it} - \alpha_{i} -
  \delta_{t} - \beta D_{it} - \zeta D_{i, t+1})^{2}.
\end{equation*}
Again,  this  least squares problem is the same as
\begin{equation*}
  (\widehat{\alpha}, \widehat{\delta}, \widehat{\zeta})
  = \argmin \sum^{n}_{i=1}\sum_{t=0}^{1}(Y_{it} - \alpha_{i} -
  \delta_{t} - \zeta D_{i, t+1})^{2}.
\end{equation*}
Then, using Result~\ref{tfe-para}, we have
\begin{align*}
\widehat{\zeta}
  &= \l(\frac{\sum_{i\colon G_i = 1} Y_{i1}}{n_{1}} -
                  \frac{\sum_{i\colon G_i = 1} Y_{i0}}{n_{1}}\r)  -
                  \l(\frac{\sum_{i\colon G_i = 0} Y_{i1}}{n_{0}} -
                  \frac{\sum_{i\colon G_i = 0} Y_{i0}}{n_{0}}\r),
\end{align*}
which is the standard DID estimator to the pre-treatment periods $t =
0, 1.$  \qed

\clearpage
\setcounter{equation}{1}

\section{Details of Double DID Estimator}
\label{sec:d-d-did}
\subsection{Properties of Double DID Estimator}
Here, we prove several important properties of the double DID
estimator based on the GMM theory \citep{hansen1982gmm}.

\begin{theorem}
  When the extended parallel trends assumption
  (Assumption~\ref{as-e-parallel}) holds, the double DID estimator
  with the optimal weight matrix (equation~\eqref{eq:optimalW} in the main paper) is
  consistent, and its asymptotic variance is smaller than or equal to
  that of the standard, extended, and sequential DID estimators, i.e., $\Var(\taudd) \leq \mbox{min}(\Var(\taud), \Var(\taus), \Var(\taue)).$
\end{theorem}

\subsection*{Proof.}
Suppose we define a moment function $m_i(\tau)$ as
\begin{equation*}
  m_i(\tau) = \begin{pmatrix}
    \tau - \widehat{\tau}_{\texttt{DID}}(i)\\[-5pt]
    \tau - \widehat{\tau}_{\texttt{s-DID}}(i)
  \end{pmatrix}
\end{equation*}
where
\begin{eqnarray*}
  \widehat{\tau}_{\texttt{DID}}(i) & = & \left(\frac{n}{n_{12}} G_i
                                         Y_{i2} - \frac{n}{n_{11}} G_i
                                         Y_{i1}\right) -
                                         \left(\frac{n}{n_{02}} (1-G_i)
                                         Y_{i2} - \frac{n}{n_{01}}
                                         (1-G_i) Y_{i1}\right) \nonumber\\
  \widehat{\tau}_{\texttt{s-DID}}(i) & = & \left\{\left(\frac{n}{n_{12}} G_i
                                         Y_{i2} - \frac{n}{n_{11}} G_i
                                         Y_{i1}\right) -
                                         \left(\frac{n}{n_{02}} (1-G_i)
                                         Y_{i2} - \frac{n}{n_{01}}
                                         (1-G_i) Y_{i1}\right)\right\}
  \nonumber \\
                                   && \hspace{0.2in} -
                                      \left\{\left(\frac{n}{n_{11}} G_i
                                      Y_{i1} - \frac{n}{n_{10}} G_i
                                      Y_{i0}\right) - \left(\frac{n}{n_{01}} (1-G_i) Y_{i1} - \frac{n}{n_{00}} (1-G_i) Y_{i0}\right)\right\}
\end{eqnarray*}
for the repeated cross-sectional setting. They can be similarly
defined in the panel data setting. Then, we can write the double DID estimator as the GMM estimator:
\begin{equation}
  \widehat{\tau}_{\texttt{d-DID}}(\mathbf{W}) = \argmin_{\tau} \left(\frac{1}{n}
    \sum_{i=1}^n m_i(\tau)\right)^\top
  \mathbf{W} \left(\frac{1}{n} \sum_{i=1}^n m_i(\tau)\right)  \label{eq:gmm-opt}
\end{equation}
where we index the double DID estimator by $\mathbf{W}$, which is a
weight matrix of dimension $2 \times 2$.

In general, the variance of the GMM estimator is given by
\begin{equation*}
  \Var(\widehat{\tau}_{\texttt{d-DID}}(\mathbf{W}))  = (M^\top \mathbf{W} M)^{-1} M^\top
  \mathbf{W} \Omega \mathbf{W}^\top M (M^\top \mathbf{W} M)^{-1}.
\end{equation*}
where $M =  \frac{1}{n} \sum_{i=1}^n \E
\left\{\frac{\partial}{\partial \tau}  m_i(\tau)\right\},$ and
\begin{equation*}
  \Omega = \begin{pmatrix}
    \Var(\widehat{\tau}_{\texttt{DID}}) &
    \Cov(\widehat{\tau}_{\texttt{DID}}, \widehat{\tau}_{\texttt{s-DID}})\\
    \Cov(\widehat{\tau}_{\texttt{DID}},
    \widehat{\tau}_{\texttt{s-DID}}) & \Var(\widehat{\tau}_{\texttt{s-DID}})
  \end{pmatrix}.
\end{equation*}
\cite{hansen1982gmm} showed in general that
$\Var(\widehat{\tau}_{\texttt{d-DID}}(\mathbf{W}))$ is minimized when
$\mathbf{W}$ is set to $\Omega^{-1}.$ We define this optimal weight
as $\mathbf{W}^\ast$
\begin{equation*}
  \mathbf{W}^\ast = \Omega^{-1} = \begin{pmatrix}
    \Var(\widehat{\tau}_{\texttt{DID}}) &
    \Cov(\widehat{\tau}_{\texttt{DID}}, \widehat{\tau}_{\texttt{s-DID}})\\
    \Cov(\widehat{\tau}_{\texttt{DID}},
    \widehat{\tau}_{\texttt{s-DID}}) & \Var(\widehat{\tau}_{\texttt{s-DID}})
  \end{pmatrix}^{-1}.
\end{equation*}

In general, the asymptotic variance of this optimal GMM estimator is given by
\begin{equation*}
  \Var(\widehat{\tau}_{\texttt{d-DID}}(\mathbf{W}^\ast))  = (M^\top
  \mathbf{W}^\ast M)^{-1}.
\end{equation*}
Because $M = \bm{1},$ the asymptotic variance of
$\Var(\widehat{\tau}_{\texttt{d-DID}}(\mathbf{W}^\ast))$ can be
explicitly written as
\begin{equation*}
  \Var(\taudd(\mathbf{W}^\ast)) = (\bm{1}^{\top} \*W^\ast \bm{1})^{-1} = \cfrac{\Var(\taud)\cdot \Var(\taus) - \Cov(\taud,
    \taus)^2}{\Var(\taud) + \Var(\taus) - 2 \Cov(\taud, \taus)}.
\end{equation*}
Finally, the standard, sequential, and extended DID estimators are all
special cases of the double DID with a specific choice of the
weight matrix as described in Table 1 of the main paper. Because
for any $\mathbf{W},$ $
\Var(\widehat{\tau}_{\texttt{d-DID}}(\mathbf{W}^\ast)) \leq
\Var(\widehat{\tau}_{\texttt{d-DID}}(\mathbf{W}))$, it implies that
$$
\Var(\taudd (\mathbf{W}^\ast)) \leq \mbox{min}(\Var(\taud), \Var(\taus), \Var(\taue)).
$$

Now, we can show the consistency of the estimator and its variance estimator.
The optimal weight matrix $\mathbf{W}^\ast$ can be estimated by its sample analog:
\begin{equation*}
  \widehat{\mathbf{W}} = \begin{pmatrix}
    \widehat{\Var}(\widehat{\tau}_{\texttt{DID}}) &
    \widehat{\Cov}(\widehat{\tau}_{\texttt{DID}}, \widehat{\tau}_{\texttt{s-DID}})\\
    \widehat{\Cov}(\widehat{\tau}_{\texttt{DID}},
    \widehat{\tau}_{\texttt{s-DID}}) & \widehat{\Var}(\widehat{\tau}_{\texttt{s-DID}})
  \end{pmatrix}^{-1}.
\end{equation*}
which is a consistent estimator of $\*W^{*}$ under the standard regularity conditions.
Therefore, by solving
the GMM optimization problem (equation~\eqref{eq:gmm-opt}), we can
explicitly write the double DID as
\begin{equation*}
  \widehat{\tau}_{\texttt{d-DID}}(\widehat{\mathbf{W}}) = \widehat{w}_1 \widehat{\tau}_{\texttt{DID}}
  + \widehat{w}_2 \widehat{\tau}_{\texttt{s-DID}}
\end{equation*}
where $\widehat{w}_1 + \widehat{w}_2 = 1$, and
\begin{align*}
  \widehat{w}_1 =  \cfrac{\widehat{\Var}(\widehat{\tau}_{\texttt{s-DID}}) -
             \widehat{\Cov}(\widehat{\tau}_{\texttt{DID}}, \widehat{\tau}_{\texttt{s-DID}})}{\widehat{\Var}(\widehat{\tau}_{\texttt{DID}})
             + \widehat{\Var}(\widehat{\tau}_{\texttt{s-DID}}) - 2\widehat{\Cov}(\widehat{\tau}_{\texttt{DID}},
             \widehat{\tau}_{\texttt{s-DID}})}, \\
\widehat{w}_2  =  \cfrac{\widehat{\Var}(\widehat{\tau}_{\texttt{DID}}) -
             \widehat{\Cov}(\widehat{\tau}_{\texttt{DID}}, \widehat{\tau}_{\texttt{s-DID}})}{\widehat{\Var}(\widehat{\tau}_{\texttt{DID}})
             + \widehat{\Var}(\widehat{\tau}_{\texttt{s-DID}}) - 2\widehat{\Cov}(\widehat{\tau}_{\texttt{DID}},
             \widehat{\tau}_{\texttt{s-DID}})}.
\end{align*}

Under the extended parallel trends assumption (Assumption~\ref{as-e-parallel}), both the standard
DID and the sequential DID estimator are consistent to the ATT.
Therefore, by the continuous mapping theorem and law of large numbers, we have
\begin{equation*}
\widehat{\tau}_{\texttt{d-DID}}(\widehat{\mathbf{W}}) \overset{p}{\to} \tau
\end{equation*}
and
\begin{equation*}
\widehat{\Var}(\widehat{\tau}_{\texttt{d-DID}}(\widehat{\*W})) \overset{p}{\to}
\Var(\widehat{\tau}_{\texttt{d-DID}}(\*W^{*})),
\end{equation*}
which complets the proof.
\qed

\subsection{Standard Error Estimation}
As described in Section~\ref{subsubsec:step2}, we use the block bootstrap.
\begin{enumerate}
  \item Estimate $\{\widehat{\tau}^{(b)}_{\texttt{DID}}, \widehat{\tau}^{(b)}_{\texttt{s-DID}}\}^{B}_{b=1}$ where $B$ indicates the total number of bootstrap iterations. We recommend the block-bootstrap where the block is taken at the level of treatment assignment.
  \item Estimate the optimal weight matrix via computing the variance-covariance matrix:
  \begin{align*}
  \widehat{\text{Var}}(\widehat{\tau}_{\texttt{DID}}) &= \frac{1}{B}\sum^{B}_{b=1}
    (\widehat{\tau}^{(b)}_{\texttt{DID}} - \overline{\widehat{\tau}}_{\texttt{DID}})^{2}\\
  \widehat{\text{Var}}(\widehat{\tau}_{\texttt{s-DID}}) &= \frac{1}{B}\sum^{B}_{b=1}
    (\widehat{\tau}^{(b)}_{\texttt{s-DID}} - \overline{\widehat{\tau}}_{\texttt{s-DID}})^{2}\\
  \widehat{\text{Cov}}(\widehat{\tau}_{\texttt{DID}}, \widehat{\tau}_{\texttt{s-DID}})
    &= \frac{1}{B}\sum^{B}_{b=1}
    (\widehat{\tau}^{(b)}_{\texttt{DID}} - \overline{\widehat{\tau}}_{\texttt{DID}})
    (\widehat{\tau}^{(b)}_{\texttt{s-DID}} - \overline{\widehat{\tau}}_{\texttt{s-DID}})
  \end{align*}
  where $\overline{\widehat{\tau}}_{\texttt{DID}} = \sum^{B}_{b=1}\widehat{\tau}^{(b)}_{\texttt{DID}}/B$,
  and $\overline{\widehat{\tau}}_{\texttt{s-DID}} = \sum^{B}_{b=1}\widehat{\tau}^{(b)}_{\texttt{s-DID}}/B$ are empirical average of two estimators.
  Finally, we obtain the estimate of the weight matrix by inverting the variance-covariance matrix (equation~\eqref{eq:optimalW} in the main text),
  $$
  \widehat{\*W} =
  \left(
    \begin{array}{cc}
    \widehat{\text{Var}}(\widehat{\tau}_{\texttt{DID}})
      & \widehat{\text{Cov}}(\widehat{\tau}_{\texttt{DID}}, \widehat{\tau}_{\texttt{s-DID}})\\
    \widehat{\text{Cov}}(\widehat{\tau}_{\texttt{DID}}, \widehat{\tau}_{\texttt{s-DID}})
      & \widehat{\text{Var}}(\widehat{\tau}_{\texttt{s-DID}})
    \end{array}
  \right)^{-1}
  $$
  \item The double DID estimator is given by equation~\eqref{eq:d-did-w} in the main
    paper.
  \item The variance of double DID estimator is then obtained via the standard efficient GMM variance formula
  \begin{equation*}
  \widehat{\text{Var}}(\widehat{\tau}_{\texttt{d-DID}}) = (\bm{1}^{\top}\widehat{\mathbf{W}} \bm{1})^{-1}.
  \end{equation*}

\end{enumerate}

\clearpage
\section{Extensions of Double DID}
\subsection{Double DID Regression}
\label{sec:d-reg}
Like other DID estimators, the double DID estimator has a nice connection
to a widely-used regression approach. Using this double DID
regression, researchers can include other pre-treatment
covariates $\bX_{it}$ to make the DID design more robust and
efficient. We provide technical details in Appendix

To introduce the regression-based double DID estimator, we begin with the basic DID.
As discussed in Appendix~\ref{subsec:did-reg}, the basic DID estimator is equivalent to a coefficient in the linear
regression of equation~\eqref{eq:linear-did}. Inspired by this connection, researchers
often adjust for additional pre-treatment covariates as:
\begin{equation}
  Y_{it}  \sim \alpha + \theta G_i +  \gamma I_t + \beta (G_i \times
  I_t) + \bX_{it}^\top \bm{\rho}, \label{eq:linear-did-cov}
\end{equation}
where we adjust for the additional pre-treatment covariates
$\bX_{it}$. A coefficient of the interaction term $\widehat{\beta}$ is a consistent estimator for the ATT when the conditional parallel trends assumption holds and the
parametric model is correctly specified.
Here, we make the parallel trends assumption \textit{conditional}
on covariates $\bX_{it}$. The idea is that even when the parallel trends
assumption might not hold without controlling for any covariates,
trends of the two groups might be parallel conditionally after adjusting for observed covariates.
For example, the conditional parallel trends
assumption means that treatment and control groups have the same
trends of the infrastructure quality after controlling for population and GDP per capita.

The sequential DID estimator is extended similarly. Based on the connection to the linear
regression of equation~\eqref{eq:linear-sdid}, we can adjust for additional pre-treatment covariates as:
\begin{equation}
  \Delta Y_{it}\sim \alpha_s + \theta_s G_i +  \gamma_s I_t + \beta_s (G_i \times
  I_t) + \bX_{it}^\top \bm{\rho}_s, \label{eq:linear-sdid-cov}
\end{equation}
where  $\Delta Y_{it}  =  Y_{it} - (\sum_{i\colon G_i=1} Y_{i,t-1})/n_{1, t-1}$ if $G_i = 1$ and $\Delta Y_{it}  = Y_{it} -
(\sum_{i\colon G_i=0} Y_{i,t-1})/n_{0, t-1} $  if $G_i = 0$. The
estimated coefficient $\widehat{\beta}_s$ is consistent for the ATT under the
\textit{conditional} parallel trends-in-trends assumption and the
conventional assumption of correct specification.

The double DID regression combines the two regression
estimators via the GMM:
\begin{equation}
  \widehat{\beta}_{\texttt{d-DID}} = \argmin_{\beta_d}
  \begin{pmatrix}
    \beta_d - \widehat{\beta}  \\[-5pt]
    \beta_d - \widehat{\beta}_s
  \end{pmatrix}^\top
  \mathbf{W}
  \begin{pmatrix}
    \beta_d - \widehat{\beta}  \\[-5pt]
    \beta_d - \widehat{\beta}_s
  \end{pmatrix} \label{eq:d-did-cov}
\end{equation}
where $\mathbf{W}$ is a weighting matrix of dimension $2
\times 2$.

Thus, as the double DID estimator with no covariates, the double DID regression has
two steps. The first step is to assess the underlying assumptions. Here,
instead of using the standard DID estimator,
we use the standard DID regression on pre-treatment
periods to assess the conditional extended parallel trends
assumption. The
second step is to estimate the ATT, while adjusting for pre-treatment covariates. Instead of using the double
DID estimator without covariates, we implement the regression-based
double DID estimator (equation~\eqref{eq:d-did-cov}).

\clearpage
\subsection{Generalized $K$-DID}
\label{sec:general}
In this section, we propose the generalized $K$-DID, which extends the
double DID in Section~\ref{sec:d-did} to arbitrary number of \textit{pre}- and \textit{post}-treatment periods
in the basic DID setting. We consider the staggered adoption
design in Section~\ref{sec:sad}.

\subsubsection{The Setup and Causal Quantities of Interest}\label{subsec:g-setup}
We first extend the setup to account for arbitrary number of
pre- and post-treatment periods. Suppose we observe outcome $Y_{it}$ for $i \in \{1, \ldots, n\}$ and
$t \in \{0, 1,\ldots, T\}$. We define the binary treatment variable to be
$D_{it} \in \{0, 1\}$. The treatment is assigned right before time period $T^\ast$,
and thus, time periods $t \in \{T^\ast,  \ldots, T\}$ are the post-treatment periods and time
periods $t \in \{0, \ldots, T^\ast-1\}$ are the pre-treatment
periods. As in Section~\ref{sec:benefit}, we denote the treatment group as $G_{i} = 1$ and
$G_{i} = 0$ otherwise. Note that $D_{it} = 0$ for $t \in \{1, \ldots, T^\ast\}$ for all units.

We are interested in the causal effect at post-treatment time $T^\ast + s$ where $s  \geq
0$. When $s = 0$, this corresponds to the contemporaneous treatment effect. By specifying
different values of $s > 0,$ researchers can study a variety of long-term causal
effects of the treatment. Formally, our quantity of interest is
the average treatment effect on the treated  (ATT) at post-treatment time $T^\ast+s$.
\begin{equation*}
  \tau(s) \equiv \E[Y_{i,T^\ast + s}(1) - Y_{i,T^\ast  + s}(0) \mid G_{i} = 1].
\end{equation*}
For example, when $s = 3$, this could mean the causal effect of the
policy after three years from its initial introduction. This definition is a
generalization of the standard ATT: when $s = 0,$ this quantity is equal to the ATT defined in equation~\eqref{eq:att}.

\subsubsection{Generalize Parallel Trends Assumptions}\label{subsec:g-assumption}
What assumptions do we need to identify the ATT at post-treatment time
$T^\ast+s$?
Here, we provide a generalization of the parallel trends assumption,
which incorporates both the standard parallel trends assumption and
the parallel trends-in-trends assumption.

\begin{assumption}[$k$-th Order Parallel Trends]
  \label{K-e-parallel} For some integer $k$ such that $1 \leq k \leq T^\ast,$
  \begin{equation*}
    \Delta_s^{k}\l(\E[Y_{i,T^\ast + s}(0) \mid G_{i} = 1]\r)  = \Delta_s^{k}\l(\E[Y_{i,T^\ast + s}(0) \mid G_{i} = 0]\r),
  \end{equation*}
\end{assumption}
where $\Delta_s^{k}$ is the $k$-th order difference operator defined
recursively  as follows. For $g \in \{0, 1\}$,
\begin{align*}
  \Delta^1_s \l(\E[Y_{i,T^\ast + s}(0) \mid G_{i} = g]\r)  \equiv
  \E[Y_{i,T^\ast + s}(0)\mid G_{i} = g] - \E[Y_{i, T^\ast-1}(0) \mid G_{i} = g],
\end{align*}
when $k = 1$ and, in general,
\begin{align*}
  & \Delta_s^k \l(\E[Y_{i,T^\ast + s}(0) \mid G_{i} = g]\r) \\
    \equiv \ &  \Delta^{k-1}_s \l(\E[Y_{i,T^\ast + s}(0) \mid  G_{i} =
    g]\r) - M^k_{s}\Delta^{k-1} \l(\E[Y_{i, T^\ast -1}(0) \mid G_{i} =
             g]\r),\\
  = \ &  \E[Y_{i,T^\ast + s}(0)\mid G_{i} = g] - \E[Y_{i,T^\ast-1}(0) \mid
      G_{i} = g] - \sum_{j=1}^{k-1} M^{j+1}_{s}\Delta^{j} \l(\E[Y_{i, T^\ast-1}(0) \mid G_{i} =
             g]\r),
\end{align*}
where $M^{\ell}_s = \prod_{j=1}^{\ell-1} (s+j)/\prod_{j=1}^{\ell-1}
j$ for $\ell \geq 2$. $\Delta^{k} \l(\E[Y_{i, T^\ast-1}(0) \mid G_{i}
=g]\r)$ is also recursively defined as $\Delta^{k} \l(\E[Y_{i,
  T^\ast-1}(0) \mid G_{i} =g]\r)  \equiv \Delta^{k-1} \l(\E[Y_{i,
  T^\ast-1}(0) \mid G_{i} =g]\r) - \Delta^{k-1} \l(\E[Y_{i,
  T^\ast-2}(0) \mid G_{i} =g]\r),$ and $\Delta^{1} \l(\E[Y_{i,
  T^\ast-m}(0) \mid G_{i} =g]\r) =    \E[Y_{i,T^\ast -m}(0)\mid G_{i}
= g] - \E[Y_{i, T^\ast-m-1}(0) \mid G_{i} = g]$ for $m = \{1, 2\}.$
The standard parallel trends assumption and the parallel-trends-in-trends assumption
are both special cases of this assumption.
The $k$-th order parallel trends assumption reduces to the standard
parallel trends assumption (Assumption~\ref{as-parallel})
when $s =1$ and $k=1$, and
to the parallel-trends-in-trends assumption
(Assumption~\ref{as-parallel-tit}) when $s=1$ and $k=2.$

To further clarify the meaning of Assumption~\ref{K-e-parallel}, we can consider a simpler but
stronger condition. In particular, the $k$-th order parallel trends assumption
(Assumption~\ref{K-e-parallel}) is implied by the following $p$-th degree polynomial model of confounding.
\begin{equation*}
  \E[Y_{it}(0) \mid G_i = 1] -  \E[Y_{it}(0)  \mid G_i = 0]  = \alpha + \sum_{p=1}^{k-1} \Gamma_p t^p,
\end{equation*}
with unknown parameters $\alpha$ and $\bm{\Gamma}$.
Here, the left hand side of the equality captures the difference between the two groups (treatment and control)
in terms of the mean of potential outcomes under the control condition.
This representation shows
that the standard parallel
trends assumption (Assumption~\ref{as-parallel}) is implied by the
time-invariant confounding; the parallel trends-in-trends assumption (Assumption~\ref{as-parallel-tit})
 is implied by the linear
time-varying confounding; and in general, the $k$-th order parallel
trends assumption is implied by the $k$-th order polynomial confounding.

\subsubsection{Estimate ATT with Multiple Pre- and Post-Treatment Periods}\label{subsec:g-estimation}
We consider the identification and estimation of the ATT at
post-treatment time $T^\ast +s$. Under the $k$-th order parallel trends assumption  (Assumption~\ref{K-e-parallel}), the
ATT is identified as follows.
\begin{equation*}
  \tau(s) =  \Delta_s^{k}\l(\E[Y_{i,T^\ast + s} \mid G_{i} = 1]\r)  -
  \Delta_s^{k}\l(\E[Y_{i,T^\ast + s} \mid G_{i} = 0]\r).
\end{equation*}
Because each conditional expectation can be consistently estimated via its sample analogue,
\begin{equation*}
  \widehat{\tau}_k(s) =  \Delta_s^{k}\l( \frac{\sum_{i\colon G_i=1}
    Y_{i,T^\ast + s}}{n_{1,T^\ast + s}}  \r)  -  \Delta_s^{k}\l(\frac{\sum_{i\colon G_i=0} Y_{i,T^\ast+s}}{n_{0,T^\ast+s}} \r)
\end{equation*}
is a consistent estimator for the ATT at time $T^\ast + s$ under the $k$-th order parallel
trends assumption. When $s=0$ and $k=1$, this estimator corresponds to the standard DID
estimator (equation~\eqref{eq:did-est}). When $s=0$ and $k=2,$ this is equal to
the sequential DID estimator (equation~\eqref{eq:s-did}). While
existing approaches \citep[e.g.,][]{angrist2008mostly, mora2012did,
  lee2016did, mora2019did} consider
each estimator separately, we propose combining multiple DID
estimators within the GMM framework.

In general, the generalized double DID combines $K$ moment conditions
where $K$ is the number of pre-treatment periods researchers use. When there are more than two pre-treatment periods, we
can naturally combine more than two DID estimators, improving upon the
double DID in Section~\ref{sec:d-did}. Formally, the generalized
double DID is defined as,
\begin{equation*}
  \widehat{\tau}(s)  = \argmin_\tau \bm{g} (\tau)^\top \widehat{\mathbf{W}} \bm{g}(\tau)
\end{equation*}
where $\bm{g} (\tau)  =  (\tau - \widehat{\tau}_1(s), \ldots, \tau -
\widehat{\tau}_{K}(s))^\top$. Based on the theory of the efficient GMM \citep{hansen1982gmm}, the optimal weight matrix is
$\widehat{\mathbf{W}} = \Var (\widehat{\tau}_{(1:K)}(s))^{-1}$
where $\Var(\cdot)$ is the variance-covariance matrix and
$\widehat{\tau}_{(1:K)}(s)  =  (\widehat{\tau}_1(s), \ldots,
\widehat{\tau}_{K}(s))^\top.$ When $T^\ast=2,$ this converges to the
standard DID estimator (equation~\eqref{eq:did-est}).
When $T^\ast=3,$ this corresponds to the basic form of the double DID estimator
(equation~\eqref{eq:d-did}). Within the GMM framework, we can select
moment conditions using the J-statistics \citep{hansen1982gmm}. We can
similarly generalize the double DID regression.

To assess the extended parallel trends assumption, we can apply the
generalized double DID to pre-treatment periods $t \in \{1,
\ldots, T^\ast-1\}$ as if the last pre-treatment period $T^\ast-1$ is the
target time period. Moments are $\bm{g} (\tau)  =  (\tau - \widehat{\tau}_1(0), \ldots, \tau -
\widehat{\tau}_{K}(0))^\top$ where $\widehat{\tau}_k(0) =
\Delta_s^{k}\l( \frac{\sum_{i\colon G_i=1} Y_{i,T^\ast-1}}{n_{1,T^\ast-1}}  \r)  -
\Delta_s^{k}\l(\frac{\sum_{i\colon G_i=0} Y_{i,T^\ast-1}}{n_{0,T^\ast-1}} \r).$
Similarly, to assess the extended parallel trends-in-trends
assumption, we can apply the generalized double DID to
pre-treatment periods with moments $\bm{g} (\tau)  =  (\tau - \widehat{\tau}_2(0), \ldots, \tau -
\widehat{\tau}_{K}(0))^\top$.

\subsection{Generalized $K$-DID for Staggered Adoption Design}
\label{sec:K-sad}
Combining the setup introduced in Section~\ref{subsec:g-setup} and the
one in Section~\ref{subsec:setup-sad}, we propose the generalized
$K$-DID for the SA design, which allows researchers to estimate
long-term causal effects in the SA design. We focus on the
SA-ATT at post-treatment time $t + s$ where $t$ is the timing of the treatment assignment
and $s  \geq 0$ represents how far in the future we want estimate the ATT
for. We first redefine the group indicator $G$ to estimate the
long-term SA-ATT at post-treatment time $t+s$. In particular, we define
\begin{equation*}
  G_{its} \ = \
  \begin{cases}
    \ \ 1 & \mbox{if  } A_i = t  \\
    \ \ 0 & \mbox{if  } A_i > t +s  \\
    \ \ -1 & \mbox{otherwise}
  \end{cases}
\end{equation*}
where $G_{its} = 1$ represents units who receive the treatment at time
$t$, and $G_{its} = 0$ indicates units who do not receive the
treatment by time $t+s$. $G_{its} = -1$ includes other units who
receive the treatment before time $t$ or receive the treatment
between $t+1$ and $t+s.$ When $s=0$, this definition corresponds to
the group indicator in equation~\eqref{eq:group-sad}.

Formally, our first quantity of interest is the
\textit{staggered-adoption average treatment effect on the treated}
(SA-ATT) at post-treatment time $t+s$.
\begin{equation*}
  \stau(s, t) \equiv \E[Y_{i,t + s}(1) - Y_{i,t  + s}(0) \mid G_{its} = 1].
\end{equation*}
By averaging over time, we can also define the
\textit{time-average staggered-adoption average treatment effect on the treated}
(time-average SA-ATT) at $s$ periods after treatment onset.
\begin{equation*}
  \ostau(s) \equiv \sum_{t \in \cT} \pi_t\stau(s, t),
\end{equation*}
where $\cT$ represents a set of the time periods for which
researchers want to estimate the ATT. The SA-ATT in period $t$, $\stau(t),$ is weighted by the proportion of units who receive the treatment at time
$t$: $\pi_t = \sum_{i=1}^n \mathbf{1}\{A_i = t\}/\sum_{i=1}^n
\mathbf{1}\{A_i \in \cT\}$.

Here, we provide a generalization of the parallel trends assumption,
which incorporates both the standard parallel trends assumption and
the parallel trends-in-trends assumption.

\begin{assumption}[$k$-th Order Parallel Trends for Staggered Adoption
  Design]
  \label{K-e-parallel-sad} For some integer $k$ such that $1 \leq k \leq
  T,$ and for $k \leq t \leq T-s,$
  \begin{equation*}
    \Delta_s^{k}\l(\E[Y_{i,t + s}(0) \mid G_{its} = 1]\r)  = \Delta_s^{k}\l(\E[Y_{i,t + s}(0) \mid G_{its} = 0]\r),
  \end{equation*}
\end{assumption}
where $\Delta_s^{k}$ is the $k$-th order difference operator defined
in Assumption~\ref{K-e-parallel}.

Under Assumption~\ref{K-e-parallel-sad}, the SA-ATT at post-treatment time $t+s$ is
identified as follows.
\begin{equation*}
  \stau(s, t) =  \Delta_s^{k}\l(\E[Y_{i,t + s} \mid G_{its} = 1]\r)  -
  \Delta_s^{k}\l(\E[Y_{i,t + s} \mid G_{its} = 0]\r).
\end{equation*}
Since conditional expectations can be consistently estimated via the sample analogue,
\begin{equation*}
  \wstau_k(s,t) =  \Delta_s^{k}\l( \frac{\sum_{i\colon G_{its}=1}
    Y_{i,t + s}}{n_{1,t + s}}  \r)  -  \Delta_s^{k}\l(\frac{\sum_{i\colon G_{its}=0} Y_{i,t+s}}{n_{0,t+s}} \r)
\end{equation*}
is a consistent estimator for the SA-ATT at post-treatment time $t+s$
under Assumption~\ref{K-e-parallel-sad}.

In general, we combine $K$ DID estimators to obtain the generalized $K$-DID for
the SA-ATT at post-treatment time $t+s$ as follows.
\begin{equation*}
  \wstau(s, t)  = \argmin_{\stau} \bm{g} (\stau)^\top \widehat{\mathbf{W}} \bm{g}(\stau)
\end{equation*}
where $\bm{g} (\stau)  =  (\stau - \wstau_1(s), \ldots, \stau -
\wstau_K(s))^\top$. The optimal weight matrix is
$\widehat{\mathbf{W}} = \Var (\wstau_{(1: K)}(s))^{-1}$
where $\wstau_{(1:K)}(s)  =  (\wstau_1(s), \ldots,
\wstau_K(s))^\top.$

To estimate the time-average SA-ATT, we first define the time-average
$k$-th order time-average DID estimator as,
\begin{equation*}
  \wostau_k(s) = \sum_{t \in \cT} \pi_t \wstau_k(s,t).
\end{equation*}

Finally, the generalized $K$-DID combines $K$ moment conditions
as follows.
\begin{equation*}
  \wostau(s)  = \argmin_{\ostau} \bm{g} (\ostau)^\top \widehat{\mathbf{\overline{W}}} \bm{g}(\ostau)
\end{equation*}
where $\bm{g} (\ostau)  =  (\ostau - \wostau_1(s), \ldots, \ostau -
\wostau_K(s))^\top$. The optimal weight matrix is
$\widehat{\mathbf{\overline{W}}} = \Var (\wostau_{(1: K)}(s))^{-1}$
where $\wostau_{(1:K)}(s)  =  (\wostau_1(s), \ldots,
\wostau_K(s))^\top.$

\clearpage
\subsection{Double DID Regression for Staggered Adoption Design}
\label{sec:sa-reg}
We now extend the double DID regression to the SA design setting.
We first extend the standard DID regression
(Appendix~\ref{sec:d-reg}) to the SA design.
In particular, to estimate the SA-ATT at time $t$,
we can fit the following regression for units who are not yet treated at time $t-1$, that is, $\{i: G_{it} \geq 0\}$.
\begin{equation*}
  Y_{iv}  \sim \alpha + \theta G_{it} +  \gamma I_v + \sbeta(t) (G_{it}
  \times I_v) + \bX_{iv}^\top \bm{\rho}, \label{eq:linear-did-cov-sad}
\end{equation*}
where $v \in \{t-1, t\}$ and the time indicator $I_v$ (equal to $1$ if
$v = t$ and $0$ if $v = t-1$). Note that $G_{it}$ defines the
treatment and control group at time $t$, and thus, it does not depend
on time index $v$. The estimated coefficient $\wsbeta(t)$ is consistent for the
SA-ATT under the conditional parallel trends assumption.

Similarly, we can extend the sequential DID regression to the SA
design. Using the connection to the linear regression of
equation~\eqref{eq:linear-sdid}, we can adjust for additional
pre-treatment covariates as:
\begin{equation*}
  \Delta Y_{iv}\sim \alpha_s + \theta_s G_{it} +  \gamma_s I_v + \sbeta_s(t) (G_{it} \times
  I_v) + \bX_{iv}^\top \bm{\rho}_s, \label{eq:linear-sdid-cov-sad}
\end{equation*}
where $v \in \{t - 1, t\}$ and $\Delta Y_{iv}  =  Y_{iv} - (\sum_{i\colon G_{it}=1}
Y_{i, v-1})/n_{1, v-1}$ if $G_{it} = 1$ and $\Delta Y_{iv}  = Y_{iv} -
(\sum_{i\colon G_{it}=0} Y_{i,v-1})/n_{0, v-1} $  if $G_{it} = 0$. The
estimated coefficient $\wsbeta_s(t)$ is consistent for the
SA-ATT under the conditional parallel trends-in-trends assumption.

Therefore, the double DID regression for the SA design combines the two regression estimators via the GMM:
\begin{equation*}
  \wsbeta_{\texttt{d-DID}}(t) = \argmin_{\sbeta_d(t)}
  \begin{pmatrix}
    \sbeta_d(t) - \wsbeta(t)  \\[-5pt]
    \sbeta_d(t) - \wsbeta_s(t)
  \end{pmatrix}^\top
  \mathbf{W}(t)
  \begin{pmatrix}
    \sbeta_d(t) - \wsbeta(t)  \\[-5pt]
    \sbeta_d(t) - \wsbeta_s(t).
  \end{pmatrix} \label{eq:d-did-cov-sad}
\end{equation*}
where the choice of the weight matrix follows the same two-step
procedure as Section~\ref{subsec:sa-d-did}. We also provide further
details in Appendix~\ref{sec:K-sad}. The optimal weight matrix $\mathbf{W}(t)$ is equal to $\Var
(\wsbeta_{(1:2)}(t))^{-1}$ where $\wsbeta_{(1:2)} (t) =  (\wsbeta(t), \wsbeta_s(t))^\top.$

To estimate the time-average SA-ATT, we extend the double DID
regression as follows.
\begin{equation*}
  \wosbeta_{\texttt{d-DID}} = \argmin_{\osbeta_d}
  \begin{pmatrix}
    \osbeta_d - \wosbeta  \\[-5pt]
    \osbeta_d - \wosbeta_s
  \end{pmatrix}^\top
  \overline{\*W}
  \begin{pmatrix}
    \osbeta_d - \wosbeta  \\[-5pt]
    \osbeta_d - \wosbeta_s
  \end{pmatrix} \label{eq:d-did-cov-o}
\end{equation*}
where
\begin{equation*}
  \wosbeta = \sum_{t \in \cT} \pi_t \wsbeta(t), \hspace{0.2in} \mbox{and} \hspace{0.2in}
  \wosbeta_s =  \sum_{t \in \cT} \pi_t \wsbeta_s(t).
\end{equation*}
The optimal weight matrix $\overline{\*W}$ is equal
to $\Var (\wosbeta_{(1:2)})^{-1}$ where $\wosbeta_{(1:2)}  =  (\wosbeta, \wosbeta_s)^\top.$

\clearpage
\section{Equivalence Approach}
\label{sec:equiv}
Here, we provide technical details on the equivalence approach we
introduced in Section~\ref{subsec:b-d-did}. In the standard hypothesis
testing, researchers usually evaluate the two-sided null hypothesis
$H_0: \delta = 0$ where $\delta =  \{\E[Y_{i1}(0) \mid G_i = 1] -  \E[Y_{i0}(0)  \mid G_i = 1]\}  -
\{\E[Y_{i1}(0) \mid G_i = 0] -  \E[Y_{i0}(0)  \mid G_i = 0]\}$ when we
are conducting the pre-treament-trends test. However, this approach
has a risk of conflating evidence for parallel trends and statistical inefficiency. For example, when sample size
is small, even if pre-treatment trends of the treatment and control
groups differ (i.e., the null hypothesis is false), a test of
the difference might not be statistically significant due to large
standard error. And, analysts might “pass” the pre-treatment-trends
test by not finding enough evidence for the difference.

The equivalence approach can mitigate this concern by flipping the
null hypothesis, so that the rejection of the null can be the evidence for parallel trends.
In particular, we consider two one-sided tests:
\begin{equation*}
  H_0: \theta \geq \gamma_U, \ \ \mbox{or} \ \ \theta \leq \gamma_L
\end{equation*}
where $(\gamma_U, \gamma_L)$ is a user-specified equivalence range. By
rejecting this null hypothesis, researchers can provide statistical
evidence for the alternative hypothesis:
\begin{equation*}
  H_0: \gamma_L < \theta < \gamma_U,
\end{equation*}
which means that $\theta$ (i.e., the difference in pre-treament-trends
across treatment and control groups) are within an interval
$[\gamma_L, \gamma_U].$

One difficulty of the equivalence approach is that researchers have to
choose this equivalence range $(\gamma_U, \gamma_L)$, which might not
be straightforward in practice. To overcome this challenge, we follow
\cite{hartman2018equi} to estimate the 95\% equivalence confidence
interval, which is the smallest equivalence range supported by the
observed data. Suppose we obtain $[-c, c]$ as the symmetric 95\%
equivalence confidence interval where $c > 0$ is some positive
constant. Then, this means that if researchers think the absolute
value of $\theta$ smaller than $c$ is substantively negligible, the
5\% equivalence test would reject the null hypothesis and provide the
evidence for the parallel pre-treatment-trends. In contrast,  if researchers think the absolute
value of $\theta$ being $c$ is substantively too large as bias in
practice, the 5\% equivalence test would fail to reject the null
hypothesis and cannot provide the evidence for the parallel
pre-treatment-trends. In sum, by estimating the equivalence confidence
interval, readers of the analysis can decide how much evidence for the
parallel pre-treatment-trends exists in the observed data. Researchers
can estimate the 95\% equivalence confidence interval by the following
general two steps. First, estimate 90\% confidence interval, which we
denote by $[b_L, b_U].$ Second, we can obtain the symmetric 95\% equivalence
confidence interval as $[-b, b]$ where we define $b = \max\{|b_L|,
  |b_U|\}.$ See \cite{wellek2010equi, hartman2018equi} for more details.

\begin{figure}[!h]
  \begin{center}
    \includegraphics[width = 0.7\textwidth]{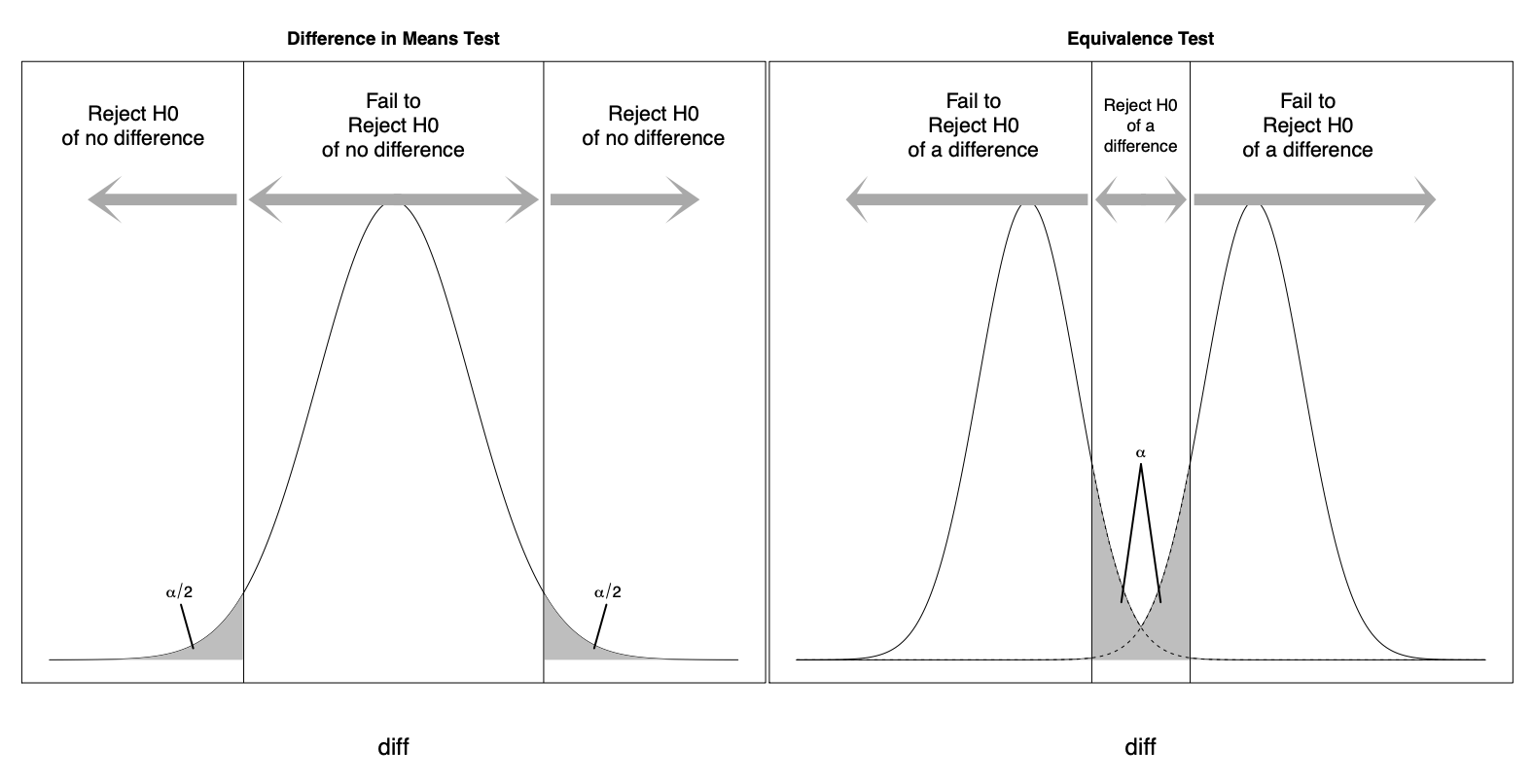}
  \end{center}
  \vspace{-0.25in} \spacingset{1}{\caption{Figure 1 from
      \cite{hartman2018equi} on the difference between the standard
      hypothesis testing and the equivalence testing.}\label{fig:equiv}}
\end{figure}

\clearpage
\section{Simulation Study}\label{sec:sim}
We conduct a simulation study to compare the performance of the
various DID estimators discussed in this paper. We demonstrate two key
results. First, the double DID is unbiased under the extended parallel
trends assumption or under the parallel trends-in-trends
assumption. Second,  the double DID has the smallest standard errors among
unbiased DID estimators. In particular, standard errors of the double
DID are smaller than those of the extended DID (i.e., the two-way fixed effects
estimator) even under the extended parallel trends assumption.

We compare three DID estimators --- the
double DID, the extended DID, and the sequential DID  --- using two
scenarios. In the first scenario, the extended parallel trends
assumption (Assumption~\ref{as-e-parallel}) holds where the
difference between potential outcomes under control $\E[Y_{it}(0) \mid G_i
= 1] - \E[Y_{it}(0) \mid G_i =0]$ is constant over time.
This corresponds to time-invariant unmeasured confounding, and we expect
that all the DID estimators are unbiased in this scenario.
The second scenario represents the parallel-trends-in-trends assumption
(Assumption~\ref{as-parallel-tit}) where unmeasured confounding varies
over time linearly. Here, we expect that the double DID and the sequential DID are unbiased,
whereas the extended DID is biased.

For each of the two scenarios, we consider the balanced panel data with $n$ units and five-time periods where treatments are assigned at the last time period.
We vary the number of units $(n)$ from $100$ to $1000$ and evaluate the quality
of estimators by absolute bias and standard errors over 2000
Monte Carlo simulations. We describe the details of the simulation setup next.

\subsection{Simulation Design}
We consider the balanced panel data with $T = 5$ ($t = \{0, 1, 2, 3,
4\}$) where the last period ($t = 4$) is treated as the post-treatment period.
We vary the number of units at each time period as $n \in \{100, 250,
500, 1000\}$. Thus, the total number of observations are $nT \in \{500, 1250,
2500, 5000\}$. We compare three estimators: the double DID, the
extended DID, and the sequential DID.

Note that we consider four pre-treatment periods here, and thus the generalized double DID is
not equal to the sequential DID even under the parallel
trends-in-trends assumption because it combines two other moments and
optimally weight them (see Appendix~\ref{sec:general}). The equivalence between the sequential DID and
the double DID holds only when there are two pre-treatment
periods. We see below that the generalized double DID improves upon the sequential DID
even under the parallel trends-in-trends assumption as they optimally weight
observations from different time periods.

We study two scenarios: one under the extended parallel trends
assumption (Assumption~\ref{as-e-parallel}) and the other under the parallel-trends-in-trends assumption
(Assumption~\ref{as-parallel-tit}). In the first scenario, the
difference between potential outcomes under control $\E[Y_{it}(0) \mid G_i
= 1] - \E[Y_{it}(0) \mid G_i =0]$ is constant over time. In
particular, we set
\begin{equation}
  \E[Y_{it}(0) \mid G_i = g] = \alpha_t + 0.05 \times g
\end{equation}
where $(\alpha_0, \alpha_1, \alpha_2, \alpha_3, \alpha_4) = (1, 2, 3,
4, 5).$ In the second scenario, we allow for linear time-varying
confounding. In particular, we set
\begin{equation}
  \E[Y_{it}(0) \mid G_i = g] = \alpha_t + 0.1\times g \times (t+1)
\end{equation}
where $(\alpha_0, \alpha_1, \alpha_2, \alpha_3, \alpha_4) = (1, 2, 3,
4, 5).$

Then, potential outcomes under control are drawn as follows.
  $Y_{it}(0) = \E[Y_{it}(0) \mid G_i] + \epsilon_{it}$  where $\epsilon_{it}$ follows the AR(1) process with autocorrelation
parameter $\rho.$ That is,
\begin{align*}
  \epsilon_{it} & =   \rho \epsilon_{i, t-1} + \xi_{it}, \\
  \epsilon_{i0}  & =  \mathcal{N}(0, 3/(1-\rho^2)),\\
  \xi_{it}  & =  \mathcal{N}(0, 3).
\end{align*}
The causal effect is denoted by $\tau$ and thus, $Y_{it}(1) = \tau +
Y_{it}(0)$ where we set $\tau =  0.2.$ Finally, $Y_{it} = Y_{it}(0)$ for
$t  \leq 3$ (pre-treatment periods) and $Y_{it} = G_i Y_{it}(1) +
(1-G_i) Y_{it} (0)$ for $t = 4$ (post-treatment period). The half of
the samples are in the treatment group ($G_i = 1$) and the other half
is in the control group ($G_i  =  0$).

In Figure~\ref{fig:sim-efficiency-diff}, we set the autocorrelation
parameter $\rho = 0.6$. This value is similar to the autocorrelation parameter used in famous simulation studies in
\cite{bertrand2004much} ($\rho = 0.8$). We pick a smaller value to
make our simulations harder as we see below. In Figure~\ref{fig:sim-add},
we also provide additional results where we consider a
full range of the autocorrelation parameters $\rho \in \{0, 0.2, 0.4, 0.6,
0.8\}$ (the same positive autocorrelation values considered in
\cite{bertrand2004much}).
Both figures show the absolute bias and the standard errors
which are defined as
\begin{align*}
\text{absolute bias} =
\bigg|
  \frac{1}{M}\sum^{M}_{m=1}(\widehat{\tau}_{m} - \tau)
\bigg|
\quad\text{and}\quad
\text{standard error} =
\sqrt{\frac{1}{M}\sum^{M}_{m=1}(\widehat{\tau}_{m} - \tau)^{2}},
\end{align*}
where $M$ is the total number of Monte Carlo iterations. Note that
this standard error is a true standard error over the sampling distribution.

\subsection{Results}
\begin{figure}[!t]
  \centering
  \includegraphics[width = 0.7\textwidth]{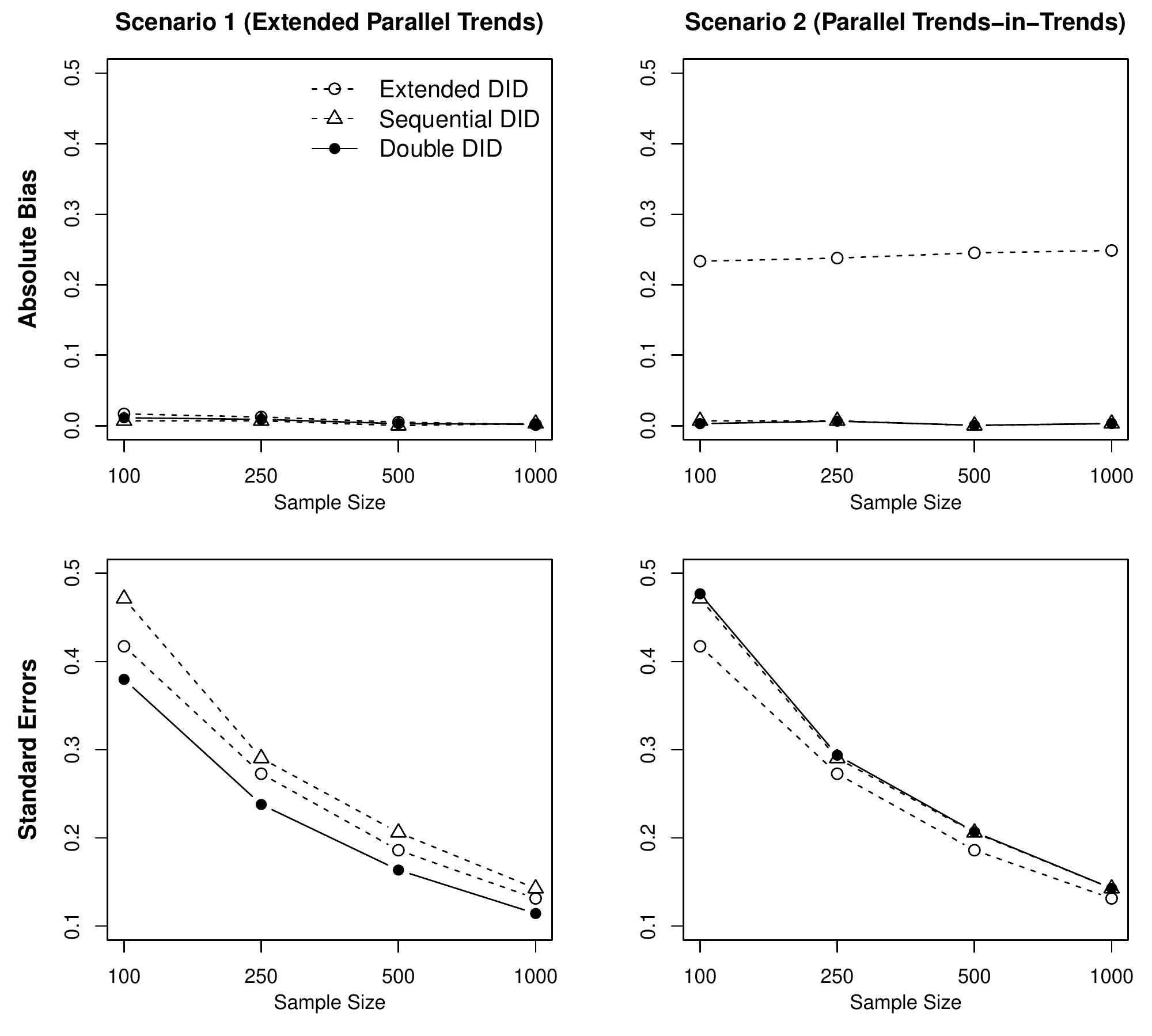}
  \vspace{-0.05in} \spacingset{1}{
    \caption{Comparing DID estimators in terms of the absolute bias
      and the standard errors. The first row shows that the double DID
      estimator (black circle with solid line) is unbiased under both
      scenarios. The second row demonstrates that the double DID has
      the smallest standard errors among unbiased DID estimators.}
    \label{fig:sim-efficiency-diff}
  }
\end{figure}

Figure~\ref{fig:sim-efficiency-diff} shows the results when the autocorrelation
parameter $\rho = 0.6$. To begin with the absolute bias, visualized in the first row,
all estimators have little bias under the
extended parallel trends assumption (Scenario 1), as expected from theoretical results.
In contrast, under the
parallel-trends-in-trends assumption (Scenario 2), the extended DID (white circle
with dotted line) is biased, while the double DID (black circle
with solid line) and the sequential DID (white triangle with dotted line) are unbiased.

The second row represents the standard errors of each estimator. Under
the extended parallel trends assumption (the first column),
the double DID estimator has the smallest
standard error, smaller than the extended DID estimator (i.e., the
two-way fixed effects estimator). This efficiency gain comes from the
fact that the double DID uses
the GMM framework to optimally weight observations from different time
periods, although the two-way fixed effects estimator uses equal weights to all
pre-treatment periods.

Under the parallel trends-in-trends assumption
(the second row; the second column), the double DID has almost the same
standard error as the sequential DID. This shows that the double DID changes
weights according to scenarios and solves a practical dilemma of the sequential DID --- it is unbiased under the
weaker assumption of the parallel trends-in-trends, but not
efficient under the extended parallel trends.

\begin{figure}[!t]
  \centering
  \includegraphics[scale=0.6]{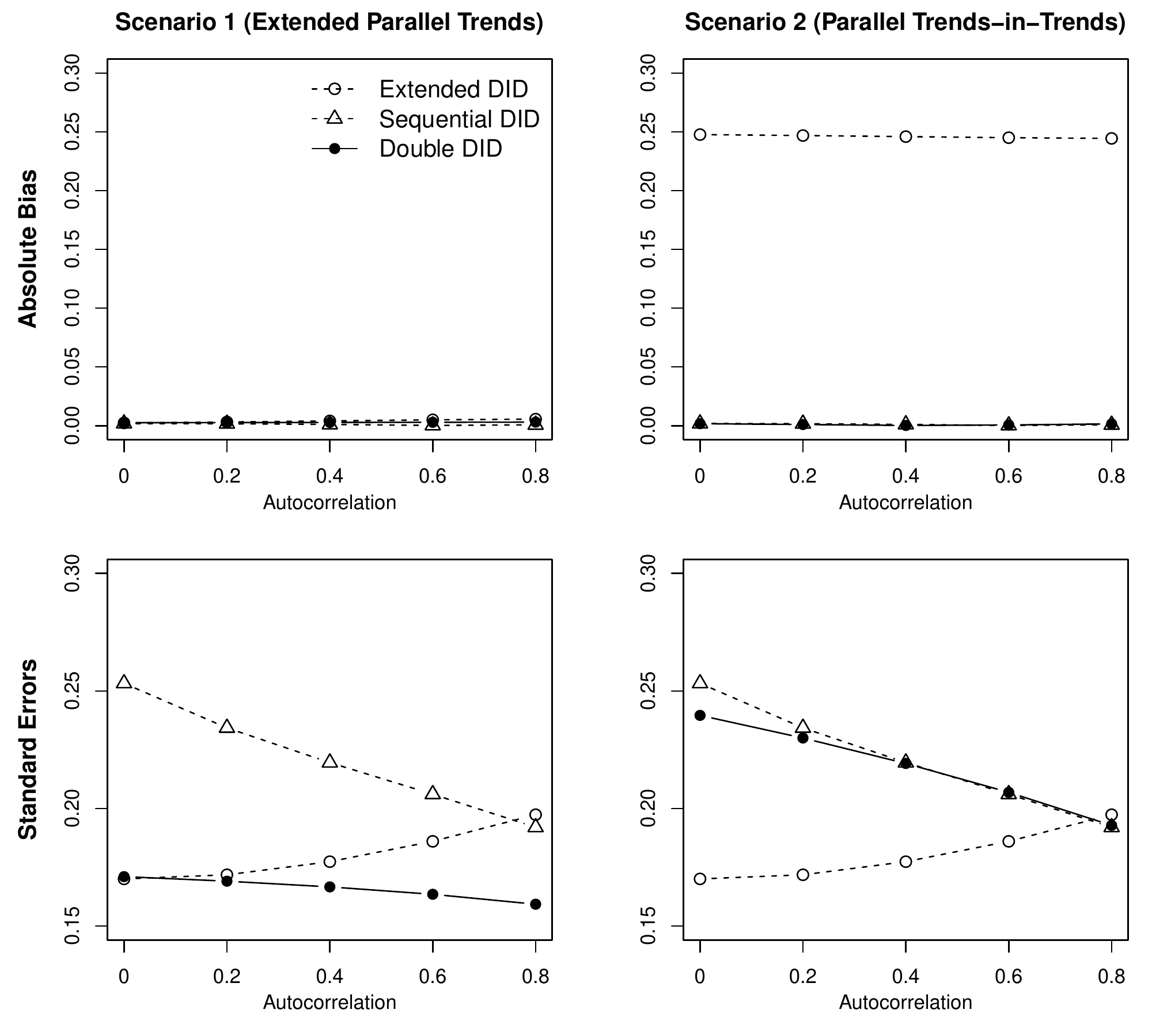}
  \spacingset{1}{\caption{Comparing DID estimators in terms of the absolute bias
      and the standard errors according to the autocorrelation of
      errors. \textit{Note}: The first row shows that the double DID
      estimator (black circle with solid line) is unbiased under both
      scenarios. The second row demonstrates that the double DID has
      the smallest standard errors among unbiased DID
      estimators. Under the extended parallel trends assumption (the
      first column), the efficiency gain relative to the extended DID (i.e., two-way fixed
      effects estimator) is large when the autocorrelation parameter
      $\rho$ is large. Under the parallel trends-in-trends assumption (the second column), the efficiency gain relative to the
      sequential DID is large when $\rho$ is small.}
  \label{fig:sim-add}}
\end{figure}

In Figure~\ref{fig:sim-add}, we provide additional results where we consider a
full range of the autocorrelation parameters $\rho \in \{0, 0.2, 0.4, 0.6,
0.8\}$ (the same positive autocorrelation values considered in
\cite{bertrand2004much}). We find that when
the autocorrelation of errors is small, standard errors of the double DID
are smaller than those of the sequential DID even under the parallel
trends-in-trends assumption.

The first row of Figure~\ref{fig:sim-add} shows that our results on the (absolute) bias do not
change regardless of the autocorrelation of errors. In particular, the double
DID is unbiased under the extended parallel trends assumption (the
first column) or under the parallel trends-in-trends assumption (the
second column).  In terms of the standard  errors (the second row), two results are important. First, under the
extended parallel trends assumption (the first column), the standard errors of
the double DID is the smallest for all the values of $\rho$ and the
efficiency gain relative to the extended DID (i.e., two-way fixed
effects estimator) is large when the there is high auto-correlations (i.e., $\rho$ is large).
Second, under the parallel trends-in-trends assumption (the second column), the standard errors of
the double DID is the smallest among unbiased DID estimators (the
extended DID is biased).  The efficiency gain relative to the
sequential DID is large when $\rho$ is small.

\clearpage
\section{Empirical Application}
\subsection{Malesky, Nguyen, and Tran (2014): DID Design}
\label{subsec:ap_app}
In Section~\ref{subsec:basic-app}, we have focused on three outcomes to illustrate the
advantage of the double DID estimator. Each outcome is defined as
follows. ``Education and Cultural Program'' (binary):
This variable takes one if there is a program that invests in culture and education in the commune. ``Tap Water''
(binary): What is the main source of drinking /cooking water for
most people in this commune? ``Agricultural
Center'' (binary): Is there any agriculture extension center in a
given commune? Please see \citet{malesky2014impact} for further details.

In this section, we provide results for all thirty outcomes analyzed
in the original paper. To assess the underlying parallel trends assumptions, we combine visualization and formal
tests, as recommended in the main text.
The assessment suggests that we can make the extended parallel trends assumption for fifteen outcomes.
Specifically, for those fifteen outcomes, p-values for the null of pre-treatment parallel trends are above 0.10 (i.e., fail to reject the null at the conventional level), and the 95\% standardize equivalence confidence interval is
contained in the interval $[-0.2, 0.2]$.
This means that the deviation from the parallel trends in the pre-treatment periods are less than 0.2 standard deviation of the control mean in 2006.

\begin{figure}[!t]
  \begin{center}
    \includegraphics[width = \textwidth]{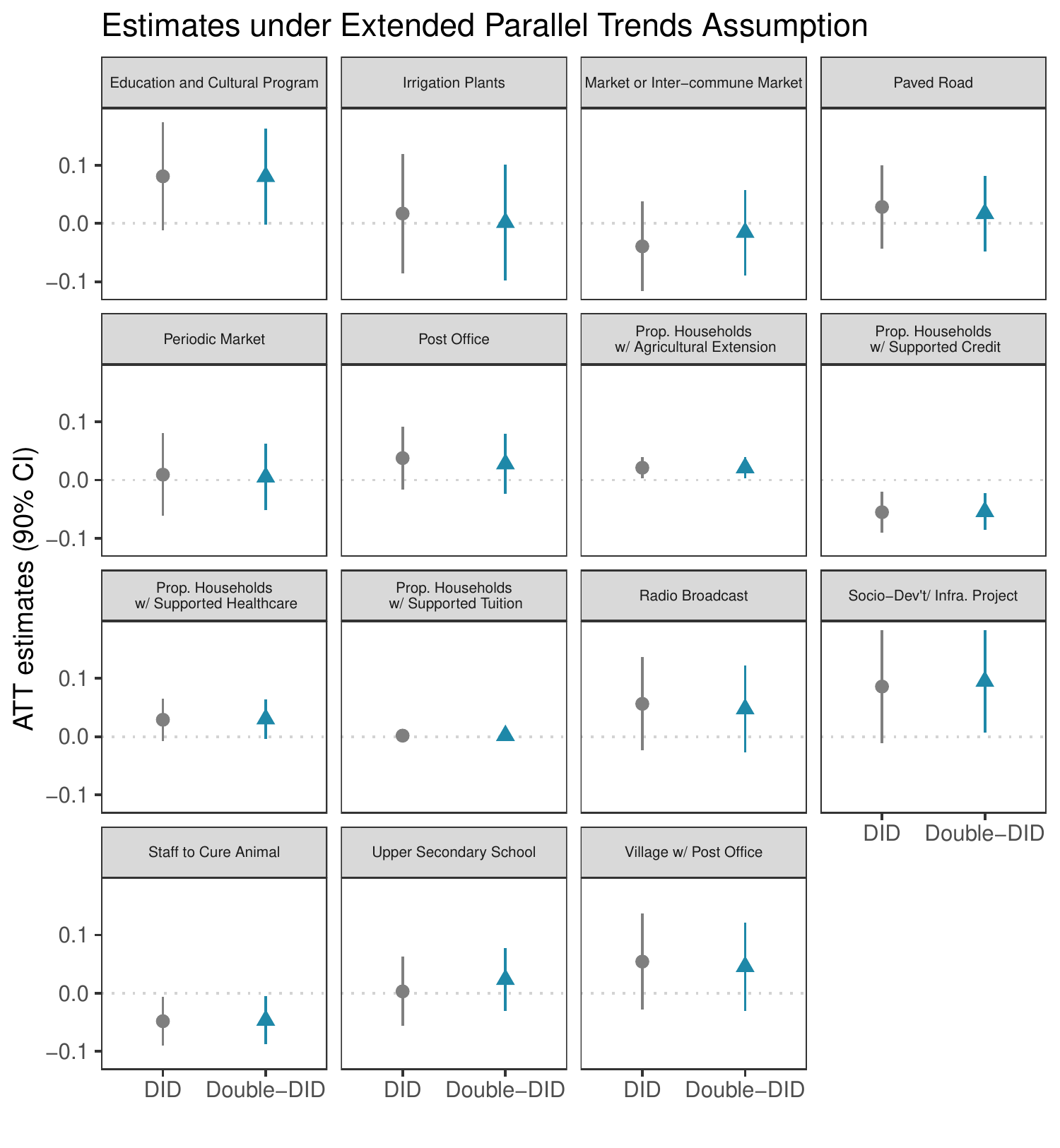}
  \end{center}
  \vspace{-0.25in} \spacingset{1}{\caption{
      Comparing Standard DID and Double DID under Extended
      Parallel Trends Assumption. The double DID estimates are
      similar to those from the standard DID, and yet, standard errors are
      smaller because the double DID effectively uses pre-treatment periods
      within the GMM.}  \label{appx_fig:parallel-trends}
      }
\end{figure}

Figure~\ref{appx_fig:parallel-trends} shows estimated treatment effects under the extended parallel trends assumption.
As in Section~\ref{subsec:basic-app}, the double DID estimates are
similar to those from the standard DID, and yet, standard errors are
smaller because the double DID effectively uses pre-treatment periods
within the GMM. Here, we only have two pre-treatment periods, but when
there are more pre-treatment periods, the efficiency gain of the
double DID can be even larger.

We rely on the parallel trends-in-trends assumption for eight outcomes out of the fifteen remaining outcomes.
These outcomes have the 95\% standardized equivalence confidence interval wider than $[-0.20, 0.20]$,
but show that treatment and control groups' pre-treatment trends have the same
sign.
The same sign of the pre-treatment trends suggests that parallel trends-in-trends assumption, which can account for the linear time-varying unmeasured confounder, can be plausible for these outcomes,
even though the stronger parallel trends assumption is possibly violated.

\begin{figure}[!t]
  \begin{center}
    \centerline{
      \includegraphics[scale=1.1]{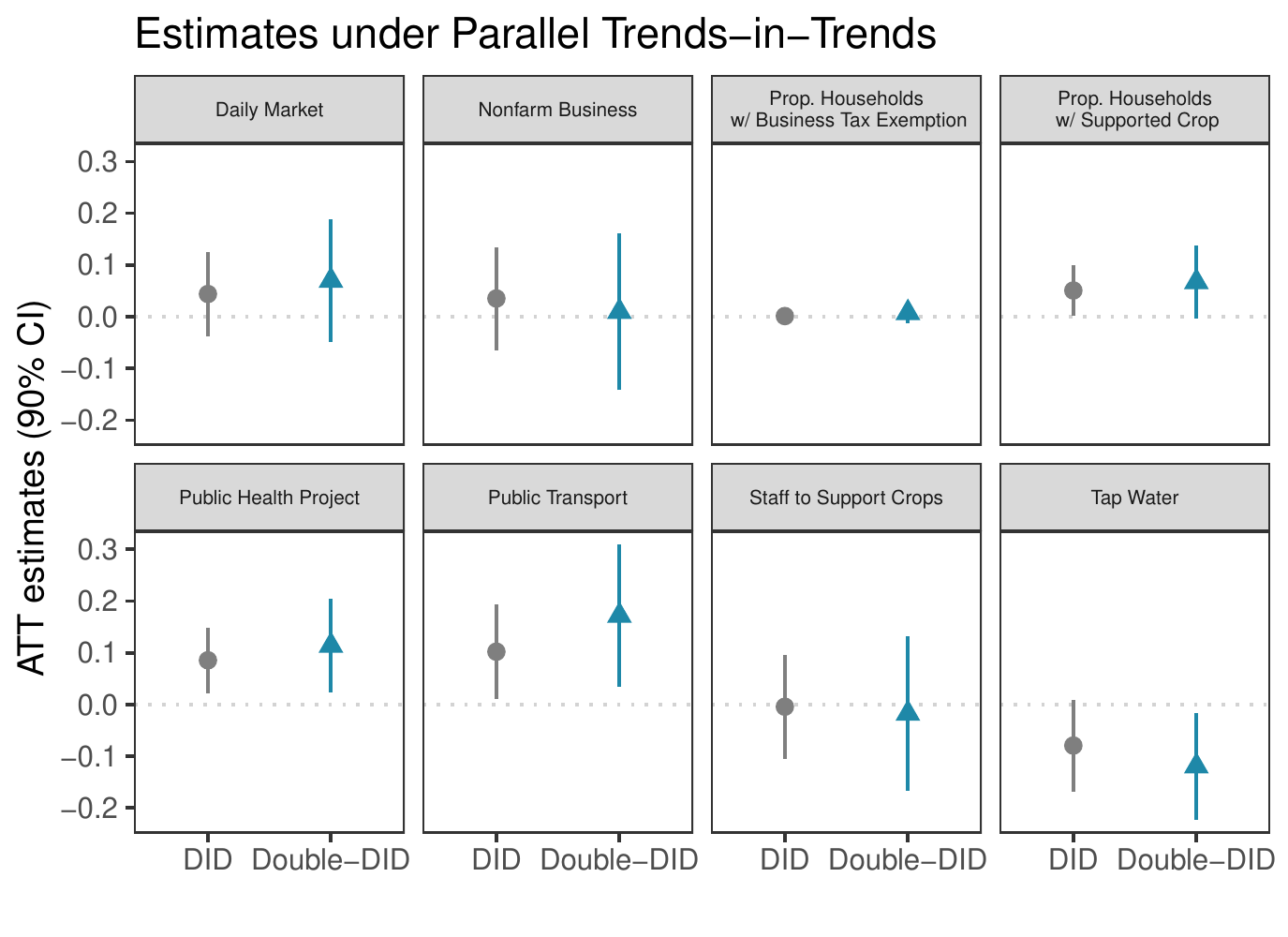}
    }
  \end{center}
  \vspace{-0.25in} \spacingset{1}{\caption{
      Comparing Standard DID and Double DID under
      Parallel Trends-in-Trends Assumption.
      The double DID estimates are often different from those of the standard DID because the extended parallel trends assumption is implausible for these outcomes.}\label{appx_fig:ptt}
      }
\end{figure}

Figure~\ref{appx_fig:ptt} shows results
under the parallel trends-in-trends assumption. As in Section~\ref{subsec:basic-app}, the double DID estimates are often  different from those of the standard DID because
the extended parallel trends assumption is implausible for these
outcomes. Importantly, standard errors of the double DID are often
larger than the standard DID.
This is because the double DID needs to
adjust for biases in the standard DID by using pre-treatment trends.

For the remaining seven outcomes of which
treatment and control groups' pre-treatment trends have the opposite
sign, it is difficult to justify either the extended parallel trends
or parallel trends-in-trends assumption without additional
information. Thus, there is no credible estimator for the ATT without
making stronger assumptions. When there are more than two
pre-treatment periods, researchers can apply the sequential DID
estimator to pre-treatment periods in order to formally assess the
extended parallel trends-in-trends assumption. We emphasize that, although
we use the equivalence range of $[-0.20, 0.20]$ as a cutoff for an
illustration, it is recommended to base this decision on substantive
domain knowledge whenever possible in practice.

\subsection{Paglayan (2019): Staggered Adoption Design}
\label{subsec:sa-app}
In this section, we apply the proposed double DID estimator to revisit
\cite{paglayan2019}, which uses the staggered adoption (SA) design to study
the effect of granting collective bargaining rights to teacher's union
on educational expenditures and teacher's salary.
\cite{paglayan2019} applies the standard two-way fixed effect models to estimate the effect of
the introduction of the mandatory  bargaining law in the US states
on the two outcome. The original author exploits the variation induced by the different introduction timing of the law:
A few states introduced the law as early as in the mid 1960's, while some
states, such as Arizona or Kentucky, never introduced the mandate.
Among the states that granted the bargaining rights, the introduction
timing varies from the mid 1960's to the mid 1980's (Nebraska was the last
state that adopted the law).

\subsubsection{Assessing Underlying Assumptions}

We apply the proposed double DID for the SA design to the panel data
consists of state-year observations.
A state is treated at a particular year, if the state passes the law or has already passed the law
of mandatory bargaining.
Following the original study, we study two outcome: Per-pupil expenditure and annual teacher salary, both are on a log scale.
There are 2,058 observations, containing 49 states (excluding Washington DC and Wisconsin, due to the short availability of the pre-treatment outcomes) and spanning from 1959 through 2000.

\begin{figure}[!t]
  \centering
  \includegraphics[width=0.8\textwidth]{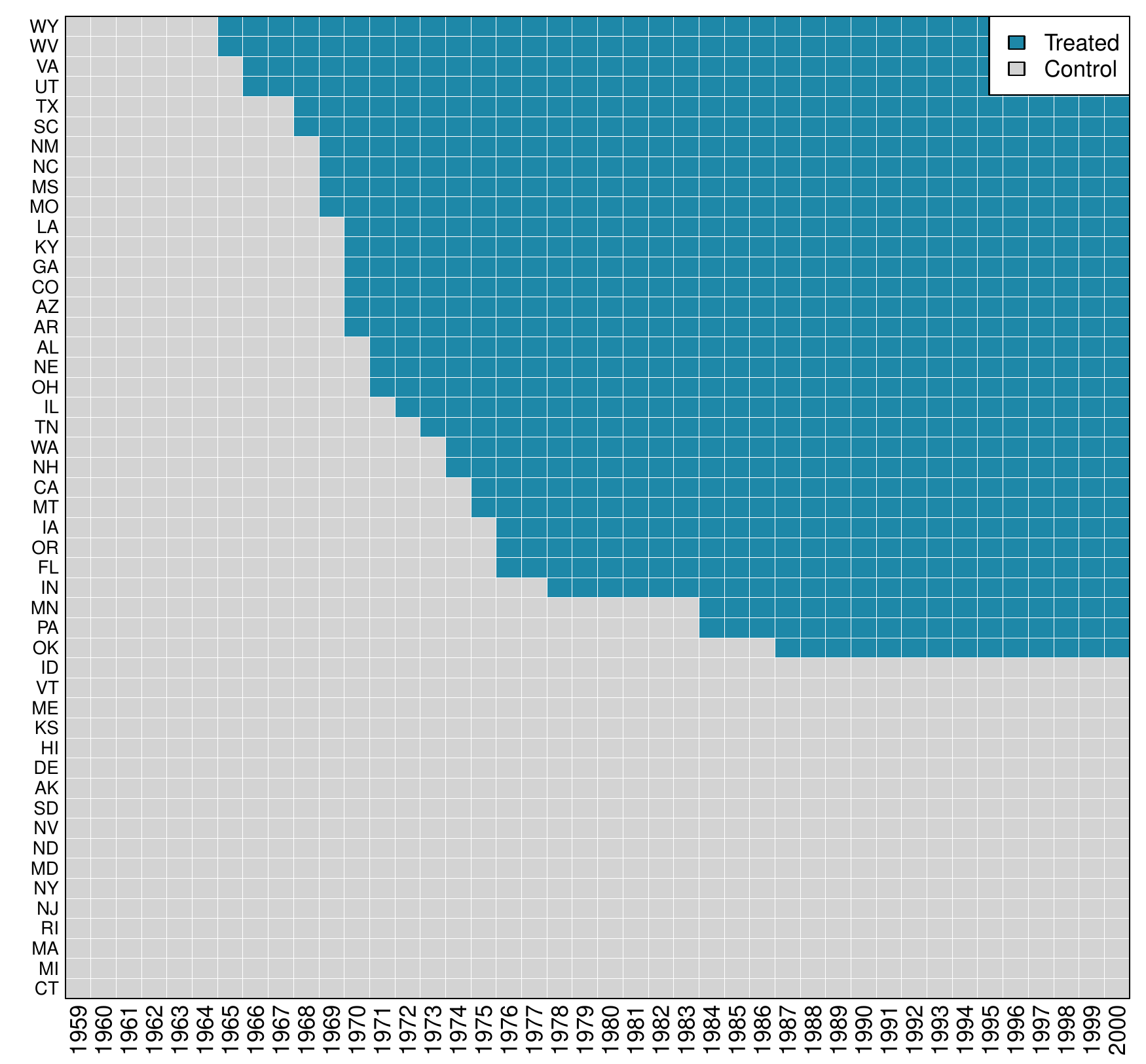}
  \spacingset{1}{
    \caption{Treatment Variation Plot. \textit{Note:} Cells in gray are
      state-year observations that are not treated (i.e., the
      mandatory bargaining law is not implemented), while cells in
      blue are observations that are under the treatment
      condition. Rows are sorted such that states that adopt the
      policy at earlier years are shown near the top, while states
      that never adopt the policy are shown near the bottom.
      The figure indicates that there are variations across states in
      adoption timings, and that some states never adopt the
      policy.}\label{fig:paglayan_variation}
  }
\end{figure}

Figure~\ref{fig:paglayan_variation} shows the variation of the treatment across states and over time.
Cells in gray indicate state-year observations that are not treated and blue cells indicate the treated observations.
We can observe that there are 14 unique treatment timings (the earliest is 1965 and the latest is 1987)
where the number of states at  each treatment timing  varies from one
to six (the average number of states at a treatment timing is 2.3). We
can also see that there is no reversal of a treatment status in that
once a state adopts the policy, the state has never abolished it during the sample period.

\begin{figure}[!t]
  \centering
  \centerline{\includegraphics[scale=0.8]{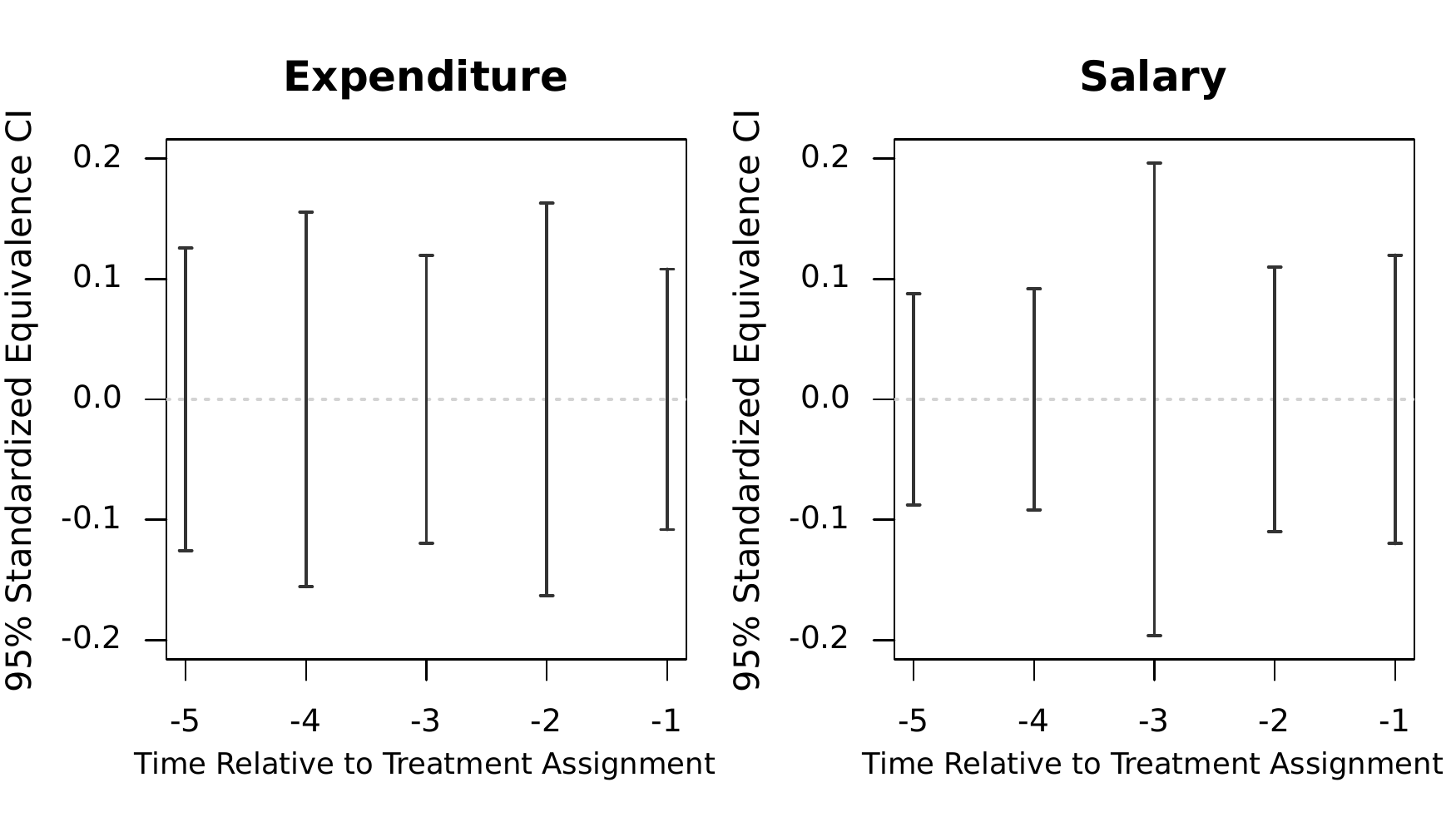}}
  \vspace{-0.25in} \spacingset{1}{
    \caption{Assessing Underlying Assumptions Using the Pre-treatment
      Outcomes (Left: logged expenditure; Right: logged teacher
      salary). \textit{Note:} We report the 95\% standardized equivalence confidence
      intervals.}
    \label{fig:paglayan_pretest}
  }
\end{figure}

We assess the underlying parallel trends assumption for the SA design
by utilizing the pre-treatment outcome. As in the pre-treatment-trends
test in the basic DID design, we apply the standard DID estimator for
the SA design to pre-treatment periods.
For example, to test the pre-treatment trends from $t - 1$ to $t$ for
units who receive the treatment at time $t$, we estimate the SA-ATT
using the outcome from $t-2$ and $t-1$ (See
Section~\ref{subsec:sa-d-did} for more details). To further facilitate
interpretation, we standardize the outcome by the mean and standard
deviation of the baseline control group, so that the effect can be
interpreted relative to the control group.

Figure~\ref{fig:paglayan_pretest} shows 95\% standardized equivalence confidence intervals
for the two outcomes of interest (See Section~\ref{subsec:b-d-did} for details on the standardization procedure).
It shows that for both outcomes, the equivalence confidence intervals
are within 0.2 standard deviation from the means of the baseline control
groups through $t - 5$ to $t - 1$. This suggests that the extended
parallel trends assumption is plausible for both outcomes.

\subsubsection{Estimating Causal Effects}
We apply the double DID for the SA design as described in Section~\ref{sec:SAD}.
The standard errors are computed by conducting the block bootstrap where the block is taken at the state level
and we take 2000 bootstrap iterations.
Analyses for the two outcomes are conducted separately.
In addition to the proposed method, we apply two existing variants of
synthetic control methods that can handle the staggered
adoption design: the generalized synthetic control method,
\texttt{gsynth} \citep{xu2017generalized}, and the augmented synthetic
control method, \texttt{augsynth} \citep{ben2019synthetic}. While the proposed double DID is better
suited for settings where there are a small to moderate number of pre-treatment
periods, we evaluate, in the setting of long pre-treatment periods, whether it can achieve comparable
performance to these variants of synthetic control methods that are
primarily designed to deal with long pre-treatment periods (see more
discussions in Section~\ref{subsec:scm}).

\begin{figure}[!t]
  \centerline{\includegraphics[scale=0.8]{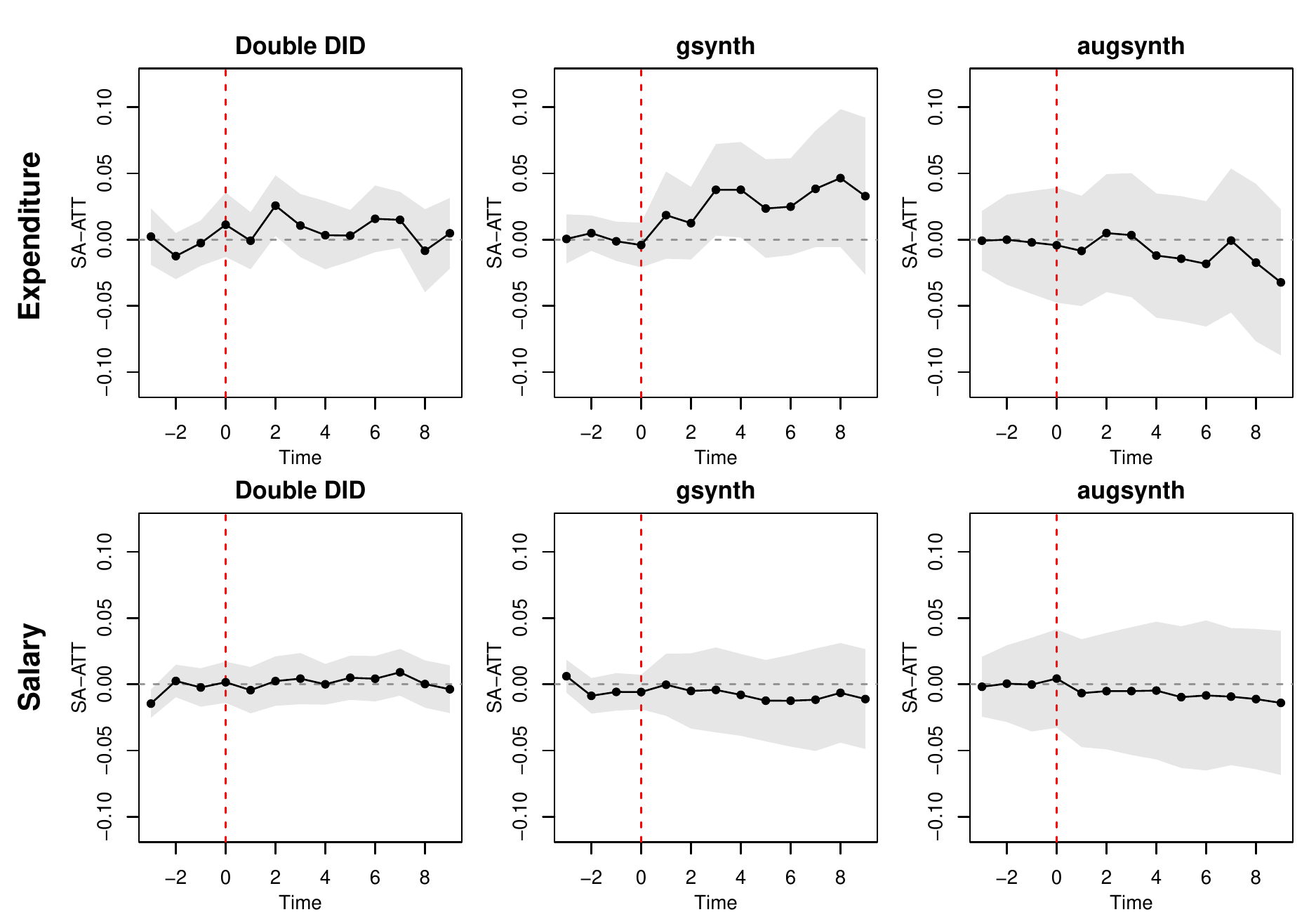}}
  \spacingset{1}{
  \caption{Plot of the Average Treatment Effect on the Treated on Two
    Outcomes. \textit{Note:} We compare estimates from the double DID,
    the generalized synthetic control method, and the augmented
    synthetic control method. The causal estimates are similar across
    methods for both outcomes and treatment effects are not
    statistically significant at the conventional 5\% level for most of the time periods.}
  \label{fig:paglayan_effects}
}
\end{figure}

Figure~\ref{fig:paglayan_effects} shows the estimates of the treatment on the per-pupil expenditure (the first row)
and the teacher's salary (the second row), where both effects are on a
log scale. We estimated the average treatment effect on the two
outcomes $\ell$ periods after the treatment assignment where $\ell =
\{0, 1, \ldots, 9\}.$ Note that $\ell = 0$ corresponds to the
contemporaneous effect. Each column corresponds to different estimators.
The first column shows the proposed double DID estimator for the staggered adoption design,
whereas the second (third) column shows estimates based on the generalized synthetic control method (the augmented synthetic control method).
We can see that estimates are similar across methods for both outcomes
and treatment effects are not statistically significant at the 5\% level for most of the time periods.
This result is consistent with the original finding of
\cite{paglayan2019} that the granting collective bargaining rights did
not increase the level of resources devoted to education.

As in this example, when there are a large number of pre-treatment
periods, it is important to apply both synthetic control methods and the proposed
double DID, and evaluate robustness across those approaches. This is
critical because they rely on different identification
assumptions. We found such robustness in this application, which
provides us with additional credibility.

\end{document}